\documentclass[aps,prb,superscriptaddress,twocolumn,longbibliography]{revtex4-2}

\usepackage[utf8]{inputenc}
\usepackage{amssymb}
\usepackage{amsmath} 
\usepackage{graphicx}
\usepackage{amsmath}  
\usepackage{mathtools}
\usepackage{array,multirow}
\usepackage{longtable}
\newcommand{\PreserveBackslash}[1]{\let\temp=\\#1\let\\=\temp}
\newcolumntype{C}[1]{>{\PreserveBackslash\centering}p{#1}} 
\usepackage{physics}
\usepackage{tikz}
\usepackage{xcolor}
\usepackage{hyperref}
\usepackage{float}
\advance\textwidth by .5cm
\advance\evensidemargin by -.33cm
\advance\oddsidemargin by -.33cm

\begin{document}

\title{Role of Fock-space correlations in many-body localization}

\author{Thibault Scoquart}
\email{thibault.scoquart@kit.edu}

\affiliation{\mbox{Institute for Quantum Materials and Technologies, Karlsruhe Institute of Technology, 76131 Karlsruhe, Germany}}
\affiliation{\mbox{Institut f\"ur Theorie der Kondensierten Materie, Karlsruhe Institute of Technology, 76131 Karlsruhe, Germany}}

\author{Igor V.~Gornyi}

\affiliation{\mbox{Institute for Quantum Materials and Technologies, Karlsruhe Institute of Technology, 76131 Karlsruhe, Germany}}
\affiliation{\mbox{Institut f\"ur Theorie der Kondensierten Materie, Karlsruhe Institute of Technology, 76131 Karlsruhe, Germany}}

\author{Alexander D.~Mirlin}

\affiliation{\mbox{Institute for Quantum Materials and Technologies, Karlsruhe Institute of Technology, 76131 Karlsruhe, Germany}}
\affiliation{\mbox{Institut f\"ur Theorie der Kondensierten Materie, Karlsruhe Institute of Technology, 76131 Karlsruhe, Germany}}

\begin{abstract}

Models of many-body localization (MBL) can be represented as tight-binding models in the many-body Hilbert space (Fock space). We explore the role of correlations between matrix elements of the effective Fock-space Hamiltonians in the scaling of MBL critical disorder $W_c(n)$ with the size $n$ of the system. For this purpose, we consider five models, which all have the same distributions of diagonal (energy) and off-diagonal (``hopping'') Fock-space matrix elements but different Fock-space correlations. These include quantum-dot (QD) and one-dimensional (1D) MBL models, their modifications (uQD and u1D models) with removed correlations of off-diagonal matrix elements, as well a quantum random energy model (QREM) with no correlations at all. Our numerical results are in full consistency with analytical arguments predicting 
$n^{3/4} (\ln n)^{-1/4} \lesssim W_c \lesssim n \ln n$
for the scaling of $W_c(n)$ in the QD model (we find $W_c \sim n$ numerically), $W_c(n) \sim  \text{const}$ for the 1D model, $W_c \sim n \ln n$ for the uQD and u1D models without off-diagonal correlations, and $W_c \sim n^{1/2} \ln n$ for QREM.  The key difference between the QD and 1D models is in the structure of correlations of many-body energies.  Removing off-diagonal Fock-space correlations makes both these models ``maximally chaotic''. Our findings demonstrate that the scaling of $W_c(n)$ for MBL transitions is governed by a combined effect of Fock-space correlations of diagonal and off-diagonal matrix elements.

\end{abstract} 

\maketitle



\section{Introduction}
\label{sec:intro}

The problem of many-body localization (MBL)  addresses quantum localization in interacting disordered systems far from the ground state (i.e., at finite energy density) \cite{gornyi2005interacting,basko2006metal}. This problem can be viewed as a many-body extension of the famous Anderson-localization problem \cite{anderson58}. In the single-particle setting, localization-delocalization phase transitions (Anderson transitions) are characterized by remarkably rich physics, which has been explored by analytical and numerical means \cite{evers08}. For the MBL problem, fully controllable analytical and numerical investigations represent highly challenging tasks. While great progress has been achieved in understanding the MBL physics (see reviews \cite{nandkishore15,Alet2018a,abanin2019colloquium,gopalakrishnan2020dynamics,tikhonov2021from,doggen2021many,Sierant2024}), many important aspects remain a subject of active current research. 

The Hamiltonian of an MBL model can be equivalently represented as a tight-binding model in the many-body Hilbert space, which we will term for brevity ``Fock space''. (For spin-1/2 models that we consider, the many-body Hilbert space is in one-to-one correspondence with that of fermions or hard-core bosons, thus justifying the ``Fock space'' terminology.) In such a representation, site energies correspond to energies of many-body basis states, while hopping matrix elements are amplitudes of transitions between these states. 
The Fock-space view on the MBL problem is highly instructive since it is closely related to one of the fundamental properties of the MBL phase---breakdown of {\color{black} (quantum)} ergodicity {\color{black} \cite{Monteiro2021}, \footnote{{\color{black} It is worth noting that there are different definitions of ergodicity in various fields of classical and quantum physics. In particular, the definition of quantum (non-)ergodicity in the context of MBL  (see a recent discussion in Ref.~\cite{Monteiro2021}) differs from that traditionally adopted in the field of spin glasses, where a system can be called non-ergodic once full equilibration takes a time exponentially long in system size $n$. Indeed, this time can still be much shorter than the Heisenberg time (also exponential in $n$), in which case the system will satisfy the condition of quantum ergodicity that we adopt.  What we call ``ergodicity'' reflects the full mixture of many-body states within an energy shell containing many states (cf. microcanonical ensemble of statistical mechanics and the eigenstate thermalization hypothesis). For a recent review of connections between quantum spin glasses and MBL, see Ref.~\cite{Cugliandolo}}}}. This can be explored by studying Fock-space observables, i.e., those related to eigenvalues and eigenstates of the many-body Hamiltonian. In the ergodic phase, eigenstates hybridize within an energy shell that contains a very large number of states. This implies, in particular,  the scaling of inverse participation ratio (IPR) corresponding to spreading of many-body eigenstates over the basis states within the energy shell as well as the Wigner-Dyson level statistics. In contrast, on the MBL side of the transition, the hybridization is typically strongly suppressed, even for states that are adjacent in energy space, which is reflected in the IPR scaling and in the Poisson level statistics. 
Within the Fock-space approach, the concept of transition between ergodicity and MBL is also applicable to many-body quantum-dot models
\cite{altshuler1997quasiparticle,jacquod1997emergence,mirlin1997localization,silvestrov1997decay,silvestrov1998chaos,gornyi2016many,gornyi2017spectral,jacquod1997emergence, georgeot1997breit, leyronas2000scaling, shepelyansky2001quantum,  PhysRevE.62.R7575, jacquod2001duality, rivas2002numerical,bulchandani2022onset,garcia-garcia2018chaotic,micklitz2019nonergodic,monteiro2020minimal,Monteiro2021,nandy2022delayed,larzul2022quenches,Herre2023}
that do not exhibit real-space localization. 
The Fock-space approach  (including the analysis of properties of many-body eigenstates, matrix elements, and resonances) has proven to be very useful for theoretical investigation of the physics around MBL transitions \cite{serbyn2017thouless,tikhonov18,mace19multifractal,tikhonov2021from,tikhonov2021eigenstate,nag2019many-body,tarzia2020many,roy2021fock,crowley2022constructive,bulchandani2022onset,creed2023probability,Ghosh2024, Sutradhar2022, Roy2023, DeTomasi2021, Sierant2024}.
Furthermore, there is remarkable progress in experimental studies of  Fock-space dynamics and of statistics of many-body energies in systems of coupled qubits across the MBL transition	\cite{smith2016many-body,roushan2017spectroscopic,xu2018emulating,lukin2019probing,Yao2023}.

The Fock-space representation of the MBL models bears analogies to Anderson localization on random regular graphs (RRG), see Ref.~\cite{tikhonov2021from} for a recent review of the RRG model and its relations to the MBL. The Anderson localization on RRG (and in some variations of this model) was studied in a number of works \cite{biroli2012difference,de2014anderson,tikhonov2016anderson,garcia-mata17,metz2017level,Biroli2017,kravtsov2018non,biroli2018,PhysRevB.98.134205,tikhonov19statistics,tikhonov19critical,PhysRevResearch.2.012020,Roy2020numerics,tikhonov2021eigenstate,garcia-mata2022critical,sierant2023universality,valba2022mobility} (see also earlier studies of a related sparse random matrix model \cite{mirlin1991universality,fyodorov1991localization}). Of special interest in the MBL context is the Anderson transition on RRG with a large coordination number, which has been explored analytically and numerically in the recent paper \cite{Herre2023}. 
 
While the analysis of Anderson localization on RRG has been very instructive for understanding the MBL physics, the actual Fock-space structure of a many-body Hamiltonian is more involved than that of the RRG model. Specifically, Fock-space matrix elements in MBL problems necessarily exhibit strong correlations, since they are built of a much smaller number of couplings entering the second-quantized Hamiltonian. Importance of these correlations was emphasized in particular in Refs.~\cite{gornyi2017spectral,ghosh2019many-body,Roy2020}. 

The goal of this paper is to explore the role of Fock-space correlations in the scaling of the critical disorder $W_c(n)$ of MBL transitions with the system size $n$. Strictly speaking, for a finite $n$, the transition from ergodicity to MBL with increasing disorder $W$ is not fully sharp, i.e., it happens in a window $\Delta W(n)$ (and thus may be termed a crossover) around a certain disorder $W_c(n)$, in analogy with conventional continuous (second-order) phase transitions. Clearly, using different observables (e.g., those related to level statistics or to eigenfunctions statistics) may lead to slightly different values of $W_c(n)$ within the $\Delta W(n)$ window. However, with increasing $n$, the relative width of this window shrinks, $\Delta W(n) / W_c(n) \to 0$ at $n \to \infty$, so that one can speak about a well-defined phase transition at $W_c(n)$ in the large-$n$ limit. Importantly, for many models one finds a very non-trivial scaling of the critical disorder $W_c$ with $n$; see, e.g., Refs.~\cite{tikhonov18,Gopalakrishnan2019a}.  Understanding this scaling---and the underlying mechanisms---in various settings is of crucial importance for understanding the physics of the evolution from ergodicity to MBL. 

\begin{figure}[t!]
\centering
\includegraphics[width = \columnwidth]{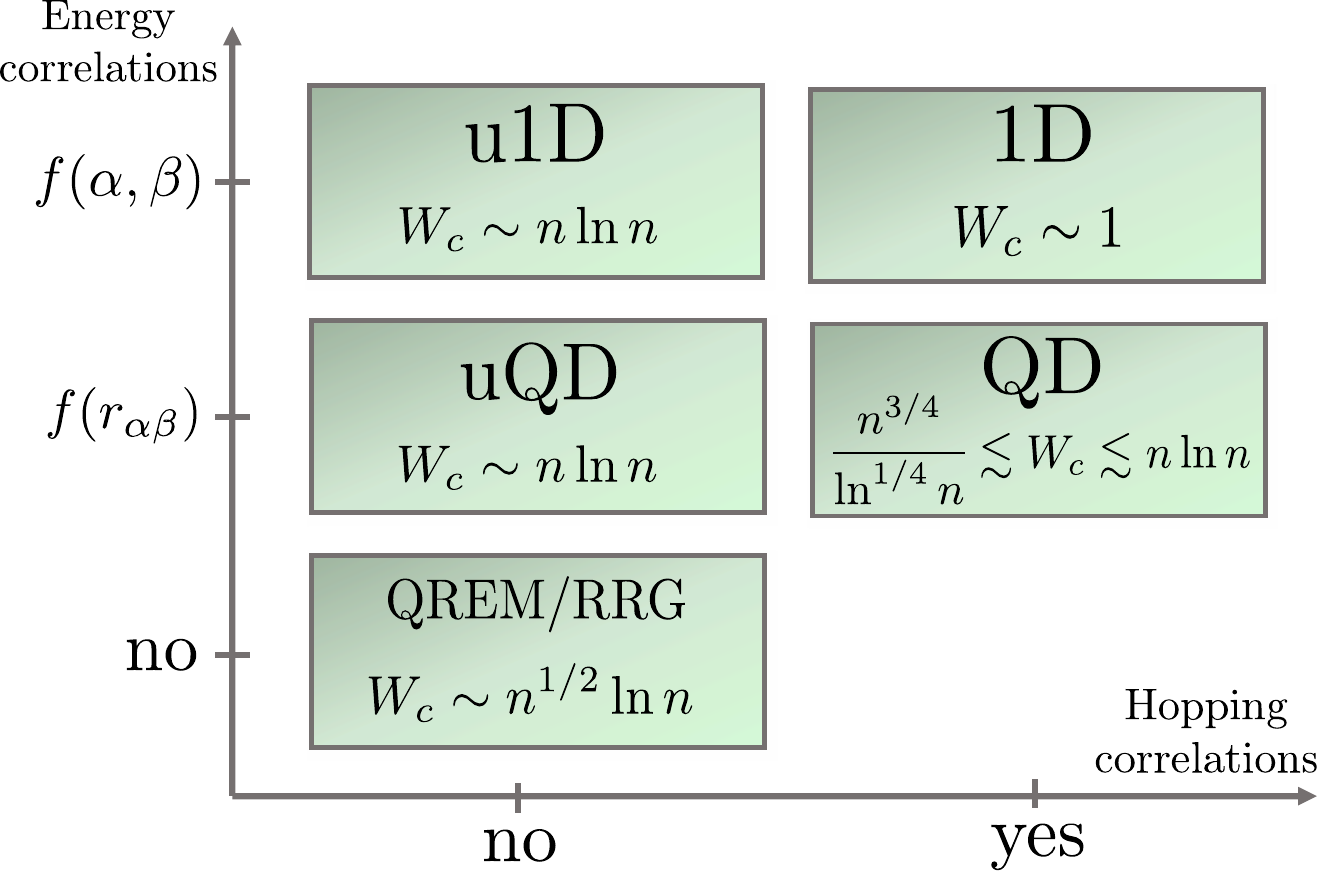}
\caption{Summary of scaling of the critical disorder $W_c(n)$ of the MBL transition for models considered in this article.  Presence and character of energy and hopping correlations {\color{black} (i.e., correlations in diagonal and off-diagonal matrix elements, respectively)}
in the Fock-space representation of the models are indicated.
{\color{black} The labels $f(r_{\alpha\beta})$ and $f(\alpha,\beta)$ on the energy-correlation axis refer, respectively, to correlations that depend only on the Hamming distance $r_{\alpha\beta}$ between Fock-space states $\alpha$ and $\beta$ and to correlations with a more complex dependence on $\alpha$ and $\beta$ (reflecting 1D spatial structure), see Sec.~\ref{sec:model_definitions} for details.}
 The analytical results for $W_c(n)$ shown in the figure (see Sec.~\ref{sec:analytical}) are supported by numerical simulations in Sections \ref{sec:numerics} and \ref{sec:numerics-models-with-corr}. For the QD model, the numerical results suggest  $W_c(n) \sim n$, in consistency with analytical bounds shown in the figure.}
\label{fig:Figure_merit}
\end{figure}

We study numerically (by exact diagonalization) five Fock-space models {\color{black} (explicitly defined in Sec.~\ref{sec:model_definitions}),} which all share the same distributions of energies and hopping matrix elements but with distinct correlation properties. Two of these models are Fock-space representations of MBL problems of two extreme geometries: a quantum dot (QD) and a one-dimensional (1D) chain. As we discuss in detail below, the crucial difference between the Fock-space representations of these two models is in correlations of diagonal matrix elements. 
Further two models---which we term ``uncorrelated quantum dot (uQD) model'' and ``uncorrelated 1D (u1D) model''---are obtained from these models by removing correlations between off-diagonal elements. Finally, we also consider a model without any correlations of matrix elements, which is a version of the quantum random energy model (QREM){\color{black}~\cite{Goldschmidt1990,laumann2014many-body,baldwin2016the_many-body}.} Exploring $W_c(n)$ in all five models, we obtain numerical results that are in full agreement with the corresponding  analytical arguments, predicting  
$n^{3/4} (\ln n)^{-1/4} \lesssim W_c \lesssim n \ln n$ for the scaling of $W_c(n)$ in the QD model (we find $W_c \sim n$ numerically), $W_c(n) \sim  \text{const}$ for the 1D model, $W_c \sim n \ln n$ for the uQD and u1D models, and $W_c \sim n^{1/2} \ln n$ for QREM,
see Fig.~\ref{fig:Figure_merit}.
Our findings provide a comprehensive picture of how Fock-space correlations of diagonal and off-diagonal matrix elements jointly govern the scaling of MBL transitions. 

We also analyze the scaling of the transition width and find numerically $\Delta W(n) / W_c(n) \sim n^{-\mu}$ with $\mu \approx 0.95 - 1.3$ for all five models. Our analytical results for QREM, uQD, and u1D models show that this is indeed the expected behavior in the range of $n$ accessible to exact diagonalization. At the same time, this is not the asymptotic large-$n$ behavior of the width that we find to be $\Delta W(n) / W_c(n) \sim n^{-3}\ln^2 n$ for these models and which is applicable to larger systems, $n>22$. 

The structure of the paper is as follows. 
In Sec.~\ref{sec:model_definitions} we define the models and analyze the Fock-space correlations of the corresponding matrix elements. In Sec.~\ref{sec:analytical} we present analytical arguments for the scaling of critical disorder in these models. Our numerical approach is explained in Sec.~\ref{sec:numerics}, where we also apply it to the model without any Fock-space correlations (i.e., QREM).
Numerical results for four models with different types of Fock-space correlations (QD, 1D, uQD, and u1D) are then presented and analyzed in Sec.~\ref{sec:numerics-models-with-corr}. Section \ref{sec:summary} contains a summary of our results, along with a discussion of prospects for future research. Some technical details are shifted to Appendices.

\section{Models and associated Fock space correlations}
\label{sec:model_definitions}

\subsection{Fock space graph representation}
\label{sec:Fock-space-rep}

In this paper, we focus on spin-1/2 models. For $n$ spins, the many-body Hilbert space (the Fock space) has a dimension $2^n$. We will use states $\ket{\alpha}$ that are eigenstates of all $\hat{S}_i^z$ operators ($i=1, \ldots, n$) as a basis of this space. The considered models have a Fock-space representation of the form
\begin{align}
\hat{H} &= \hat{H}_0 + \hat{H}_1\nonumber\\
&=  \sum\limits_{\alpha = 1}^{2^n} E_\alpha \ket{\alpha}\bra{\alpha}  + \sum\limits_{\substack{\alpha,\beta = 1\\ \alpha\neq\beta}}^{2^n} T_{\alpha\beta} \ket{\alpha}\bra{\beta} \,.
\label{eq:Fock_space_rep}
\end{align}
The basis states $\ket{\alpha}$, which can be presented as spin strings of the type $\ket{\uparrow, \downarrow, \dots, \uparrow}$, are eigenstates of the part $\hat{H}_0$ of the Hamiltonian, $\hat{H}_0 \ket{\alpha} = E_\alpha \ket{\alpha}$, with many-body energies $E_\alpha$.
The MBL transition in the models under consideration is driven by the strength $W$ of the random Zeeman field in $z$ direction, so that the basis states $\ket{\alpha}$ become also eigenstates of the full Hamiltonian $\hat{H}$ in the limit $W\to \infty$.  Further, the part $\hat{H}_1$ of the Hamiltonian includes terms involving spin-flip processes, with 
$T_{\alpha\beta} \equiv \bra{\alpha}\hat{H}_1\ket{\beta}$ being a transition matrix element between spin configurations $\alpha$ and $\beta$. The Hamiltonian $\hat{H}$ equivalently describes a tight-binding model for a fictitious single particle living on the Fock space of our model of interest, with on-site energies $E_\alpha$ and hopping amplitudes $T_{\alpha\beta}$. 

{\color{black} Geometrically, the Fock space of the models under consideration can be viewed as formed by $N=2^n$ vertices of an $n$-dimensional hypercube. 
Thus, generically, models under consideration can be 
viewed as single-particle localization problems on graphs that have the form of a hypercube in $n$ dimensions. The $2^n$ nodes of the graph 
(which are at the vertices of the hypercube)
are characterized by on-site energies $E_\alpha$, and the hopping amplitudes $T_{\alpha\beta}$ are associated with graph edges. We define the Hamming distance $r_{\alpha\beta}$ between states $\alpha$ and $\beta$
as the number of differing spins in these states.
The Hamiltonians $\hat{H}$ that we will consider in this paper will include only single-spin-flip processes, so that edges of the graph will be identical to edges of the hypercube. Correspondingly, the Hamming distance $r_{\alpha\beta}$  will be equal to the length of the shortest path (number of edges of the graph) linking these two nodes on the graph.}


Since we are interested in disordered models, the energies $E_\alpha$ and the amplitudes $T_{\alpha\beta}$ are random variables, and we describe the physics by observables averaged over the corresponding ensemble of disorder realizations. Thus, for a given spin model, the mapping to the Fock-space representation (\ref{eq:Fock_space_rep}) is completely determined by specifying the joint distribution of the set of random variables $\{E_\alpha\}$ and $\{T_{\alpha\beta}\}$. This multivariate distribution, which includes information about fluctuations and correlations of matrix elements of the Fock-space Hamiltonian, thus fully controls the physics of the system. For the models that we consider here, the distributions are 
 multivariate Gaussian and are fully determined by two covariance matrices
\begin{align}
(C_{E})_{\alpha\beta} &\equiv  \langle E_\alpha E_\beta\rangle \,, \\
(C_{T})_{\alpha\beta\mu\nu} &\equiv  \langle T_{\alpha\beta}^* T_{\mu\nu} \rangle.
\label{eq:covariance_matrices_definition}
\end{align}

All models that we study in this paper are characterized by the same values of variances  $(C_{E})_{\alpha\alpha}$
and $(C_{T})_{\alpha\beta\alpha\beta}$ of energies $E_\alpha$ and ``hopping'' matrix elements $T_{\alpha\beta}$. At the same time, they differ by the form of covariances $(C_{E})_{\alpha\beta}$ (for $\alpha \ne \beta$) and  $(C_{T})_{\alpha\beta\mu\nu}$ (for 
$\{\alpha,\beta\} \ne \{\mu,\nu\}$). This allows us to explore implications of correlations for the scaling of the MBL transition. More specifically:

\begin{itemize}
\item  Both genuine many-body models that we consider (quantum dot and 1D) have non-trivial Fock-space energy correlations $(C_{E})_{\alpha\beta}$. A crucial difference between them is that for the quantum-dot model $(C_{E})_{\alpha\beta}$
depends only on the Hamming distance between the states it couples,
\begin{align}
(C_{E})_{\alpha\beta} = f(r_{\alpha\beta}) \,.
\label{eq:CE-Hamming}
\end{align}
This property does not hold for the 1D model, for which
$(C_{E})_{\alpha\beta}$ has a more complicated dependence 
on the difference between spin configuration $\alpha$ and $\beta$, reflecting the structure of the model in real space. 
Comparing the results for both models (between themselves, and also with the fully uncorrelated QREM-type model) allows us to investigate the role of energy correlations $(C_{E})_{\alpha\beta}$ in the scaling of the MBL transition $W_c(n)$.

\item Furthermore, both many-body models are characterized by non-trivial correlations of ``hopping'' matrix elements in Fock space. Again, the corresponding correlation function $(C_{T})_{\alpha\beta\mu\nu}$ depends only on the Hamming distance for the quantum-dot model and reflects the 1D spatial structure in the case of a 1D model. For comparison, we study also two Fock space models 
(that we call ``uncorrelated quantum dot'' and ``uncorrelated 1D'' for brevity) that have the same 
 $(C_{E})_{\alpha\beta}$ as the respective many-body model but have no correlations between hopping matrix elements [i.e., $(C_{T})_{\alpha\beta\mu\nu} =0$ 
$\{\alpha,\beta\} \ne \{\mu,\nu\}$]. 
In addition, we study a quantum random energy model without correlations at all, 
whether in diagonal or in off-diagonal matrix elements. This allows us to explore separately the impact  of  $(C_{E})_{\alpha\beta}$ and 
$(C_{T})_{\alpha\beta\mu\nu}$ correlations.

\end{itemize}

In the rest of Sec.~\ref{sec:model_definitions}, we introduce five models mentioned above (quantum dot, 1D, ``uncorrelated quantum dot'', ``uncorrelated 1D'', and quantum random energy model), which are studied in this paper and serve as a basis for our comparative analysis of the effect of Fock-space correlations on the scaling of the MBL transition. Since all of these models are restricted to single-spin flip processes, every spin configuration $\ket{\alpha}$ is connected to $n$ other states through non-zero matrix elements $T_{\alpha\beta}$. As a consequence, the associated Fock-space representation is a regular graph with connectivity $n$. For each of these models, we specify the mapping to the Fock-space representation (\ref{eq:Fock_space_rep}) by computing the distributions of $E_\alpha$ and $T_{\alpha\beta}$ and their correlation properties, encapsulated in the covariance matrices $C_E$ and $C_T$.

\subsection{Quantum dot (QD) model}
\label{sec:QD-model}

\subsubsection{Definition of the model}
\label{sec:QD-model-def}

We define a single-spin-flip quantum dot model, with interactions between every pair $(i,j)$ of spins. For brevity, we call it ``QD model'' below and use the superscript ``QD'' for the corresponding Hamiltonian and covariance matrices. 
The Hamiltonian of the model reads:
\begin{align}
&\hat{H}^\text{QD} =  \hat{H}^\text{QD}_0 + \hat{H}^\text{QD}_1 \,, 
\label{eq:QD}
\\
&\hat{H}^\text{QD}_0= \sum\limits_{i = 1}^{n} \epsilon_i\hat{S}_i^z + \frac{2}{\sqrt{n}}\sum\limits_{i,j = 1}^{n} V_{ij}^{z}\hat{S}_i^z \hat{S}_j^z \,,
\label{eq:QD-H0}
\\
&\hat{H}^\text{QD}_1 = \frac{1}{\sqrt{n}}\smashoperator{\sum}\limits_{\substack{i,j = 1\\ a \in \{x,y\}}}^{n}  V_{ij}^{a} \left( \hat{S}_i^z \hat{S}_j^a + \text{H.c.}\right),
\label{eq:QD-H1}
\end{align}
where the spin-$\smash{\frac{1}{2}}$ operators are defined as $\smash{\hat{S}_i^a = \frac{1}{2} \sigma_i^a}$ with $i = 1,\dots,n$ and $\sigma_i^a$ the Pauli matrices. The single-particle energies $\epsilon_i$ are uncorrelated random variables uniformly distributed in $\left[-W, W\right]$, where $W$ sets the disorder strength and is a parameter that is used to drive the system through the MBL transition. The interaction couplings $\smash{V_{ij}^{a}}$ (with $a\in\{x,y,z\}$) are uncorrelated real Gaussian random variables with 
\begin{align}
&\langle V^a_{ij} \rangle = 0 \,, \label{eq:V_mean}\\
&\langle V^a_{ij} V^b_{kl} \rangle = 0 \ \mathrm{for}\ a\neq b\,,
\label{eq:VaVb}\\
&\langle V^z_{ij} V^z_{kl} \rangle = \delta_{ik}\delta_{jl} \,, \label{eq:Vz_variance}\\
&\langle V^x_{ij} V^x_{kl} \rangle = \langle V^y_{ij} V^y_{kl} \rangle = 2 \delta_{ik}\delta_{jl} \,. \label{eq:Vxy_variance}
\end{align}
Everywhere in the paper, $\left\langle \ldots \right\rangle$ denotes the average over the statistical ensemble (i.e., over the distribution of $\{\epsilon_i\}$ and  $\{V_{ij}^a\}$ in the present context).

It is worth mentioning that our QD model can be obtained from the spin quantum-dot model studied
in Ref.~\cite{Herre2023} by removing from the Hamiltonian all terms 
corresponding to two-spin-flip processes. A  similar model with random all-to-all interactions and single-spin flips was also considered in Ref.~\cite{bulchandani2022onset}.

\subsubsection{Fock-space representation}

We proceed now as discussed in Sec.~\ref{sec:Fock-space-rep} and represent the QD model as a Fock-space tight-binding model in the basis of eigenstates $\ket{\alpha}$ of $\hat{H}^\text{QD}_0$ 
(i.e., of all $\hat{S}_i^z$).
The many-body energies $E_\alpha$ are expressed as 
\begin{align}
E_\alpha \equiv \bra{\alpha}\hat{H}^\text{QD}\ket{\alpha} =  \frac{1}{2}\sum\limits_{i=1}^n \epsilon_i s_i^{(\alpha)} + \frac{1}{2\sqrt{n}}\sum\limits_{i,j=1}^n V_{ij}^z s_{ij}^{(\alpha)} ,
\label{eq:QD_on_site_energies}
\end{align}
with $s_i^{(\alpha)} \equiv \bra{\alpha}\sigma_i^z\ket{\alpha} = \pm 1$, and $s_{ij}^{(\alpha)} \equiv s_i^{(\alpha)} s_j^{(\alpha)} = \pm 1$.
Further, the matrix element  $T_{\alpha\beta} \equiv \bra{\alpha}\hat{H}^\text{QD}\ket{\beta}$ is non-zero if and only if $\ket{\alpha}$ and $\ket{\beta}$ differ by a single spin flip. Specifically, for a pair of states that differ only by a sign of spin $s_k$ 
(i.e., for $\ket{\beta}=\ket{\bar{\alpha}_k}\equiv\sigma_k^x\ket{\alpha}$), we have
\begin{align}
T_{\alpha\beta} &= 
\bra{\alpha}\hat{H}^\text{QD}_1\ket{\bar{\alpha}_k}=\frac{1}{2\sqrt{n}} \sum\limits_{\substack{i = 1 \\ i\neq k}}^{n}\left( V_{ik}^x s_{i}^{(\alpha)} + \mathrm{i}\, V_{ik}^y s_{ik}^{(\alpha)} \right).
\label{eq:QD_matrix_elements}
\end{align}
Note that the terms with $V_{ik}^y$ are purely imaginary and depend on the sign of spin $k$. 

The QD model (\ref{eq:QD}) thus maps to the Fock-space representation (\ref{eq:Fock_space_rep}), with on-site energies and hopping matrix elements given by Eqs.~(\ref{eq:QD_on_site_energies}) and (\ref{eq:QD_matrix_elements}), respectively.
Below, we analyze the correlations of these matrix elements.

\subsubsection{Energy distributions and correlations}
\label{sec:IIB3}

By virtue of the central limit theorem, for $n \gg 1$, the many-body energies $E_\alpha$, given by Eq.~\eqref{eq:QD_on_site_energies}, obey a multivariate Gaussian distribution. Let us first consider the individual distributions of $E_\alpha$. The first term in Eq.~\eqref{eq:QD_on_site_energies} is a Gaussian random variable $\smash{ \sim \mathcal{N}(0,{nW^2/12})}$ (here we use the conventional notations for a Gaussian distribution, with two arguments denoting the mean value and the variance). The second term obeys the Gaussian distribution $\smash{\mathcal{N}(0,n/4)}$. 
Thus, we get
\begin{align}
E_\alpha \sim  \mathcal{N}\left(0,  \frac{nW^2+ 3n}{12}\right).
\label{eq:distrib_energies_QD}
\end{align}

We turn now to the full covariance matrix that encodes information about energy correlations.
According to Eq.~(\ref{eq:QD_on_site_energies}), we obtain
\begin{align}
(C_E^{\rm QD})_{\alpha\beta} = \left\langle E_\alpha E_\beta \right\rangle = \frac{W^2}{12} \sum\limits_{i=1}^n  s_i^{(\alpha)} s_i^{(\beta)} + \frac{1}{4n}\sum\limits_{i,j=1}^n s_{ij}^{(\alpha)}s_{ij}^{(\beta)},
\label{eq:correl_energies_intermediate}
\end{align}
where we used $\smash{\left\langle \epsilon_i \epsilon_{j} \right\rangle = \delta_{ij} W^2/3 }$ and $\langle V_{ij}^z V_{kl}^z \rangle = \delta_{ik}\delta_{jl}$ [see Eq.~(\ref{eq:Vz_variance})]. As discussed above, we introduce the Hamming distance $r_{\alpha\beta}$ between two many-body states $\alpha$ and $\beta$ as the minimum number of spins to be flipped in order to transform  $\alpha$ into $\beta$.
It is not difficult to see that both sums in Eq.~(\ref{eq:correl_energies_intermediate}) can be expressed as functions of $r_{\alpha\beta}$. For the first sum, this is fully straightforward, as $\smash{s_i^{(\alpha)} s_i^{(\beta)} = 1}$ if the spin $i$ is in the same state in both states $\alpha$ and $\beta$, and $\smash{s_i^{(\alpha)} s_i^{(\beta)} = -1}$ otherwise. We thus get  
\begin{align}
\sum\limits_{i=1}^n  s_i^{(\alpha)} s_i^{(\beta)}  = n- 2r_{\alpha\beta}.
\end{align}
To compute the second sum in Eq.~\eqref{eq:correl_energies_intermediate}, we denote as $\mathcal{D}$ the set of spins $i$ such that 
$s_i^{(\alpha)} = - s_i^{(\beta)}$. Clearly, $\mathcal{D}$ contains $r_{\alpha\beta}$ elements.
It is easy to see that $\smash{s_{ij}^{(\alpha)} s_{ij}^{(\beta)} = 1}$ when $i\in \mathcal{D}$ and $j \in \mathcal{D}$ or, else, when $i\notin \mathcal{D}$ and $j \notin \mathcal{D}$. These terms thus provide a contribution $(n-r_{\alpha\beta})^2+r_{\alpha\beta}^2$ to the sum. 
For the remaining terms, which correspond to the cases where $i\in \mathcal{D}$ and $j \notin \mathcal{D}$ or, else, $i\notin \mathcal{D}$ and $j \in \mathcal{D}$, we have $\smash{s_{ij}^{(\alpha)} s_{ij}^{(\beta)} = -1}$. These terms contribute $-2r_{\alpha\beta}(n-r_{\alpha\beta})$. Thus, we obtain
\begin{align}
\sum\limits_{i,j=1}^n s_{ij}^{(\alpha)}s_{ij}^{(\beta)} = (n-2r_{\alpha\beta})^2,
\end{align}
Combining the two terms in Eq.~(\ref{eq:correl_energies_intermediate}), we find
the following expression for the covariance matrix:
\begin{align}
(C_E^{\rm QD})_{\alpha\beta} &=  n\left[\frac{W^2}{12} \left(1- \frac{2r_{\alpha\beta}}{n}\right) + \frac{1}{4}\left(1-\frac{2r_{\alpha\beta}}{n}\right)^2\right].
\label{eq:correl_energies_final}
\end{align}

\subsubsection{Distribution and correlations of hopping matrix elements}
\label{sec:QD-hoppings-corr}

Obviously, matrix elements 
$T_{\alpha\beta}$
obey a multivariate complex Gaussian distribution with zero mean. In particular, it is easily seen from Eq.~(\ref{eq:QD_matrix_elements}) that individual $T_{\alpha\beta}$ (those that are non-zero, i.e., connect two states that differ by a single spin flip) are Gaussian complex random variables such that (at $n \gg 1$)
\begin{align}
T_{\alpha\beta} \sim \mathcal{N}\left(0, \frac{1}{2} \right) + \mathrm{i}\,\mathcal{N}\left(0,\frac{1}{2}\right),
\label{eq:QD_distrib_T}
\end{align} 
where we used Eq.~\eqref{eq:Vxy_variance}.
Correspondingly, diagonal elements of the covariance matrix $C_T^\text{QD}$ read 
\begin{align}
(C_T^\text{QD})_{\alpha\beta\alpha\beta} = \left\langle |T_{\alpha\beta}|^2 \right\rangle = \frac{n-1}{n} \simeq 1 \,.
\label{eq:QD_squared_avg_matrix_elts}
\end{align}
Later, we will also need the average absolute value of transition amplitudes $|T_{\alpha\beta}|$,  
\begin{align}
\langle |T_{\alpha\beta}| \rangle = \frac{\sqrt{\pi}}{2}.
\label{eq:QD_mean_transition_amplitude}
\end{align}

To extend Eq.~\eqref{eq:QD_squared_avg_matrix_elts} to the full covariance matrix $C_T^{\rm QD}$, we note that  $(C_{T}^{\rm QD})_{\alpha\beta\mu\nu} \equiv  \langle T_{\alpha\beta}^* T_{\mu\nu} \rangle$ is non-zero only if the states $\{\alpha, \beta\}$ are connected by a single flip of spin $s_k$ and the states $\{\mu,\nu\}$ are connected by a single flip of the same spin $s_k$, with $s_k^{(\alpha)} = s_k^{(\mu)}$. Geometrically, this means that the edges $(\alpha \to \beta)$ and $(\mu \to \nu)$ of the hypercube are collinear,
{\color{black} with the same direction, and parallel to $k$ axis. (We recall that $T_{\alpha\beta}$ are complex, Eq.~\eqref{eq:QD_distrib_T}, so that the order of indices $\alpha, \beta$ does matter, and the same for $\mu, \nu$.)} In this case, a simple calculation yields
\begin{equation}
(C_{T}^{\rm QD})_{\alpha\beta\mu\nu} = \frac{1}{n}
\sum\limits_{\substack{i = 1 \\ i \neq k}}^{n}
s_i^{(\alpha)} s_i^{(\mu)} = \frac{n-1-2r_{\alpha\mu}}{n} \simeq 1 - \frac{2r_{\alpha\mu}}{n} \,.
\label{eq:QD-CT-full}
\end{equation}

We have thus fully determined the statistics of matrix elements of the Fock-space Hamiltonian of the QD model. Both the many-body energies $E_\alpha$ and the ``hopping'' matrix elements $T_{\alpha\beta}$ are characterized by correlated multivariate Gaussian distributions. The covariance matrix for the energies is given by Eq.~(\ref{eq:correl_energies_final}); its elements depend only on the Hamming distance $r_{\alpha\beta}$ on the Fock space graph. 
Non-zero elements of the covariance matrix of hoppings are given by  Eq.~\eqref{eq:QD-CT-full}; they also depend on the relative position of two states $\alpha$ and $\mu$ only via the Hamming distance $r_{\alpha\mu}$. 

\subsection{Uncorrelated quantum dot (uQD) model}
\label{sec:uQD-model}

In the Fock-space representation of the QD model,
there are strong correlations between energies $E_\alpha$ (Sec.~\ref{sec:IIB3}) and also strong correlations between the hopping matrix elements $T_{\alpha\beta}$ (Sec.~\ref{sec:QD-hoppings-corr}). To explore the role of the latter correlations, we define a modified model in the Fock space, in which the energy covariance matrix $C_E$ has exactly the same form as in the QD model, Eq.~\eqref{eq:correl_energies_final}, and, at the same time, the off-diagonal matrix elements of $C_T$ are set to zero:
\begin{align}
(C_T^\text{uQD})_{\alpha\beta\mu\nu} \equiv (C_T^\text{QD})_{\alpha\beta\alpha\beta}\, 
\delta_{\alpha\mu}\delta_{\beta\nu} \simeq  \delta_{\alpha\mu}\delta_{\beta\nu},
\end{align}
where $\alpha$ and $\beta$ are connected by one spin flip. 
In this way, we remove correlations between the hopping matrix elements in the Fock-space version of the QD model. 

We term the model obtained in this way the ``uncorrelated QD model'' or, still shorter, ``uQD model''. Let us reiterate, however, that this model on the Fock-space (hypercube) graph is uncorrelated only in the sense of absence of correlations between the hopping matrix elements $T_{\alpha\beta}$ defined on edges of the graph. At the same time, the energies $E_\alpha$ in this model retain strong correlations of the QD model given by Eq.~\eqref{eq:correl_energies_final}, 
\begin{equation}
(C_E^{\rm uQD})_{\alpha\beta} = (C_E^{\rm QD})_{\alpha\beta} \,.
\end{equation}

It is also worth emphasizing that, while the uQD model has a simple definition in terms of a tight-binding model in the Fock space, it does not correspond to any ``conventional'' spin model (i.e., a model involving $p$-spin interactions, where $p$ is fixed in the large-$n$ limit). Indeed, it is easy to see that, in order to obtain a Fock-space model like the uQD model, one would need to include in the spin Hamiltonian terms involving coupling of all $n$ spins. This comment also applies to the uncorrelated 1D model introduced below in Sec.~\ref{sec:u1D-model}.

\subsection{One-dimensional spin chain (1D model)}
\label{sec:1D-model}

We define now a 1D single-spin-flip model and derive its Fock-space representation and associated correlations. We will see how the 1D spatial structure is reflected in Fock-space correlations, which are not more functions of solely the Hamming distance on the graph, at variance with the QD model.

\subsubsection{Definition of the model}
\label{sec:1D-model-def}

The 1D model that we consider is a length-$n$ spin chain with periodic boundary conditions, governed by the Hamiltonian:
\begin{align}
&\hat{H}^\text{1D} =  \hat{H}^\text{1D}_0 + \hat{H}^\text{1D}_1 \,,\\
&\hat{H}^\text{1D}_0= \sum\limits_{i = 1}^{n} \epsilon_i\hat{S}_i^z + 2 \sum\limits_{i = 1}^{n} V_{i,i+1}^{z}\hat{S}_i^z \hat{S}_{i+1}^z \,, \\
&\hat{H}^\text{1D}_1 = 2\smashoperator{\sum}\limits_{\substack{i = 1\\a \in \{x,y\}}}^{n}  V_{i, i+1}^{a}\hat{S}_i^z \hat{S}_{i+1}^a.
\label{eq:1D}
\end{align}
As in the QD model, the spin-$\smash{\frac{1}{2}}$ operators are defined as $\smash{\hat{S}_i^\alpha = \frac{1}{2} \sigma_i^a}$ with Pauli matrices $\sigma_i^a$, and the single-particle energies $\epsilon_i$ are uncorrelated random variables uniformly distributed in $\left[-W, W\right]$. Further, the interaction couplings $\smash{V_{i,i+1}^{a}}$ are uncorrelated Gaussian random variables with the statistics determined by Eqs.~(\ref{eq:V_mean})-(\ref{eq:Vxy_variance}), where now only the couplings with $j=i+1$ and $l=k+1$ enter, i.e.,
\begin{align}
&\langle V^a_{i,i+1} \rangle = 0 \,,
\label{eq:1D-V-mean} 
\\
&\langle V^z_{i,i+1} V^z_{k,k+1} \rangle = \delta_{ik}\,, 
\label{eq:1D-Vz-variance}
\\
&\langle V^x_{i,i+1} V^x_{k,k+1} \rangle = \langle V^y_{i,i+1} V^y_{k,k+1} \rangle = 2 \delta_{ik} \,.
\label{eq:1D-Vxy-variance}
\end{align}

\subsubsection{Fock-space representation}

\par In analogy with the QD model, $\hat{H}^\text{1D}$ can be straightforwardly mapped to the Fock space representation (\ref{eq:Fock_space_rep}) in the basis $\ket{\alpha} $ of eigenstates of $\hat{H}^\text{1D}_0$. The many-body energies $E_\alpha$ are found to be 
\begin{align}
E_\alpha \equiv \bra{\alpha}\hat{H}^\text{1D}\ket{\alpha} =  \frac{1}{2}\sum\limits_{i=1}^n s_i^{(\alpha)}\epsilon_i + \frac{1}{2}\sum\limits_{i=1}^n s_{i, i+1}^{(\alpha)} V_{i,i+1}^z \,,
\label{eq:1D_on_site_energies}
\end{align}
with the same notations as above, $s_i^{(\alpha)} \equiv \bra{\alpha}\sigma_i^z\ket{\alpha} = \pm 1$ and $s_{i,i+1}^{(\alpha)} \equiv 
s_i^{(\alpha)} s_{i+1}^{(\alpha)} = \pm 1$. The ``hopping'' matrix elements $T_{\alpha\beta} \equiv \bra{\alpha}\hat{H}^\text{1D}\ket{\beta}$ are non-zero if and only if $\ket{\alpha}$ and $\ket{\beta}$ are connected by a single spin flip. For a pair of states $\ket{\alpha}$ and  $\ket{\beta}=\ket{\bar{\alpha}_k}$ that differ only by the sign of the spin $s_k$ in a position $k$, we have
\begin{align}
T_{\alpha\beta} &=  2\bra{\alpha}\left[ V_{k-1,k}^x\hat{S}_{k-1}^z\hat{S}_k^x+ V_{k-1,k}^y\hat{S}_{k-1}^z\hat{S}_k^y \right]\ket{\bar{\alpha}_k}\nonumber\\
&= \frac{1}{2} \left( V_{k-1,k}^x s_{k-1}^{(\alpha)} + \mathrm{i}\, V_{k-1,k}^y s_{k-1,k}^{(\alpha)}  \right).
\label{eq:1D_matrix_elements}
\end{align}
As in the case of the QD model, the Fock space can be seen as a hypercube graph with $2^n$ nodes and connectivity $n$, with energies $E_\alpha$ associated with the graph vertices and single-spin-flip amplitudes $T_{\alpha\beta}$ associated with the edges.

\subsubsection{Energy distributions and correlations}
\label{subsec:1D_model_def_energy_correls}

It follows from Eq.~\eqref{eq:1D_on_site_energies}
that individual energies $E_\alpha$ obey exactly the same Gaussian distribution
\eqref{eq:distrib_energies_QD} as for the QD model.
Let us calculate the correlations. Using Eqs.~\eqref{eq:1D-V-mean} and \eqref{eq:1D-Vz-variance}, we get for the energy covariance matrix
 $(C_E^{\rm 1D})_{\alpha\beta} = \left\langle E_\alpha E_\beta \right\rangle$:
\begin{align}
(C_E^{\rm 1D})_{\alpha\beta} & = \frac{W^2}{12} \sum\limits_{i=1}^n  s_i^{(\alpha)} s_i^{(\beta)} + \frac{1}{4}\sum\limits_{i=1}^n s_{i,i+1}^{(\alpha)}s_{i,i+1}^{(\beta)} \nonumber \\
& = \frac{W^2}{12} (n - 2 r_{\alpha\beta}) + 
\frac{1}{4} (n - 2 q_{\alpha\beta}).
\label{eq:1D-correl_energies_intermediate}
\end{align}
Here $r_{\alpha\beta}$ is the Hamming distance defined above, and we have introduced the notation $q_{\alpha\beta}$ for the number of sites $i$ such that $s_{i,i+1}^{(\alpha)} = - s_{i,i+1}^{(\beta)}$.

Crucially, the covariance \eqref{eq:1D-correl_energies_intermediate} depends on the 
Fock-space path between the states $\alpha$ and $\beta$ {\it not} solely via the Hamming distance $r_{\alpha\beta}$. The emergence of $q_{\alpha\beta}$ in Fock-space correlations, Eq.~\eqref{eq:1D-correl_energies_intermediate}, reflects the 1D real-space geometry of the model. This difference between 
Eqs.~\eqref{eq:correl_energies_final} and \eqref{eq:1D-correl_energies_intermediate} is crucial for a different scaling of the MBL transition $W_c(n)$ in the QD and 1D model, as we discuss below.

It is instructive to consider an average of the covariance \eqref{eq:1D-correl_energies_intermediate} over all pairs of spin configurations $(\alpha,\beta)$ with a given Hamming distance $r_{\alpha\beta}$. (A similar calculation for a different 1D model was performed in Ref.~\cite{Roy2020}.)  
We will denote such an average by an overbar; by construction, it gives a function of $r_{\alpha\beta}$. It is not difficult to see that, for a given $r_{\alpha\beta}$, the probability that $s_{i,i+1}^{(\alpha)} = - s_{i,i+1}^{(\beta)}$ for a given site $i$ is
\begin{equation}
\frac{r_{\alpha\beta} (n-r_{\alpha\beta})}{n(n-1)} \simeq \frac{r_{\alpha\beta} (n-r_{\alpha\beta})}{n^2} \,,
\end{equation}
so that
\begin{equation}
\overline{q_{\alpha\beta}} = 2n \,\frac{r_{\alpha\beta}}{n}
\left( 1 - \frac{r_{\alpha\beta}}{n} \right).
\end{equation}
Thus, we find from Eq.~\eqref{eq:1D-correl_energies_intermediate}
\begin{align}
\overline{C_E^\text{1D}}(r_{\alpha\beta})  = n\left[\frac{W^2}{12} \left(1- \frac{2r_{\alpha\beta}}{n}\right) + \frac{1}{4}\left(1-\frac{2r_{\alpha\beta}}{n}\right)^2\right].
\label{eq:correl_energies_1D_average}
\end{align}
We observe that the right-hand side of 
Eq.~\eqref{eq:correl_energies_1D_average}
is identical to that of the QD covariance, Eq.~\ref{eq:correl_energies_final}. 
Thus, while the 1D spatial geometry is reflected in  the Fock-space covariance matrix $C_E^\text{1D}$ in a specific way, this structure is totally washed out if one considers the 
covariance averaged over directions in Fock space, $\overline{C_E^\text{1D}}(r_{\alpha\beta})$, which is the same for QD and 1D models.
It is thus of paramount importance to address $(C_E)_{\alpha\beta}$ rather than its average $\overline{C_E}(r_{\alpha\beta})$ when discussing MBL properties. In particular, one cannot deduce the scaling of $W_c(n)$ on the basis of solely $\overline{C_E}(r_{\alpha\beta})$, at variance with Ref.~\cite{Roy2020}: many-body models with the same $\overline{C_E}(r_{\alpha\beta})$ may exhibit a totally different scaling $W_c(n)$.

\subsubsection{Correlations of hopping matrix elements}

In analogy with the QD model, the Fock-space hoppings $\{T_{\alpha\beta}\}$ of the 1D model obey a  multivariate complex Gaussian distribution with zero mean. 
From Eq.~\eqref{eq:1D_matrix_elements}, we find,
using Eq.~\eqref{eq:1D-Vxy-variance},
that the matrix elements $T_{\alpha\beta}$
exhibit the same Gaussian distribution
\eqref{eq:QD_distrib_T} as in the QD model.

As in the QD model, an element  $(C_{T}^{\rm 1D})_{\alpha\beta\mu\nu} \equiv  \langle T_{\alpha\beta}^* T_{\mu\nu} \rangle$ of the covariance matrix is different from zero only if the states $\{\alpha, \beta\}$ are connected by a single flip of spin $s_k$ and the states $\{\mu,\nu\}$ are connected by a single flip of the same spin $s_k$, with $s_k^{(\alpha)} = s_k^{(\mu)}$. In this situation, we find
\begin{equation}
(C_{T}^{\rm 1D})_{\alpha\beta\mu\nu} =
s_{k-1}^{(\alpha)}  s_{k-1}^{(\mu)} \,.
\label{eq:1D-CT}
\end{equation}
As in the case of energy correlations, we see that 
 $(C_{T}^{\rm 1D})_{\alpha\beta\mu\nu}$ is not a function of Hamming distance, in contrast to the covariance $C_T$ of the QD model, 
Eq.~\eqref{eq:QD-CT-full}. This is again a manifestation of the spatial geometry of the 1D model. If one averages over Fock-space directions between the states $\alpha$ and $\mu$ (at fixed $r_{\alpha\mu}$), this information gets lost (in the same way as for $C_E$), and the result is exactly the same as for the QD model [Eq.~\eqref{eq:QD-CT-full}]:
\begin{equation}
\overline{(C_{T}^{\rm 1D})_{\alpha\beta\mu\nu}} =
1 - 2 \frac{r_{\alpha\mu}}{n} \,.
\end{equation}

\subsection{Uncorrelated one-dimensional (u1D) model}
\label{sec:u1D-model}

In the same spirit as we modified the QD model into the uQD model, we define now an ``uncorrelated 1D model'' (in brief, u1D model) by removing the correlations of hopping matrix elements $T_{\alpha\beta}$ from the 1D model. Thus, the u1D model has exactly the same energy covariance matrix $C_E$ as the 1D model,
\begin{equation}
(C_E^{\rm u1D})_{\alpha\beta} = (C_E^{\rm 1D})_{\alpha\beta} \,,
\end{equation}
which is given by Eq.~\eqref{eq:1D-correl_energies_intermediate}, and the hopping covariance matrix $C_T$ obtained from $C_T^{\rm 1D}$, Eq.~\eqref{eq:1D-CT}, by setting all off-diagonal elements [$(\alpha, \beta) \ne (\mu,\nu)$] to zero,
\begin{equation}
(C_{T}^{\rm u1D})_{\alpha\beta\mu\nu} =
(C_{T}^{\rm 1D})_{\alpha\beta\mu\nu} \delta_{\alpha \mu}\delta_{\beta\nu} \,,
\label{eq:u1D-CT}
\end{equation}
so that the only non-zero elements of  $(C_{T}^{\rm u1D})$ are
\begin{equation}
(C_{T}^{\rm u1D})_{\alpha\beta\alpha\beta} = 1 \,,
\label{eq:u1D-CT-1}
\end{equation}
where $\alpha$ and $\beta$ are connected by a single spin flip.

\begin{table*}[t!]
\begin{ruledtabular}
\begin{tabular}{lll}
Model&
\multicolumn{1}{c}{Energy correlations \(C_E\)} &
\multicolumn{1}{c}{Hopping correlations \(C_T\)}\\
\colrule\\
QD &  Determined by Hamming distance,  $(C_E^{\rm QD})_{\alpha\beta} = f(r_{\alpha\beta}) $, Eq.~(\ref{eq:correl_energies_final}) &  Determined by Hamming distance, Eq.~(\ref{eq:QD-CT-full}) \\
uQD & Determined by Hamming distance, ${(C_E^{\rm uQD})_{\alpha\beta} } = f(r_{\alpha\beta}) $ , Eq.~(\ref{eq:correl_energies_final}) &  Diagonal $C_T$ (no correlations) \\\\
1D &  Reflect spatial structure,  $(C_E^{\rm 1D})_{\alpha\beta} \neq f(r_{\alpha\beta}) $, Eq.~(\ref{eq:1D-correl_energies_intermediate}) &  Reflect spatial structure, Eq.~(\ref{eq:1D-CT}) \\
u1D & Reflect spatial structure, ${(C_E^{\rm u1D})_{\alpha\beta} }\neq f(r_{\alpha\beta}) $, Eq.~(\ref{eq:1D-correl_energies_intermediate}) &  Diagonal $C_T$ (no correlations)\\\\
QREM/RRG & Diagonal $C_E$ (no correlations) & Diagonal $C_T$ (no correlations) \\
\end{tabular}
\end{ruledtabular}
\caption{Summary of the Fock-space correlation properties for the models presented in Sec.~\ref{sec:model_definitions}. }
\label{tab:correl_summary}
\end{table*}


\subsection{Fully discarding correlations: Quantum random energy model (QREM)}
\label{sec:QREM}

All four models discussed above (QD, uQD, 1D, and u1D) have the same variances of energies, $(C_E)_{\alpha\alpha}$ and of hopping matrix elements,
$(C_T)_{\alpha\beta\alpha\beta}$ but differ by the forms of covariance matrices $(C_E)_{\alpha\beta}$ and $(C_T)_{\alpha\beta\mu\nu}$. For a more complete understanding of the role of Fock-space correlations, it is natural to consider also a model with the same variances and without any correlations (i.e., with all off-diagonal elements of covariance matrices set to zero). This is an Anderson tight-binding model on the $n$-dimensional hypercube graph, Eq.~(\ref{eq:Fock_space_rep}), with uncorrelated random energies on vertices
\begin{align}
E_\alpha \sim  \mathcal{N}\left(0,  \frac{nW^2+ 3n}{12}\right)
\label{eq:distrib_energies_QD-repeated}
\end{align}
and with uncorrelated random hopping matrix elements on hypercube edges
\begin{align}
T_{\alpha\beta} \sim \mathcal{N}\left(0, \frac{1}{2} \right) + \mathrm{i}\,\mathcal{N}\left(0,\frac{1}{2}\right).
\label{eq:QD_distrib_T-repeated}
\end{align} 
This model is analogous to the quantum random-energy model (QREM) studied {\color{black} in the context of MBL} in Refs.~\cite{laumann2014many-body,baldwin2016the_many-body}, with the only difference that hopping matrix elements 
$T_{\alpha\beta}$ were constant in Refs.~\cite{laumann2014many-body,baldwin2016the_many-body} and are uncorrelated Gaussian random variables in our case. {\color{black} A spin-based microscopic realization of the QREM appeared in the context of spin glasses~\cite{Cugliandolo} as an Ising infinite-range model with $p$-spin interaction in a transverse field in the limit $p\to \infty$~\cite{Goldschmidt1990}. In what follows, we will refer to a fully uncorrelated model defined on a hypercube by Eqs.~(\ref{eq:Fock_space_rep}), (\ref{eq:distrib_energies_QD-repeated}), and (\ref{eq:QD_distrib_T-repeated}) as ``QREM''. }

{\color{black} The QREM is also closely related to the RRG model, which is obtained by substituting the hypercube structure of the Fock space with random regular graphs with the same connectivity $m+1=n$.} In the limit of large $n$, the position of the localization transition on this graph should be the same as in the RRG model, which has been solved analytically {\color{black} (see the discussion at the beginning of Sec.~\ref{sec:RRG_scaling}).} We present analytical results and expectations for all five models (QREM, uQD, u1D, QD, and 1D; {\color{black}see Table~\ref{tab:correl_summary} for the summary of their Fock-space correlation properties}) in the next section.

\section{Analytical considerations}
\label{sec:analytical}

In this section, we discuss analytical predictions for the critical disorder $W_c(n)$ and the width $\Delta W(n) / W_c(n)$ of the MBL transitions in the models defined in Sec.~\ref{sec:model_definitions}. We begin with the QREM, since the absence of correlations simplifies its analytical treatment, thus making it a convenient starting point. After this, we consider the uQD and u1D models that involve energy correlations, and finally, the QD and 1D models that have both energy and hopping correlations.

\subsection{Fully uncorrelated model: QREM and Anderson localization on RRG}
\label{sec:RRG_scaling}

We begin with the model of QREM type, Sec.~\ref{sec:QREM}, without any correlations of energies $E_\alpha$ and hoppings $T_{\alpha\beta}$ that obey the Gaussian distributions 
\eqref{eq:distrib_energies_QD-repeated} and
\eqref{eq:QD_distrib_T-repeated}. We omit some technical details of the analysis here; they can be found in Appendix \ref{app:RRG-QREM}. 

The behavior of $W_c(n)$ at large $n$ in QREM should be the same as in the RRG model with the same coordination number $n$, the same distributions of $E_\alpha$ and $T_{\alpha\beta}$, and the same system volume $N=2^n$. {\color{black} Indeed, the two models essentially differ only by that the $n$-dimensional hypercube (on which the QREM is defined) contains more short-scale loops than a typical RRG (that is locally tree-like). However, such loops are rare also in the QREM model, so that their contributions to the observables get parametrically suppressed in the large-$n$ limit (see the discussion on the ``single resonance approximation'' in  Refs.~\cite{baldwin2016the_many-body, laumann2014many-body})}. We thus begin by considering the RRG model with a large coordination number, {\color{black} which can be solved analytically}.

The scaling of the Anderson localization transition in the RRG model is well understood  
\cite{tikhonov19statistics,tikhonov2021from,Herre2023}.  The ``standard'' RRG model considered in most of previous works (we will use a subscript ``RRG-0'' for the corresponding observables) is characterized by connectivity $m+1$, hopping matrix elements $T=1$, and the box distribution on $[-W/2, W/2]$ of random energies $E_i$. In the limit of large Hilbert-space dimension (number of vertices of the graph), $N \to \infty$, the critical disorder $W_c$ of this model in the middle of the band (energy $E=0$) is a solution of the equation
\begin{equation}
W_c = 4m \ln (W_c/2) \,,
\label{eq:RRG-maintext-Wc-equation}
\end{equation}
the same as for the corresponding model on an infinite Bethe lattice \cite{abou1973selfconsistent,mirlin1991localization}. As we are interested here in a large RRG connectivity, we will not make distinction between connectivity $m$ and $m+1$. 

For a finite (but large) $N$, the transition point $W_c^{\rm RRG-0}(m,N)$ is shifted towards smaller $W$. It can be found from an equation 
\begin{equation} 
N_\xi(W) = N \,,
\label{eq:RRG-maintext-Nxi-W-N}
\end{equation}
where $N_\xi(W)$ is the correlation volume \cite{Herre2023}, see 
Eqs.~\eqref{eq:RRG-N-xi-general-1}, \eqref{eq:RRG-N-xi-general-2} of Appendix \ref{app:RRG}. In the above notations, the solution of Eq.~\eqref{eq:RRG-maintext-Wc-equation} is $W_c^{\rm RRG-0}(m,\infty)$. 

We are interested here in a more general RRG model, with distribution $\gamma(E)$ of uncorrelated energies $E_\alpha$ 
(characterized by a single energy scale $W$) and with some distribution of uncorrelated hopping amplitudes $T_{\alpha\beta}$. 
One should then perform a substitution \cite{Herre2023}
\begin{equation}
1/W \: \longmapsto \: \gamma(0) \langle | T | \rangle \,,
\label{eq:RRG-maintext-substitution-W}
\end{equation}
where $\langle |T| \rangle$ is the average value of $|T_{\alpha\beta}|$. Since the product $\gamma(0) \langle | T | \rangle$ is proportional to $1/W$, the transformation \eqref{eq:RRG-maintext-substitution-W} amounts essentially to rescaling of the disorder $W$ (and correspondingly of $W_c$).
In particular, Eq.~\eqref{eq:RRG-Wc-equation} takes the form 
\begin{align}
1 = 4m \langle |T|\rangle \gamma(0) \ln \frac{1}{\langle |T|\rangle \gamma(0)} \,;
\label{eq:self_cons_critical_disorder_Herre}
\end{align}
its solution is $W_c^{\rm RRG}(m,\infty)$. To find $W_c^{\rm RRG}(m,N)$, one should solve 
Eq.~\eqref{eq:RRG-maintext-Nxi-W-N} for $W$, with $N_\xi(W)$ given by a transformed version of Eqs.~\eqref{eq:RRG-N-xi-general-1}, \eqref{eq:RRG-N-xi-general-2}:
\begin{equation}
N_\xi = \frac{1}{m} \exp \left\{2\pi x^{-1} \ln [(\gamma(0) \langle | T | \rangle)^{-1}]\right\},
\label{eq:RRG-maintext-rescaled-Nxi-1}
\end{equation}
where $x$ is a solution of the equation
\begin{equation}
\frac{\sin x}{x} = \frac{f(W)}{f(W^{\rm RRG}_c(m,\infty))}\,, \ \ \ 
f(W) = \frac{W}{\ln [(\gamma(0) \langle | T | \rangle)^{-1}] }.
\label{eq:RRG-maintext-rescaled-Nxi-2}
\end{equation}

For our QREM, and thus for the associated RRG model, $\langle |T| \rangle = \sqrt{\pi}/2$, see Eq.~\eqref{eq:QD_mean_transition_amplitude}, and
\begin{equation} 
\gamma(E) = \frac{1}{\sigma \sqrt{2\pi}} \, e^{-E^2/2\sigma^2}\,; \qquad \sigma^2 = \frac{nW^2}{12}\,,
\label{eq:gamma-E}
\end{equation}
see Eq.~\eqref{eq:distrib_energies_QD-repeated}. 
Here we have neglected the second term in the variance in Eq.~\eqref{eq:distrib_energies_QD-repeated} since it is much smaller than the first term under the condition $W \gg 1$. (The critical disorder $W_c$ satisfies this condition, as we will see shortly.)  This yields 
\begin{equation}
\gamma(0) \langle | T | \rangle = (3/2n)^{1/2} W^{-1} \,.
\label{eq:RRG-maintext-rescaling-factor}
\end{equation}
Further, we make a substitution $m \longmapsto n$ for the coordination number. Equation
\eqref{eq:self_cons_critical_disorder_Herre}
for $W_c^{\rm RRG}(n,\infty)$ then becomes
\begin{equation}
W_c = 4 \sqrt{\frac{3}{2}} \: n^{1/2} \ln\left[ 
(2n/3)^{1/2} W_c \right].
\label{eq:RRG-maintext-Wc-rescaled-equation}
\end{equation} 
The leading large-$n$ asymptotics of the solution of this equation is 
\begin{equation}
W^{\rm RRG}_c (n,\infty) \simeq 4 \sqrt{\frac{3}{2}} \: n^{1/2} \ln n \,.
\label{eq:RRG-Wc-asympt}
\end{equation}
Solving Eq.~\eqref{eq:RRG-maintext-Wc-rescaled-equation} iteratively, one observes that a relative correction to Eq.~\eqref{eq:RRG-Wc-asympt} scales with $n$ as $\ln \ln n / \ln n$, i.e. it decays with $n$ very slowly. 

We recall now that, for the QREM, the system volume $N$ is related to the coordination number $n$ via $N=2^n$, so that
\begin{equation}
W_c^{\rm QREM} (n) = W_c^{\rm RRG} (n, 2^n) \,.
\label{eq:QREM-maintext-relation-RRG}
\end{equation}
In the large-$n$ limit, the exponential growth of $N=2^n$ ensures that $\lim_{n\to \infty} W_c^{\rm QREM} (n) / W_c^{\rm RRG} (n, \infty) = 1$. At the same time, for moderately large $n$, this ratio may differ appreciably from unity. 

\begin{figure}[t!]
  \centering
  \includegraphics[width=\columnwidth]{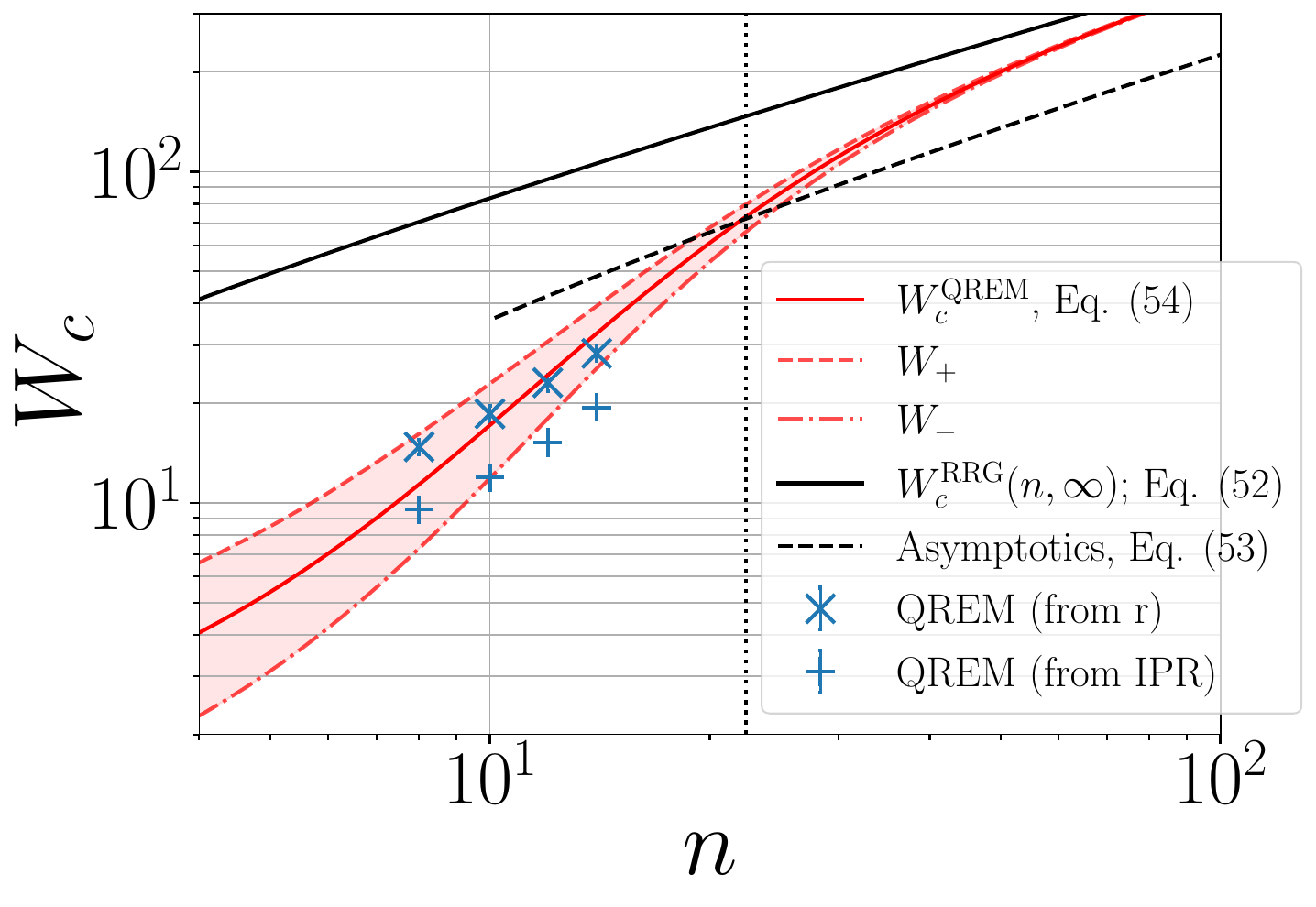}
  \caption{Localization transition in QREM. Full red line: critical disorder $W_c^{\rm QREM}(n)$, Eq.~\eqref{eq:QREM-maintext-relation-RRG}, as a function of the number $n$ of spins. 
  The finite-size transition region $[W_-(n), W_+(n)]$ around $W_c^{\rm QREM}(n)$ is shown by shading.
  Full black line: the asymptotics $W_c^{\rm RRG}(n,\infty)$ given by a solution of Eq.~\eqref{eq:RRG-maintext-Wc-rescaled-equation}. Dashed black line: the leading large-$n$ asymptotics \eqref{eq:RRG-Wc-asympt}.
  Vertical dotted line: estimated lower border of the critical regime, $n^{\rm crit} \approx 22$, defined by the condition
  $W_c^{\rm QREM}(n) = W_c^{\rm RRG}(n,\infty)/2$. Symbols: numerical exact-diagonalization data for $W_c^{\rm QREM}(n)$ from level statistics and from IPR, see Sec.~\ref{subsec:numerics_QREM}. 
  }
  \label{fig:Wc_QREM_theory}
\end{figure}

In Fig.~\ref{fig:Wc_QREM_theory}, we plot the analytical curve $W_c^{\rm QREM}(n) = W_c^{\rm RRG}(n,2^n)$ as obtained 
from Eqs.~\eqref{eq:RRG-maintext-Nxi-W-N},
\eqref{eq:RRG-maintext-rescaled-Nxi-1},
\eqref{eq:RRG-maintext-rescaled-Nxi-2}, and
\eqref{eq:RRG-maintext-rescaling-factor}.
We also show the asymptotics $W_c^{\rm RRG}(n,\infty)$
given by a solution of Eq.~\eqref{eq:RRG-maintext-Wc-rescaled-equation}
as well as the leading large-$n$ asymptotics \eqref{eq:RRG-Wc-asympt}.
The vertical dotted line in the figure represents an estimated border $n^{\rm crit} \approx 22$ of the critical regime, $n > n^{\rm crit}$, in which $W_c^{\rm QREM}(n) > (1/2) W_c^{\rm RRG}(n,\infty)$. We find that, in the critical regime, $W_c^{\rm QREM}(n)$ approaches its large-$n$ asymptotics $W_c^{\rm RRG}(n,\infty)$ according to [see Eq.~\eqref{eq:QREM-Wc-finite-size-shift}]
\begin{equation}
\frac{W_c^{\rm RRG}(n,\infty) - W_c^{\rm QREM}(n)}{W_c^{\rm RRG}(n,\infty)} \simeq \frac{2\pi^2}{3 \ln^2 2} \: \frac{\ln^2 n}{n^2} \,.
\label{eq:QREM-maintext-Wc-finite-size-shift}
\end{equation}

\begin{figure}[t!]
  \centering
  \includegraphics[width=0.5\columnwidth]{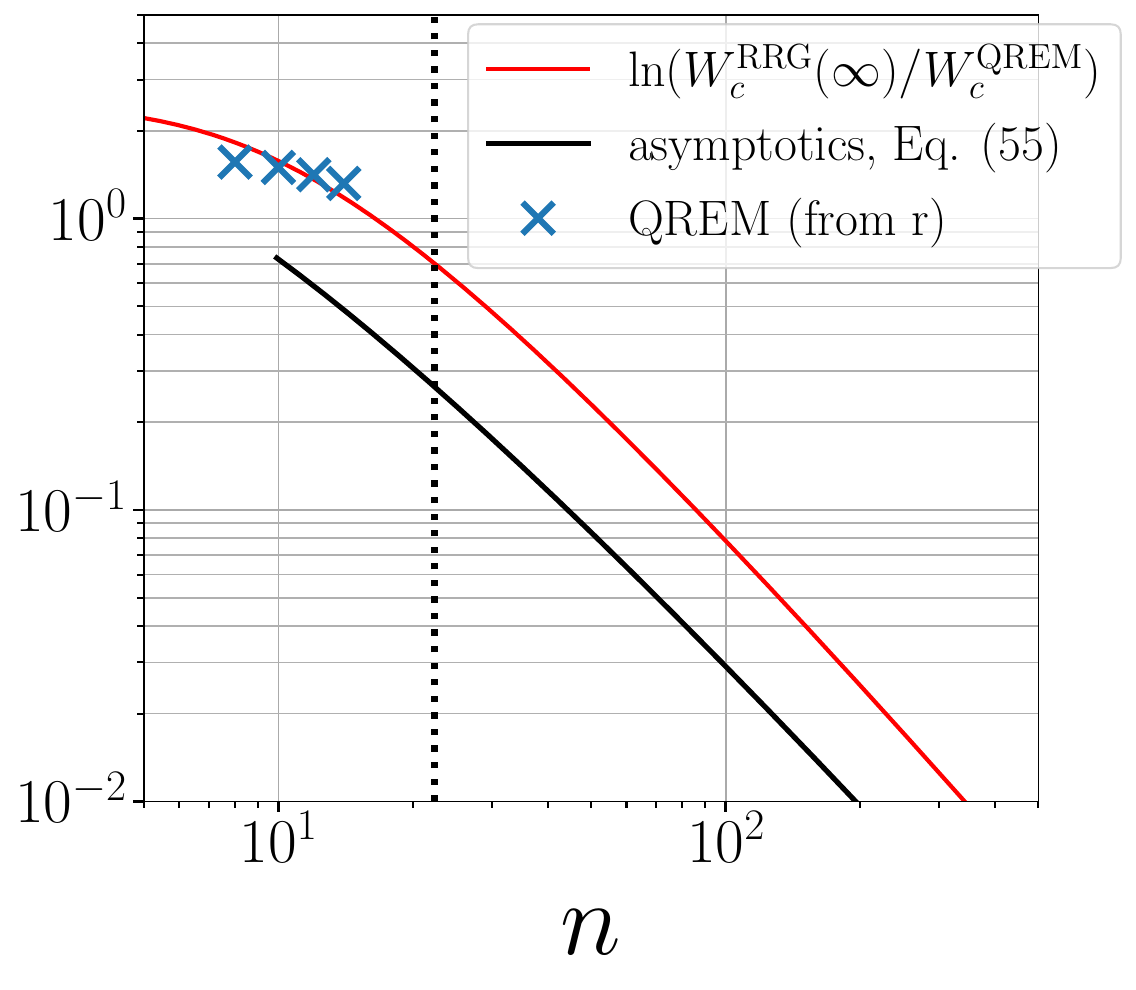}\hfill
  \includegraphics[width=0.5\columnwidth]{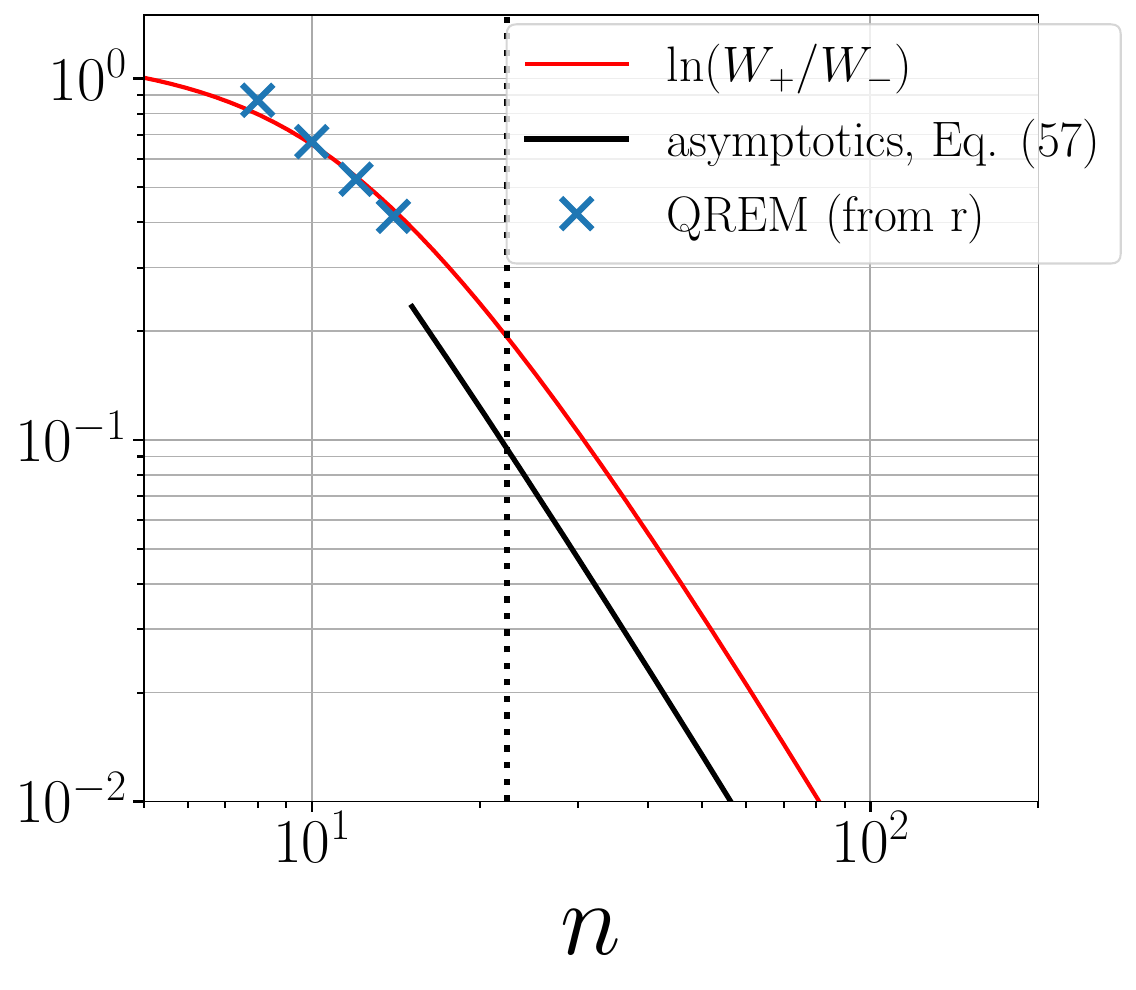}
  \caption{Finite-size effects in localization transition in QREM. {\it Left panel}: $\ln (W_c^{\rm RRG}(n,\infty)/ W_c^{\rm QREM}(n))$ representing finite-size shift of the transition.  The large-$n$ asymptotics \eqref{eq:QREM-maintext-Wc-finite-size-shift} is shown by a black line.  {\it Right panel}:  Width of the transition $\ln(W_+(n)/W_-(n))$ and its finite-size asymptotics  
  \eqref{eq:QREM-maintext-Wc-finite-size-width}.
    Symbols: numerical exact-diagonalization data for $W_c^{\rm QREM}(n)$ from level statistics (mean gap ratio $r$), see Sec.~\ref{subsec:numerics_QREM}. 
  Vertical dotted lines represent the lower border  $n^{\rm crit} \approx 22$ of the critical regime, in which
  $W_c^\text{QREM}(n) > W_c^\text{RRG}(n,\infty)/2$.
  }
  \label{fig:Wc_QREM_aymptotics}
\end{figure}

We analyze now the finite-size width of the QREM localization transition. For this purpose, we recall that observables that are used to detect the transition (such as the gap ratio $r$ of level statistics or the IPR $P_2$) have $N/N_\xi$ as a scaling parameter
(``volumic'' scaling) for $W < W_c(n, \infty)$ in the RRG model \cite{tikhonov2016anderson,garcia-mata17,biroli2018,tikhonov19statistics,tikhonov2021from,garcia-mata2022critical}. Thus, to estimate the disorder interval $[W_-(n,N),\, W_+(n,N)]$, in which the transition takes place (e.g., the level statistics evolves from a nearly-Wigner-Dyson form to a nearly-Poisson form), we define $W_-(n,N)$  and $W_+(n,N)$ via 
\begin{equation}
N_\xi(W_-(n,N)) = b_-N \,, \qquad
N_\xi(W_+(n,N)) = b_+N \,,
\label{eq:QREM-maintext-finite-size-W-plus-minus}
\end{equation}
where $b_+$ and $ b_-$ are numerical constants, with $b_+ > b_-$. These results can be translated to the QREM by setting $N=2^n$, see Appendix \ref{app:RRG}. In the critical regime, $n>n^{\rm crit} \approx 22$ we find for the transition width $\Delta W^{\rm QREM}(n) = W_+(n,N) - W_-(n,N)$  [see Eq.~\eqref{eq:QREM-Wc-finite-size-width}]
\begin{equation}
\frac{\Delta W^{\rm QREM}(n)}{W_c^{\rm QREM}(n)} =
\frac{4\pi^2}{3\ln^2 2} \: \ln(b_+/b_-) \: \frac{\ln^2 n}{n^3}\,.
\label{eq:QREM-maintext-Wc-finite-size-width}
\end{equation}
Comparing Eqs.~\eqref{eq:QREM-maintext-Wc-finite-size-width} and
\eqref{eq:QREM-maintext-Wc-finite-size-shift}, one observes that the transition sharpens with increasing $n$ faster (by an additional factor of $1/n$) than its finite-size shift decays, which can be traced back to an exponential growth of the volume with length in the RRG model and QREM, see a detailed discussion in Appendix \ref{app:RRG}. The transition interval $[W_-(n),W_+(n)]$ is shown by shading  in Fig.~\ref{fig:Wc_QREM_theory}.
We also show the transition width $\ln (W_+(n) / W_-(n))$ along with the corresponding asymptotics Eq.~(\ref{eq:QREM-maintext-Wc-finite-size-width}) in the right panel of Fig.~\ref{fig:Wc_QREM_aymptotics}.

Ahead of a detailed discussion of the results of our numerical simulations in Sec.~\ref{sec:numerics}, we include in Figs.~\ref{fig:Wc_QREM_theory} and  \ref{fig:Wc_QREM_aymptotics} exact-diagonalization results for the QREM.  
{\color{black} A good agreement between the analytical predictions and the numerical data is observed, both for the position of the transition and for its width. In consistency with the analytical prediction, the numerical data approach, with increasing $n$, the large-$n$ asymptotics. It is seen, however, that the slope of the numerical data is smaller than that of the analytically derived curve \eqref{eq:QREM-maintext-relation-RRG} in this range of $n$. These deviations 
can be attributed to finite-size corrections to our analytical treatment. We recall that the analytical curve for $W_c^{\rm QREM}$  in Fig.~\ref{fig:Wc_QREM_theory} is obtained by using 
Eq.~\eqref{eq:QREM-maintext-relation-RRG} 
in combination with Eqs.~\eqref{eq:RRG-maintext-Nxi-W-N},
\eqref{eq:RRG-maintext-rescaled-Nxi-1},
\eqref{eq:RRG-maintext-rescaled-Nxi-2}, and
\eqref{eq:RRG-maintext-rescaling-factor} for $W_c^{\rm RRG}$, which implies several sources for finite-size corrections. First, the analytical result for $W_c^{\rm RRG} (m, 2^n)$ is derived for $m \gg 1$ and $n \gg 1$ but then used, to compare with the numerics, for a moderately large $n$ and $m = n+1 \approx n$. Second, while the equality between the critical disorder in the QREM and RRG models, Eq.~\eqref{eq:QREM-maintext-relation-RRG} becomes exact in the large-$n$ limit, corrections to it are expected at finite $n$. 
We note that the numerical data in Fig.~\ref{fig:Wc_QREM_theory} exhibit a curvature corresponding to an increase of slope with $n$. Extrapolating this trend, we expect that the values of $W_c^{\rm QREM}(n)$ will stay close to our analytical curve also for larger $n$. 
}

We will return to a more detailed analysis of the numerical data for the QREM in Sec.~\ref{subsec:numerics_QREM}.

\subsection{uQD and u1D models: RRG-like approximation with energy correlations}
\label{sec:uQD-u1D-analytics}

Let us now consider the uQD and u1D models, {\color{black} defined in Sec.~\ref{sec:uQD-model} and ~\ref{sec:u1D-model} respectively,} as the limiting case of the QD and 1D models where Fock space hopping correlations have been removed: $\smash{(C_T^\text{uQD/u1D})_{\alpha\beta\mu\nu} =  \delta_{\alpha\mu}\delta_{\beta\nu}}$.
The difference from the QREM and the RRG model is that the energies $E_\alpha$ are now strongly correlated. In particular, for two states $\alpha$ and $\beta$ that are connected by a single spin flip (i.e., $r_{\alpha\beta}=1$), the energy difference satisfies $|E_\alpha-E_\beta| \le W$, while in the QREM (and in the associated RRG model) the typical energy difference is
$|E_\alpha-E_\beta| \sim n^{1/2} W$. 
A parametrically smaller difference $|E_\alpha-E_\beta|$ resulting from energy correlations in the uQD and u1D models favors resonances and, therefore, is expected to enhance delocalization. 
We will see that this expectation is indeed correct. 

We argue now that the large-$n$ asymptotics for $W_c(n)$ in the uQD and u1D models can be found by using an RRG-like approximation formulated in Ref.~\cite{Herre2023}. More specifically, this approximation was developed in Ref.~\cite{Herre2023} as an upper bound for $W_c(n)$ in genuine quantum-dot models; our point now is that it yields the correct asymptotics for the models of uQD and u1D type, with random uncorrelated Fock-space hoppings. 

The idea of this RRG-like approximation with energy correlations is the following. On the delocalized side of the transition, i.e., for $W$ 
smaller than $W_c$ but not too far from $W_c$, an eigenstate will delocalize over a narrow energy shell (but still containing a large number of states) around the given energy $E$. As everywhere in this paper, we choose $E=0$ (center of the band) for definiteness. Since only basis states with energies $E_\alpha$ close to zero are relevant, we should replace $\gamma(0)$ in Eq.~\eqref{eq:self_cons_critical_disorder_Herre} by $\gamma_1(0)$, where $\gamma_1(E)$
is a distribution of energies $E_\beta$ of states directly coupled to a state $\alpha$ with $E_\alpha=0$. This yields an equation for $W_c(n)$ in a model with energy correlations. 

In our models, states $\beta$ directly connected to the state $\alpha$ differ from $\alpha$ by flipping a single spin $\hat{S}_k$, so that
\begin{align}
E_\beta = E_\alpha \pm \epsilon_k =\pm \epsilon_k \,.
\end{align}
(Here we have neglected terms in $E_\beta - E_\alpha$ resulting from the $V^z$ interaction since they are of order unity and thus much smaller than the typical value of $\epsilon_k$ for $W \gg 1$. We have checked numerically that keeping these terms indeed leads only to a very small shift of $W_c(n)$, see
Appendix~\ref{app:RRG_with_energy_correls}.)
Therefore, $\gamma_1(E)$ is equal to the distribution of $\epsilon_k$, which we have chosen to be uniformly distributed over the interval $[-W, W]$,
\begin{align}
\gamma_1(E) = \frac{1}{2W}, \quad -W < E < W \,,
\label{eq:gamma1-E}
\end{align}
and thus $\gamma_1(0) = 1/2W$.  In combination with 
$\langle |T| \rangle = \sqrt{\pi}/2$, this yields
\begin{equation}
\gamma_1(0) \langle |T| \rangle = (\sqrt{\pi}/4) W^{-1} \,,
\end{equation}
which is to be substituted for $\gamma(0) \langle |T|\rangle$ in the RRG formulas of Sec.~\ref{sec:RRG_scaling}. Performing this substitution in
Eq.~\eqref{eq:self_cons_critical_disorder_Herre},
we finally obtain the equation for $W_c(n)$ in the uQD and u1D models:
\begin{align}
W_c = \sqrt{\pi}\: n\ln \frac{4W_c}{\sqrt{\pi}} \,.
\label{eq:self_cons_critical_disorder_RRG}
\end{align}
The leading asymptotic behavior of the solution of this equation reads:
\begin{align}
W_c^{\rm uQD}(n), \: W_c^{\rm u1D}(n) \simeq \sqrt{\pi}\, n\ln n \,.
\label{eq:Wc_RRG_scaling}
\end{align} 
Equations \eqref{eq:self_cons_critical_disorder_RRG} 
and \eqref{eq:Wc_RRG_scaling} are uQD and u1D counterparts of QREM equations
\eqref{eq:RRG-maintext-Wc-rescaled-equation} and \eqref{eq:RRG-Wc-asympt}. In notations of Sec.~\ref{sec:RRG_scaling}, the solution of 
Eq.~\eqref{eq:self_cons_critical_disorder_RRG} is $W_c^{\rm RRG}(n,\infty)$, where the superscript ``RRG'' now means the RRG-like approximation corresponding to the uQD and u1D models, which gives the large-$n$ asymptotics for $W_c(n)$ in these models. The RRG-like approximation further predicts that, for moderately large $n$, the actual values of $W_c^{\rm uQD}(n)$ and $W_c^{\rm u1D}(n)$ should deviate from their asymptotic form according to Eq.~\eqref{eq:QREM-maintext-Wc-finite-size-shift}
(where the superscript ``QREM'' should now be replaced by ``uQD'' or ``u1D''). Via the same token, the width of the transition should be described by Eq.~\eqref{eq:QREM-maintext-Wc-finite-size-width}, again with the same replacement of the superscript. 

We provide now a more formal argument in favor of validity of the RRG-like approximation leading
to Eqs.~\eqref{eq:self_cons_critical_disorder_RRG}
and \eqref{eq:Wc_RRG_scaling}. Let us fix some positive number $c < 1$ and keep in every realization of the model only vertices satisfying $|E_\alpha| < cW$. For a small $c$ and sufficiently large $n$ (such that $cn \gg 1$), we will then get a graph with a coordination number $n' = cn$ and with the distribution $\gamma'(E)$ being a box distribution on $[-cW, cW]$, so that $\gamma'(0) = 1 /2cW$. Furthermore, for a small $c$ the energies $E_\alpha$ will be almost uncorrelated. We can thus use the RRG formula \eqref{eq:self_cons_critical_disorder_Herre} to determine the critical disorder in the resulting model, which yields
\begin{align}
W_c = \sqrt{\pi}\: n\ln \frac{4cW_c}{\sqrt{\pi}} \,.
\label{eq:Wc-RRG-corr-c}
\end{align}
This is identical to Eq.~\eqref{eq:self_cons_critical_disorder_RRG} up to a factor $c$ inside the logarithm. 
Therefore, Eq.~\eqref{eq:self_cons_critical_disorder_RRG} holds, with uncertainty only in the numerical coefficient of order unity in the argument of the logarithm. Clearly, this coefficient is of no importance for the asymptotic behavior; in particular, it does not affect the leading large-$n$ asymptotics \eqref{eq:Wc_RRG_scaling} (and, in fact, also dominant subleading corrections to it). 

Comparing Eq.~\eqref{eq:Wc_RRG_scaling} with 
the result \eqref{eq:RRG-Wc-asympt} for the uncorrelated RRG model, we see that energy correlations lead to a parametric enhancement of {\color{black}$W_c(n)$}, i.e., to a parametrically larger ergodic phase, in full agreement with a qualitative argument in the beginning of Sec.~\ref{sec:uQD-u1D-analytics}. We note that this result is opposite to 
the conclusion of Refs.~\cite{Roy2020numerics,Roy2020}. The disorder $W$ that is argued to be the localization transition point in Refs.~\cite{Roy2020numerics,Roy2020} is, in fact, unrelated to the localization transition; rather, it corresponds to a crossover between different regimes deeply within the ergodic phase \footnote{{\color{black} Specifically, this is a crossover between the random-matrix-theory regime
(where eigenstates spread over all Fock-space sites)
and the self-consistent golden rule regime (also ergodic but with eigenstates spreading over energy windows parametrically smaller than the total disorder-induced bandwidth), see Ref.~\cite{Herre2023}.
These correspond to regimes I and II of Ref.~\cite{Monteiro2021}, whereas the MBL transition is located on the border of regimes III and IV in terminology of that paper.}}.
 It is also worth noting that the RRG model with energy correlations as considered in Ref.~\cite{Roy2020numerics} is actually ill-defined since the corresponding covariance matrix is not positive definite for almost any RRG realization, see Appendix \ref{app:RRG-corr} for detail. Contrary to this, our uQD and u1D models are defined on a hypercube graph, which makes them well-defined. 

The expected applicability of the  RRG-like approximation to the uQD and u1D models is crucially related to the fact that hopping amplitudes in these models are independent random variables. This suppresses interference between different paths on the graph arising in higher orders of the perturbation theory and
having the same initial and final state. Such interference effects tend to reduce $W_c$ (i.e., to promote localization) \cite{gornyi2017spectral} and are of central importance in genuine many-body models, like the QD and 1D models that we are going to discuss. In the absence of interference, contributions of different paths can be viewed as independent, which is at heart of the RRG-like approximation.

\subsection{QD model}
\label{sec:QD-analytics}

We now turn to analytical predictions for the QD model, {\color{black} defined in Sec.~\ref{sec:QD-model-def}.} Presentation in this subsection {\color{black}largely}  uses Ref.~\cite{gornyi2017spectral}, with modifications for the QD model under consideration.
We also refer the reader to Ref.~\cite{gornyi2017spectral} for a discussion of relations to earlier analytical works on localization in many-body quantum-dot models \cite{altshuler1997quasiparticle,mirlin1997localization,jacquod1997emergence,silvestrov1997decay,silvestrov1998chaos,gornyi2016many}. 
The theoretical considerations presented below are based on an analysis of resonances within a perturbative expansion with respect to the term $\hat{H}^\text{QD}_1$, 
Eq.~\eqref{eq:QD}, in the QD Hamiltonian. 

Consider a basis state $\alpha$ with energy $E_\alpha=0$ and another state $\mu$ separated by a Hamming distance $r_{\alpha\mu}$. Clearly, we will have an admixture of the state $\mu$ to the state $\alpha$ in the order $r_{\alpha\mu}$ of the perturbation theory. There will be a contribution to the corresponding amplitude from any Hilbert-space path of length $r_{\alpha\mu}$ that connects these two states:
$\alpha \to \beta \to \gamma \to \ldots \to \lambda \to \mu$.
(It is easy to see that there are exactly $r_{\alpha\mu}!$ such paths.) 
The dimensionless coupling $\eta_{\rm path}$ that is associated with such a path and controls the resulting hybridization between $\alpha$ and $\mu$ is given by
\begin{equation}
\eta_{\rm path} = \frac{T_{\alpha\beta} T_{\beta\gamma} \ldots T_{\lambda\mu} }{E_\beta E_\gamma \ldots E_\mu}\,.
\label{eq:eta-path}
\end{equation}
The total $\eta$ is given by a sum of $\eta_{\rm path}$  over the paths from $\alpha$ to $\mu$. 
If $\eta \gtrsim 1$, the states  $\alpha$ and $\mu$ are in resonance and strongly hybridize. In the opposite limit, $\eta \ll 1$, the hybridization is negligibly weak.

{\color{black} For the uQD model (and also u1D model), numerators of $\eta_{\rm path}$ for different paths are uncorrelated (since hopping matrix elements are uncorrelated random variables), so that the interference between different paths yields a random sign and thus does not yield a systematic cancellation (``no interference'').} As a result, it is not essential for counting resonances that many paths from $\alpha$ end up in the same state $\mu$. This justifies the RRG-like approximation discussed in Sec.~\ref{sec:uQD-u1D-analytics}.
On the other hand, in a genuine many-body model, like the QD model, interference between the paths is essential. In particular, if one totally discards contributions of diagonal ($\hat{S}_i^z\hat{S}_j^z$) interactions to energies in the denominator of Eq.~\eqref{eq:eta-path}, the sum of $r_{\alpha\mu}!$ terms \eqref{eq:eta-path} is identically equal to a single term of a similar type. This cancellation suppresses hybridization and therefore delocalization. The diagonal interactions strongly reduce the effect of this cancellation by reshuffling the energies $E_\nu$ of the basis states. It follows that 
$W_c^{\rm uQD}(n)$ given by
the RRG-like approximation of Sec.~\ref{sec:uQD-u1D-analytics} provides an upper bound for the critical disorder $W_c^{\rm QD}(n)$ of the QD model:
\begin{equation}
W_c^{\rm QD}(n) \le W_c^{\rm uQD}(n) \sim n \ln n \,,
\label{Wc-QD-upper-boundary}
\end{equation}
where the large-$n$ asymptotics of $W_c^{\rm uQD}(n)$ is given by Eqs.~\eqref{eq:self_cons_critical_disorder_RRG}, 
\eqref{eq:Wc_RRG_scaling}.

To obtain a lower bound for $W_c^{\rm QD}(n)$, let us analyze, up to what Hamming distance (equivalently, order of perturbation theory) can we proceed with finding resonances for a typical basis state $\alpha$ in a typical realization of random QD Hamiltonian with disorder $W$. For $W < 1$, the energy difference between the state $\alpha$ and a state $\beta$ connected to it by a single spin flip (i.e., such that $r_{\alpha\beta}=1$) satisfies $|E_\alpha-E_\beta| < 1$, i.e., is typically smaller than the hopping $T_{\alpha\beta}$. Thus, the state $\alpha$ is in resonance with all $n$ states directly coupled to it. We can proceed via such first-order resonances up to the largest Hamming distance $n$, so that the system is in the ergodic phase. Thus, $W_c(n) \gtrsim 1$.
This lower bound can be, however, strongly improved. 

For this purpose, consider a disorder $W \sim n/p$, where $1 \ll p \ll  n$. Now, a state $\alpha$ has typically $\sim p$ first-order resonances, i.e., it is resonantly connected to $\sim p$ direct neighbors on the Fock-space graph. These resonances ensure a strong hybridization of the state $\alpha$ with at least $\sim 2^p$ other many-body states. 
The idea now (see an analogous discussion for a different QD model in Appendix B of Ref.~\cite{gornyi2017spectral})
is that, already for a relatively small $p$, these $\sim 2^p$ states form an ergodic  ``resonant subsystem''. Furthermore, this resonance subsystem becomes very efficient in making the whole system ergodic. As shown in Appendix \ref{app:QD-lower-bound}, this mechanism of ergodization becomes operative when $p$ reaches the value
\begin{equation}
p \sim n^{1/4} \ln^{1/4} n \,. 
\end{equation}
Substituting this value into $W=n/p$, we obtain the lower bound for critical disorder,
\begin{equation}
W_c^{\rm QD}(n) \gtrsim n^{3/4} (\ln n)^{-1/4} \,.
\label{Wc-QD-lower-boundary}
\end{equation}
Combining Eqs.~\eqref{Wc-QD-upper-boundary} and \eqref{Wc-QD-lower-boundary}, we conclude that
\begin{equation}
n^{3/4} (\ln n)^{-1/4} \lesssim W_c^{\rm QD}(n)
\lesssim n \ln n \,.
\label{Wc-QD-final}
\end{equation}

\subsection{1D model}
\label{sec:analytics-1D}

For 1D many-body systems with a short-range interaction, it was found in Refs.~\cite{gornyi2005interacting,basko2006metal,Ros2015a} from the analysis of a perturbative expansion that
\begin{equation}
W_c^{\rm 1D} (n) \sim 1 \,,
\label{eq:1D-Wc-n}
\end{equation}
in the sense that the large-$n$ asymptotics of $W_c(n)$ does not depend on $n$. A formal proof of Eq.~\eqref{eq:1D-Wc-n} was provided in Refs.~\cite{imbrie16a,Imbrie2016JSP} under a physically plausible assumption of a limited level attraction. In Refs.~\cite{roeck17,Thiery2017a}, the effect of exponentially rare regions of anomalously weak disorder (``ergodic spots'') on MBL transitions was studied. It was found, that in spatial dimensions $d > 1$  such ergodic spots lead to ``avalanches'', making the whole system ergodic. As a result, the critical disorder $W_c(n)$ grows without bounds when the system size $n$ increases. Since an exponentially large system is needed to find an ergodic spot, the growth of $W_c(n)$ in $d>1$ geometry is slower than any power law \cite{gopalakrishnan2019instability,doggen2020slow}. At the same time, in 1D geometry, the avalanche mechanism does not modify the result \eqref{eq:1D-Wc-n}, in agreement with Refs.~\cite{imbrie16a,Imbrie2016JSP}. 

We briefly comment on numerical studies of the MBL transition in 1D systems. Most of the numerical works dealt with an XXZ spin-chain model in a random field. For a choice of parameters that has become standard, exact-diagonalization studies of systems of length $n \approx 20$ yielded a finite-size estimate of the critical disorder $W_c^{\rm XXZ}(20) \approx 4$ \cite{luitz2015many,Laflorencie2020chain}. It was also found in numerical simulations that finite-size effects are rather strong in this model: $W_c^{\rm XXZ}(n)$ exhibits a sizeable drift towards larger values when $n$ increases.
In particular, the matrix-product-state study of systems of the length $n=50$ and $n=100$ \cite{Doggen2018a,doggen2021many} yielded an estimate for the critical disorder $W_c^{\rm XXZ}(50) \approx W_c^{\rm XXZ}(100) \approx 5.5$, substantially larger than $W_c^{\rm XXZ}(20)$.
During the last couple of years, the drift of $W_c(n)$ and the related physics observed in numerical simulations of 1D models have been addressed in many papers 
\cite{Sierant2020b,Weiner2019slow,Kiefer2020a,Kiefer2020b,Kiefer2021a,Sierant2022challenges,Evers2023internal,Luitz2020absence,Ghosh2022resonance,Panda2020a,Suntajs2020a,Sels2021dynamical,abanin2019distinguishing,Sierant2020a,Morningstar2022avalanches,crowley2022constructive,long2023phenomenology,chavez2023ultraslow,biroli2023largedeviation}. In particular, these works addressed very slow (but still detectable) dynamics that is observed for numerically studied system lengths $n$ at disorder $W$ above the finite-size estimate $W_c(n)$. Many works pointed out that the observed behavior is consistent with a finite value of $W_c$ in the thermodynamic limit, Eq.~\eqref{eq:1D-Wc-n}, and, moreover, is not unexpected. Indeed, it is known that the finite-size critical disorder $W^{\rm RRG}_c(m,N)$  of an RRG model with a fixed {\color{black} coordination number $m+1$} exhibits a strong drift when the Hilbert-space dimension $N$ increases. Already for the smallest $m+1=3$ the magnitude of the drift $W^{\rm RRG}_c(m,\infty) / W^{\rm RRG}_c(m,N_{\rm ED})$ (where $N_{\rm ED}$ are system sizes that can be studied by exact diagonalization) is $\sim 1.2 - 1.3$, and it becomes as big as $\sim 5$ for $m+1=20$ \cite{Herre2023}. (We recall that, for the RRG model, we have the luxury of knowing exactly $W^{\rm RRG}_c(m,\infty)$.)
Since 1D many-body systems may be expected to have stronger fluctuations than the RRG model (in particular, due to effects of rare spatial spots), a finite-size drift up to $W^{\rm XXZ}_c(\infty) \approx 10$
(which corresponds to $W^{\rm XXZ}_c(\infty) / W^{\rm XXZ}_c(20) \sim 2.5$) as suggested by several numerical studies (see, e.g., Refs.~\cite{Sierant2022challenges,Evers2023internal,biroli2023largedeviation}) would not be too surprising.

While one expects to see universal features of the MBL physics in different 1D systems, the magnitude of finite-size effects may depend on the specific model. If one could identify 1D models in which finite-size effects are weaker than in the XXZ model, this could be useful for numerical studies of the MBL transitions. A very recent paper \cite{Sierant2023stability} suggests that finite-size effects in some Floquet models may be less severe. 
{\color{black}
Our numerical studies of the single-spin-flip 1D model    defined in Sec.~\ref{sec:1D-model-def} show that it is a promising candidate for exploring the 1D MBL transition by computational means. Further work is needed to see whether it may have particular advantages over other 1D models in this respect.}

Summarizing, we presented in Sec.\ref{sec:analytical} analytical predictions for the scaling of critical disorder $W_c(n)$ in all the models considered in this paper. Derivation of most of these results involves some approximations. Furthermore, the analytical considerations assume large $n$, and it is not a priori clear how well the values of $n$ accessible in numerical simulations satisfy the large-$n$ assumption. It is thus of crucial importance to study these models numerically and to compare the results between themselves and with the analytical predictions. Our computational (exact-diagonalization) results and their analysis are presented in the next two sections.

\section{Numerical approach. Setting the stage with QREM}
\label{sec:numerics}

In this Section and in Sec.~\ref{sec:numerics-models-with-corr}, we present and analyze exact-diagonalization numerical results for the ergodicity-to-MBL transition in the models defined in Sec.~\ref{sec:model_definitions}, with a particular focus on the scaling of the critical disorder $W_c(n)$.   In addition, we study also the scaling of the transition width $\Delta \ln W (n)$.  After explaining in Sec.~\ref{subsec:numerical_implementation}  how the models are implemented, we specify in  Sec.~\ref{sec:numerics-Wc} the observables that are studied to explore the transition, to locate $W_c(n)$, and to determine the transition width.  After this, we analyze numerical results for the QREM in Sec.~\ref{subsec:numerics_QREM}. Out of all the models that we consider, the QREM is best understood analytically (due to its direct connection to the RRG model). Thus, a comparison of the numerical results for the QREM with the corresponding analytical predictions provides a benchmark for numerical investigations of other models by the same methods, which are the subject of Sec.~\ref{sec:numerics-models-with-corr}.

\subsection{Numerical implementation}
\label{subsec:numerical_implementation}

The QD and 1D models are genuine many-body models and are defined directly by their many-body Hamiltonians given in Sec.~\ref{sec:QD-model-def}
and Sec.~\ref{sec:1D-model-def}, respectively. 
Implementation of these models is straightforward: we generate random samples for the on-site energies $\epsilon_i$ and the interaction couplings $V_{ij}^a$ and build the Hamiltonian matrix from the definition.

The uQD and u1D models, as well as the QREM, are defined by their
Fock space representation (\ref{eq:Fock_space_rep}), as specified in Sec.~\ref{sec:uQD-model}, Sec.~\ref{sec:u1D-model}, and
Sec.~\ref{sec:QREM}, respectively. To implement these models, we first generate recursively the Fock-space graph structure recursively. Starting from the configuration $\ket{\downarrow,\downarrow,\dots, \downarrow}$, we find the $n$ connected states by flipping one spin at a time. This procedure is then repeated on these $n$ states, and repeated again until all $2^n$ nodes (Fock-space states) and $2^{n-1}n$ edges have been found. We then generate a sample of uncorrelated $\smash{\left\{\epsilon_i \right\}}$ uniformly distributed on $\left[ -W, W \right]$, as well as a sample of uncorrelated $\smash{\left\{V_{ij}^z\right\}}$ normally distributed with zero mean and variance unity, and associate the on-site energy $E_\alpha$ to each node by computing them using Eq.~(\ref{eq:QD_on_site_energies}) for the uQD model and Eq.~(\ref{eq:1D_on_site_energies}) for the u1D model. 
For the QREM, energies $E_\alpha$ are generated as independent random Gaussian variables according to Eq.~\eqref{eq:distrib_energies_QD-repeated}.
Finally, for each of these three models, we generate independent random matrix elements  $T_{\alpha\beta} \sim \mathcal{N}(0,1/2) + i\mathcal{N}(0,1/2)$ for all edges of the graph.

\begin{figure}[t!]
  \includegraphics[width=0.49\columnwidth]{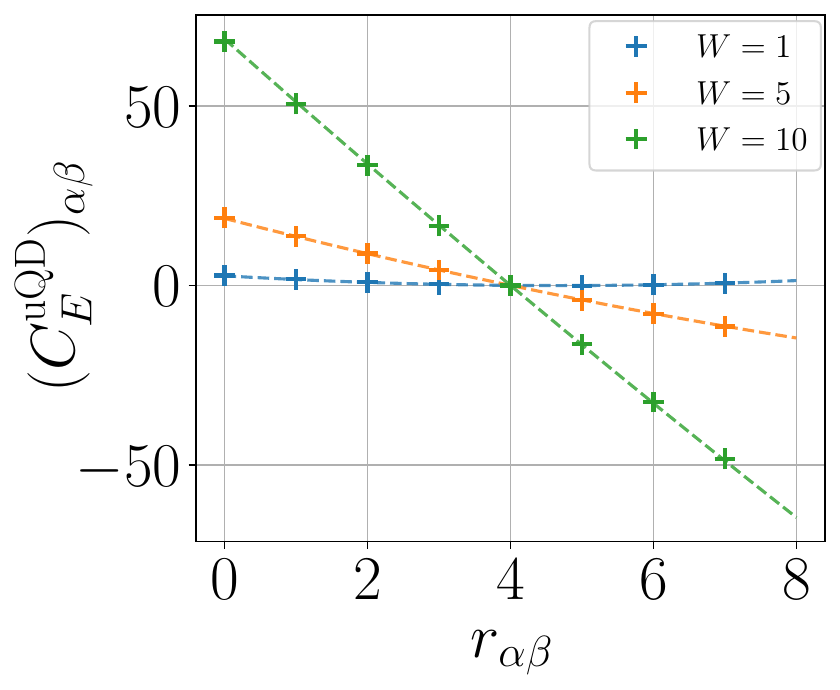}\hfill
  \includegraphics[width=0.49\columnwidth]{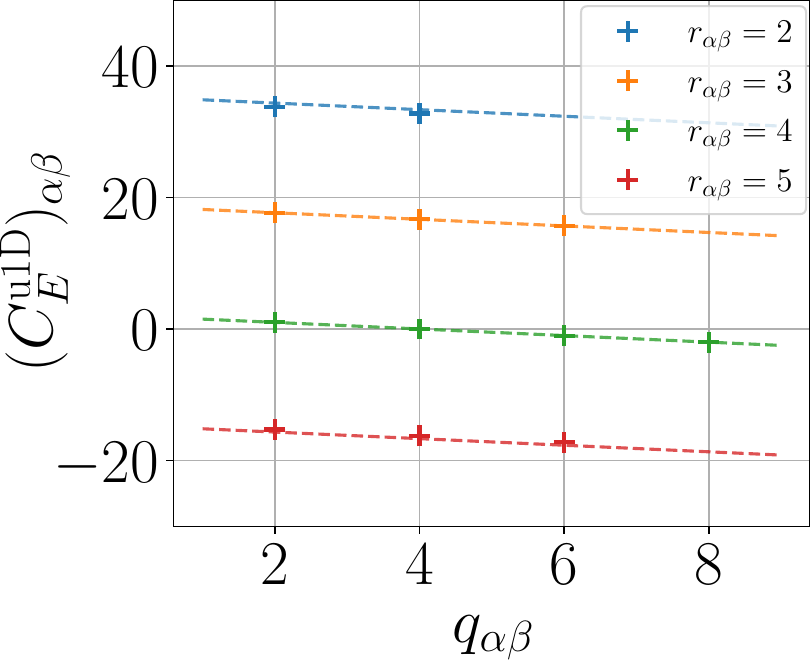}
  \caption{Energy correlations $(C_E)_{\alpha\beta}$ on the Fock-space graph for the uQD and u1D models with $n=8$ spins. 
  {\it Left panel:} $(C_E^{\rm uQD})_{\alpha\beta}$
  as a function of the Hamming distance $r_{\alpha\beta}$, for disorder strengths $W=1$, 5, and 10. The numerical data (crosses) is averaged over $10000$ disorder realizations. The analytical result (\ref{eq:correl_energies_final}) with the corresponding parameters is shown by dashed lines.
  {\it Right panel:} $(C_E^{\rm u1D})_{\alpha\beta}$
  as a function of $q_{\alpha\beta}$ defined below Eq.~\eqref{eq:1D-correl_energies_intermediate} for disorder $W=10$ and for various values of $r_{\alpha\beta}$. Dashed lines represent the analytical result \eqref{eq:1D-correl_energies_intermediate}.
  }
  \label{fig:energy_correlations_numerics}
\end{figure}

To check that our implementation is correct, we plot in Fig.~\ref{fig:energy_correlations_numerics} numerically evaluated energy correlations $(C_E^{\rm uQD})_\alpha\beta$  and
$(C_E^{\rm u1D})_{\alpha\beta}$
on the Fock-space graph, for a system size $n=8$, averaged over $10000$ disorder realizations. The left panel presents the results for $(C_E^{\rm uQD})_{\alpha\beta}$ as a function of the Hamming distance $r_{\alpha\beta}$ for several values of the disorder strengths $W$. Very good agreement with the analytical formula (\ref{eq:correl_energies_final}) (shown by dashed lines) is seen. On the right panel, we show the energy correlations for the u1D model at fixed $W=10$, for various values of $r_{\alpha\beta}$, as a function of $q_{\alpha\beta}$, defined below Eq.~\eqref{eq:1D-correl_energies_intermediate}. The agreement with the analytical prediction \eqref{eq:1D-correl_energies_intermediate} (dashed lines) is excellent, thus validating our implementation of the model.

\subsection{Observables}
\label{sec:numerics-Wc}

To study the transition, we compute the $2^n/10$ eigenvalues ${\cal E}_\sigma$ and associated eigenvectors  $\ket{\psi_\sigma}$ close to the center of the band ${\cal E} = 0$ for various system sizes $n\in \{8,10,12,14\}$ and for a wide range of disorder strengths.
By using this data, we evaluate two observables: the average gap ratio $r$ characterizing the energy spectrum and the average inverse participation ratio (IPR) characterizing eigenfunctions.

We use the eigenenergies ${\cal E}_\sigma$ to compute the consecutive level spacings $\delta_\sigma = {\cal E}_\sigma-{\cal E}_{\sigma+1}$  and to determine the mean adjacent gap ratio $r$,
\begin{align}
r = \left\langle \frac{\min(\delta_\sigma ,\delta_{\sigma-1})}{ \max(\delta_\sigma, \delta_{\sigma-1})} \right\rangle,
\label{eq:r}
\end{align}
with the averaging performed over disorder realizations and over the index $\sigma$ labeling $2^n/10$ eigenvalues around the band center. 
The number of disorder realizations for each of the models is specified below in captions to the figures where the data is presented. 

It has been shown \cite{Oganesyan2007, Pal2010a, luitz2015many, giraud2022probing, Atas2013} that $r$ is very useful for locating the localization transition. 
On the localized side, it takes the value $r_{\rm P} \simeq 0.3863$ characteristic for the Poisson statistics, while on the ergodic side, {\color{black} it is given by $r_\text{GUE} \simeq 0.5996$} for models with symmetry of the Gaussian unitary ensemble (GUE) \cite{Atas2013, giraud2022probing}. 
The fact that $r$ has the known and distinct limiting values of order unity in both phases makes it a very convenient observable for determining the position and the width of the transition in finite systems.
In the thermodynamic limit, $n\to \infty$, the mean gap ratio $r$ would exhibit a jump between these two values at the critical point, $W=W_c$. For a finite $n$, this discontinuity is smeared to a crossover. The location of this crossover yields a finite-size estimate $W_c(n)$ for the critical point, and the width of the crossover corresponds to the width $\Delta \ln W(n)$ of the critical regime. We define a finite-size estimate $W_c(n)$ as a 
{\color{black} value of $W$ at which $r(W) = (r_{\rm GUE} + r_{\rm P})/2 \simeq 0.493$ \footnote{{\color{black} This definition of the finite-size critical disorder allows us to determine $W_c(n)$ numerically in an unbiased way, i.e., without using any information on the scaling of $W_c$ with $n$. 
This is essential in models in which $W_c$ exhibits a power-law growth with $n$, as in most of the models considered in this paper.}}}. 
%
%
Further, the transition  interval  $[W_-(n),\, W_+(n)]$ is obtained from 
$r(W_-) = 0.8 \, r_{\rm GUE} + 0.2 \, r_{\rm P} $ and 
$r(W_+) = 0.2 \, r_{\rm GUE} + 0.8 \, r_{\rm P} $.
We characterize the width of the transition by $\ln(W_+(n)/W_-(n))$.

In addition to the gap ratio $r$, we calculate
the average IPR, which is a well-known observable in the context of localization transitions \cite{evers08,tikhonov2021from}. The IPR is defined as 
\begin{align}
P_2 = \left\langle \sum\limits_{\alpha = 1}^{2^n} |\bra{\alpha}\ket{\psi_\sigma}|^4 \right\rangle,
\label{eq:P_2}
\end{align}
where the sum goes over the vertices $\alpha$ of the graph, and the averaging is performed both over the eigenvectors $\ket{\psi_\sigma}$ and over disorder realizations. Generally, the dominant factor in the scaling of $P_2$ with the Hilbert space volume $N= 2^n$ at large $n$ is of power-law type, $P_2 \propto N^{-\tau(W)}$. 
On the ergodic side of the transition, $W<W_c$, one has $\tau(W)=1$.
At the transition point, $\tau(W)$ exhibits a jump from this ergodic value to $\tau(W_c) < 1$. For the RRG model (or QREM), one has
\cite{tikhonov2021from} $\tau(W_c) = 0$, and the same behavior can be expected for models properly described by an RRG-like approximation (such as our uQD and u1D models). For 1D models, and, more generally, for models with a structure in real space, one finds $0 < \tau(W_c) < 1$, with $\tau(W)$ decreasing $\propto W^{-1}$ in the MBL phase \cite{luitz2015many,gornyi2017spectral,tikhonov18,mace19multifractal}. The discontinuity of $\tau(W)$ and $W=W_c$ suggests to use the derivative to locate the transition \cite{Herre2023},
\begin{equation}
\alpha(W) = \frac{{\rm d} \ln P_2}{{\rm d}\ln W}  \,.
\label{eq:alpha-W-def}
\end{equation}
In the limit $n \to \infty$, this derivative diverges at $W=W_c$. For a finite $n$, this divergence is smeared and one finds a maximum in the dependence $\alpha(W)$, whose position can serve as a finite-size approximation of the critical disorder. This numerical approach was verified in Ref.~\cite{Herre2023} by using RRG models (with various coordination numbers) for which $W_c(n\to\infty)$ was found analytically.

\subsection{Numerical results for the QREM}
\label{subsec:numerics_QREM}

\begin{figure}[t!]
  \centering
  \includegraphics[width=0.9\columnwidth]{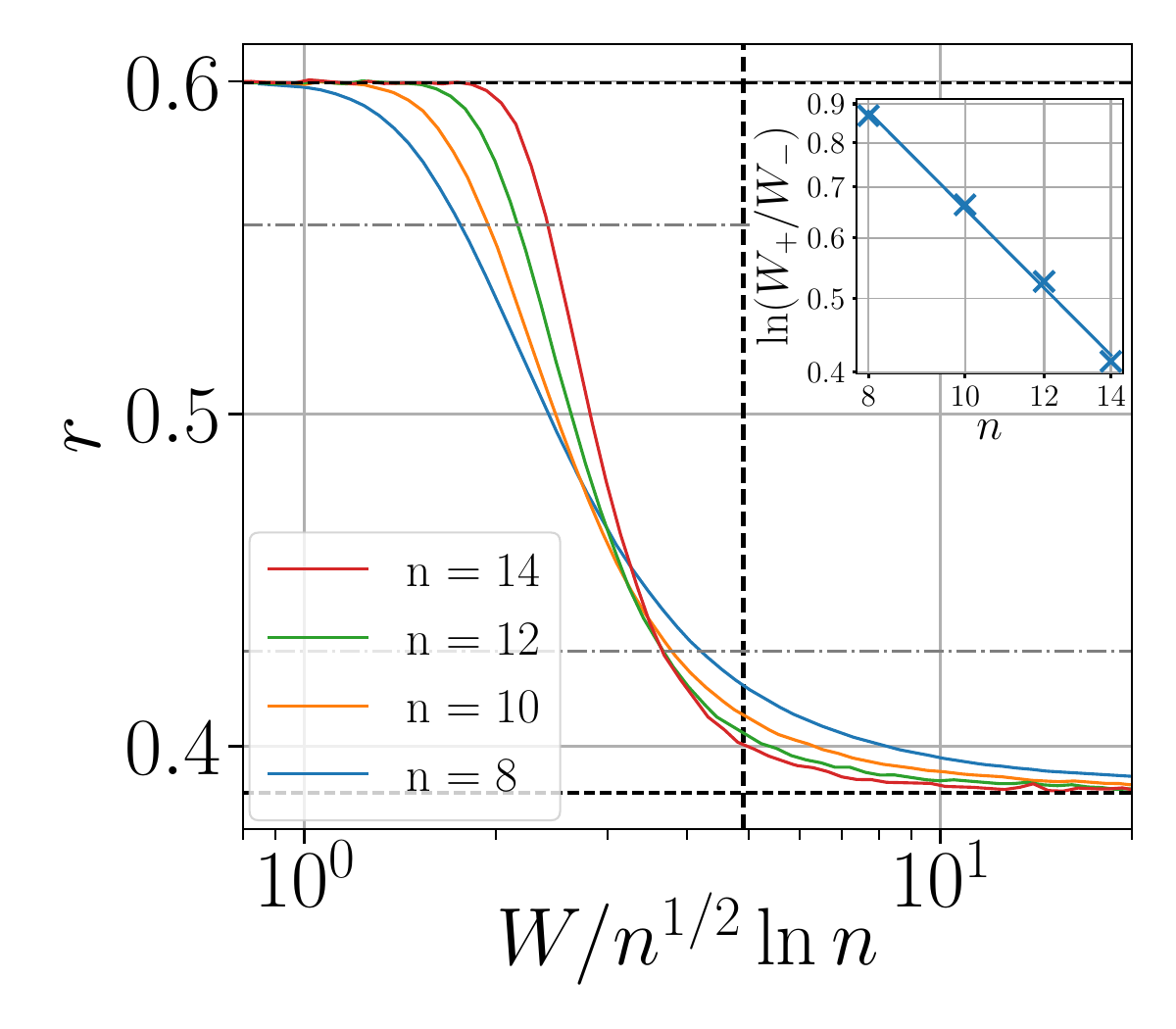}
  \caption{Localization transition in QREM via the level statistics. {\it Main panel:} Mean adjacent gap ratio $r$ as a function of rescaled 
  disorder $W/n^{1/2}\ln n$. The Wigner-Dyson and Poisson values, $r_\text{GUE} \simeq 0.5996$ and $r_{\rm P} \simeq 0.3863$, are marked by horizontal dashed lines. The horizontal dash-dotted lines are $r_- = 0.8 \, r_{\rm GUE} + 0.2 \, r_{\rm P} $ and $r_+ = 0.2 \, r_{\rm GUE} + 0.8 \, r_{\rm P} $ that serve to define $W_-$ and $W_+$  (via $r(W_-)=r_-$ and $r(W_+)=r_+$) used to calculate the transition width shown in the right panel.
  The vertical dashed line
marks the analytical prediction for $W_c(n)$ in the limit $n\to \infty$, see Eq.~\eqref{eq:RRG-Wc-asympt}.
  {\it Inset:} transition width $\ln (W_+/W_-)$ as a function of $n$. The straight line representing a power-law fit {\color{black}
  $\ln (W_+/W_-) \sim n^{-1.32}$}
  is a guide to the eye. The data is averaged over 320000, 20000, 2500, and 400 disorder realizations for $n=8$, 10, 12, and 14, respectively.
  }
  \label{fig:rW_QREM}
\end{figure}

In Fig.~\ref{fig:rW_QREM}, we show the data for the mean gap ratio $r$ in the QREM.  The disorder in this plot is rescaled as $W/n^{1/2} \ln n$, in accordance with the analytically predicted large-$n$ scaling of $W_c^{\rm QREM}(n)$, Eq.~\eqref{eq:RRG-Wc-asympt}.
It is seen that this rescaling leads to {\color{black} a good collapse} of numerical values of $W_c$ [values of $W$ at which {\color{black} $r=(r_{\rm GUE}+r_{\rm P})/2 \simeq 0.493$}], i.e., that the asymptotic $n^{1/2} \ln n$ scaling is observed with a good accuracy already at relatively small values of $n=8 - 14$.
{\color{black} After rescaling, a relatively weak finite-size drift of the critical disorder towards its large-$n$ asymptotics (vertical dashed line) remains, in full correspondence with the data slowly approaching the large-$n$ asymptotics  in  Fig.~\ref{fig:Wc_QREM_theory}.  This finite-size drift bears analogy with that of $W_c$ in RRG models at fixed coordination number and for increasing system size, see, e.g., Ref.~\cite{Herre2023}. }
It is also seen that the transition rapidly becomes sharper with increasing $n$. The transition width as a function of $n$ is shown in the inset of Fig.~\ref{fig:rW_QREM}. The data is well fitted by the power law {\color{black}$ \ln(W_+/W_-) \sim n^{-1.32}$.}  
The analytical result,
Eq.~\eqref{eq:QREM-maintext-Wc-finite-size-width}, predicts a still faster sharpening of the transition $\sim n^{-3}\ln^2 n $ in the large-$n$ limit. This difference is, however, fully expected as can be seen in the right panel of Fig.~\ref{fig:Wc_QREM_aymptotics}. If we describe the analytical $n$-dependence of the transition width by a flowing effective exponent 
\begin{equation}
\mu(n) = - \frac{\partial \ln \ln (W_+/W_-)}{ \partial \ln n} \,,
\label{eq:transition-width-mu-n}
\end{equation}
we find {\color{black}$\mu(n) \approx 1.32$} for values of $n$ corresponding to our numerical simulations. 

\begin{figure}[t!]
  \centering
  \includegraphics[width=0.49\columnwidth]{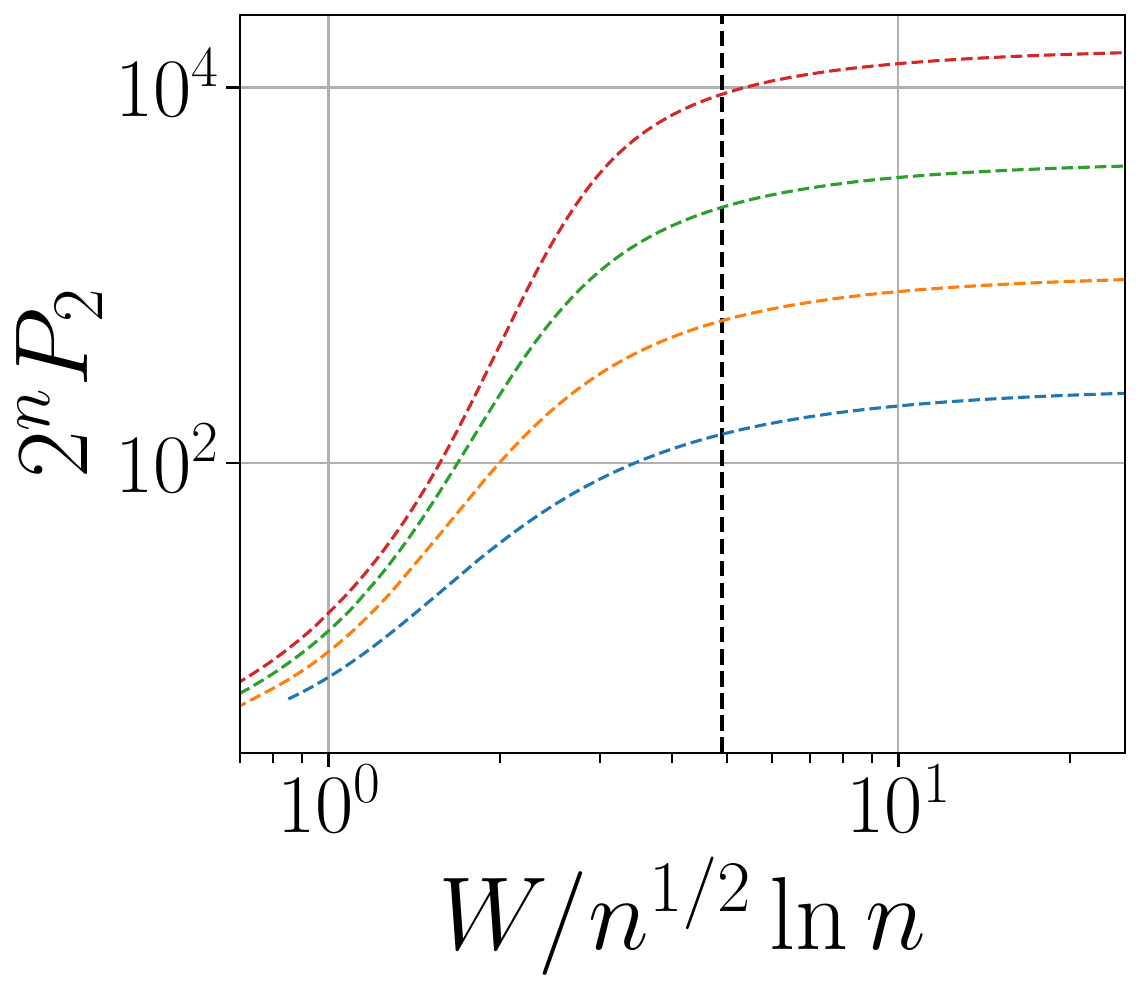}
  \includegraphics[width=0.49\columnwidth]{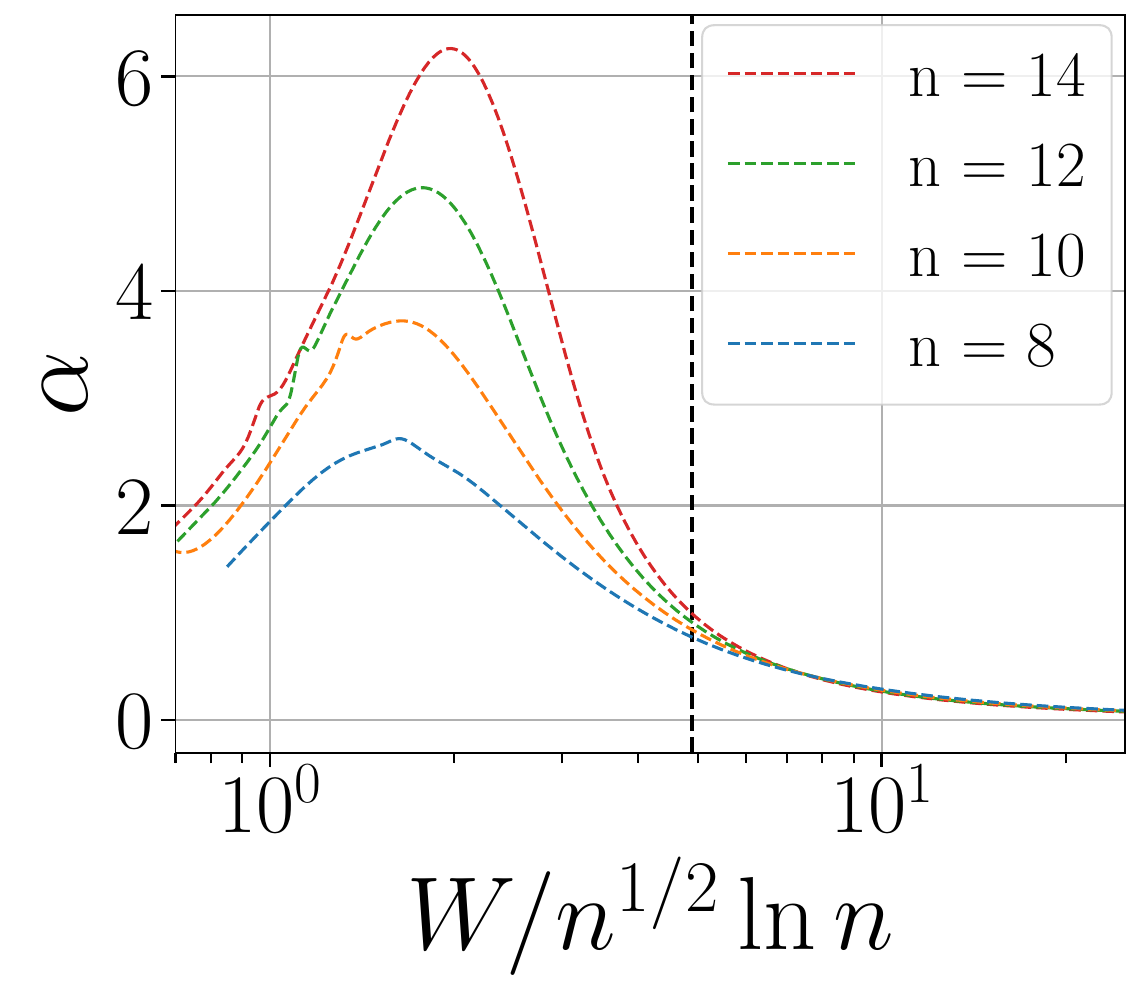}
  \caption{Transition in QREM via average IPR $P_2$ of eigenstates. {\it Left panel:} $2^n P_2$ as a function of rescaled disorder $W/n^{1/2}\ln n$.   
  {\it Right panel:} Logarithmic derivative $\alpha$, Eq.~\eqref{eq:alpha-W-def}. The maximum of the dependence $\alpha(W)$ serves as a finite-size approximation of the critical disorder $W_c(n)$. The vertical dashed lines in both panels
mark the analytical prediction for $W_c(n)$ in the limit $n\to \infty$, see Eq.~\eqref{eq:RRG-Wc-asympt}. The data is averaged over 320000, 20000, 2500, and 400 disorder realizations for $n=8$, 10, 12, and 14, respectively.
  }
  \label{fig:IPR_QREM}
\end{figure}

Figure \ref{fig:IPR_QREM} presents the data for the IPR (left panel) and for its logaithmic derivative $\alpha$, Eq.~\ref{eq:alpha-W-def} (right panel), again as functions of rescaled disorder $W/n^{1/2} \ln n$. These results provide additional support to the analytically predicted scaling $W_c(n) \sim n^{1/2} \ln n$.

\begin{figure}[t!]
  \centering
  \includegraphics[width=0.9\columnwidth]{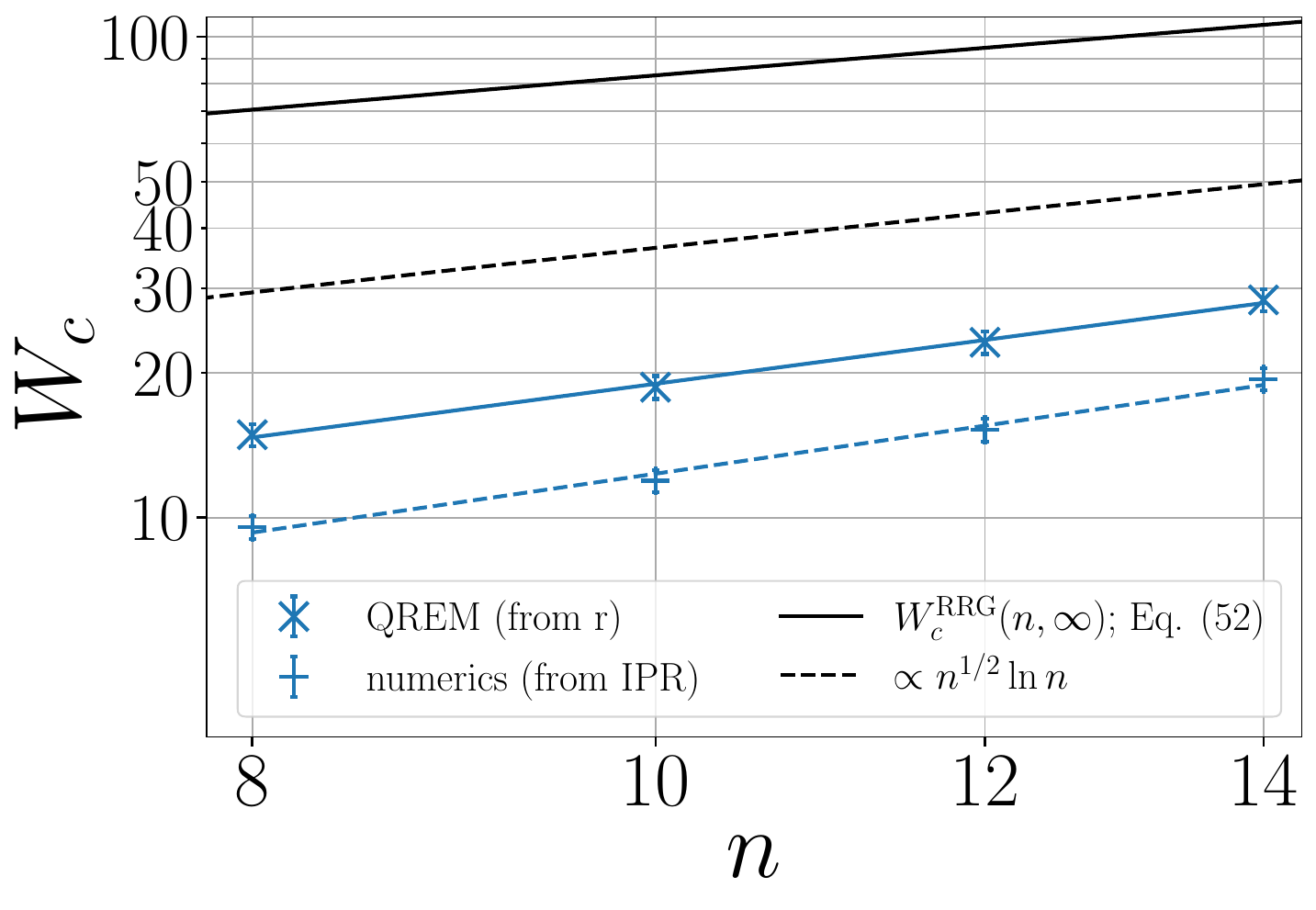}
  \caption{ Scaling of critical disorder of localization transition in QREM (log-log plot). Symbols are numerical values of $W^{\rm QREM}_c(n)$ based on data for mean gap ratio $r$,  Fig.~\ref{fig:rW_QREM}, and for logarithmic derivative $\alpha$ of the IPR $P_2$, Fig.~\ref{fig:IPR_QREM}. Straight lines (corresponding to power-law fits) are guide to the eye. 
  Dashed black line visualizes the $n^{1/2} \ln n$ scaling.
  Full black line is the large-$n$ asymptotics $W_c^{\rm RRG}(n,\infty)$  given by the solution of Eq.~\eqref{eq:RRG-maintext-Wc-rescaled-equation}.
  }
  \label{fig:Wc_QREM}
\end{figure}

In Fig.~\ref{fig:Wc_QREM} we show the values of $W_c^{\rm QREM}(n)$ as obtained from the data for level statistics (gap ratio $r$) and for eigenfunctions (logarithmic derivative $\alpha$ of IPR $P_2$). A good agreement with the scaling $W_c(n) \sim n^{1/2} \ln n$ is evident. We also show in this figure the predicted large-$n$ asymptotics of $W_c^{\rm QREM}(n)$, i.e., $W_c^{\rm RRG}(n,\infty)$  given by the solution of Eq.~\eqref{eq:RRG-maintext-Wc-rescaled-equation}. 
It is seen that the numerically extracted values are a few times below the asymptotical curve and approach it with increasing $n$. This is in excellent agreement with analytical expectations as is manifest from Fig.~\ref{fig:Wc_QREM_theory} which provides a comparison of numerical values of $W_c^{\rm QREM}(n)$ with the analytical result \eqref{eq:QREM-maintext-relation-RRG} that includes deviations from the asymptotic curve related to a finite volume $N=2^n$ of the QREM Hilbert space. We also observe that $W_c(n)$ extracted from the IPR data exhibits somewhat larger finite-size corrections in comparison to the critical disorder obtained from the level-spacing data. The difference between these two estimates of the finite-size critical disorder is of the order of the transition width, and we expect that they become closer and merge with further increasing $n$, in correspondence with the sharpening of the localization transition. 

Since we largely use QREM as a benchmark in this work (see the beginning of Sec.~\ref{sec:numerics}), it is instructive to briefly summarize our key findings with respect to the localization transition in this model as studied numerically for $n=8 - 14$, including a comparison with analytical predictions:

\begin{itemize}

\item The analytical scaling $W_c(n) \sim n^{1/2} \ln n$ is nicely observed in numerical data, despite the fact that the values of $n$ that can be studied by exact diagonalization are not so large. Because of the finite size $N=2^n$ of the QREM Hilbert space, the prefactor in front of $n^{1/2} \ln n$ is appreciably below the one predicted for $n \to \infty$, slowly approaching the asymptotic value with increasing $n$, as also predicted analytically.

\item The transition width quickly shrinks with increasing $n$, confirming that there is a sharp transition in the large-$n$ limit. The flowing exponent $\mu(n)$, Eq.~\eqref{eq:transition-width-mu-n}, characterizing sharpening of the transition with $n$ is numerically {\color{black}$\mu \approx 1.32$.} This is exactly in the range of $\mu(n)$ analytically expected for this range of $n$ but is substantially below the predicted asymptotic value $\mu(n \to \infty) = 3$.

\item Finite-size values of critical disorder $W_c^{\rm QREM}(n)$ obtained from the level statistics (gap ratio $r$) data and from the maximum of the logarithmic derivative $\alpha$ of IPR show nearly identical dependence on $n$. At the same time, the value obtained from IPR is somewhat lower (with a difference of the order of transition width), i.e., it exhibits a larger finite-size deviation.

\end{itemize}

Armed with (and encouraged by) these results for QREM, we are now ready to proceed with the presentation and analysis of numerical data for the genuine many-body models---QD and 1D---as well their counterparts with uncorrelated Fock-space hoppings---uQD and u1D.

\section{Numerics for models with Fock-space correlations}
\label{sec:numerics-models-with-corr}

In Sec.~\ref{subsec:comparison_QD_uQD}, we present and discuss numerical results for the QD and uQD models, while
Sec.~\ref{subsec:comparison_1D_u1D} contains an analogous discussion of the 1D and u1D models.  Finally, in Sec.~\ref{subsec:comparison_QD_1D} we compare and analyze numerical findings for all the models. 

\subsection{Numerical results for the QD and uQD models}
\label{subsec:comparison_QD_uQD}

In this subsection, we analyze and compare our numerical results for the MBL transition in QD and uQD models {\color{black} (see definitions in Secs.~\ref{sec:QD-model-def} and~\ref{sec:uQD-model}).} 

\begin{figure*}[t!]
  \includegraphics[width=0.33\textwidth]{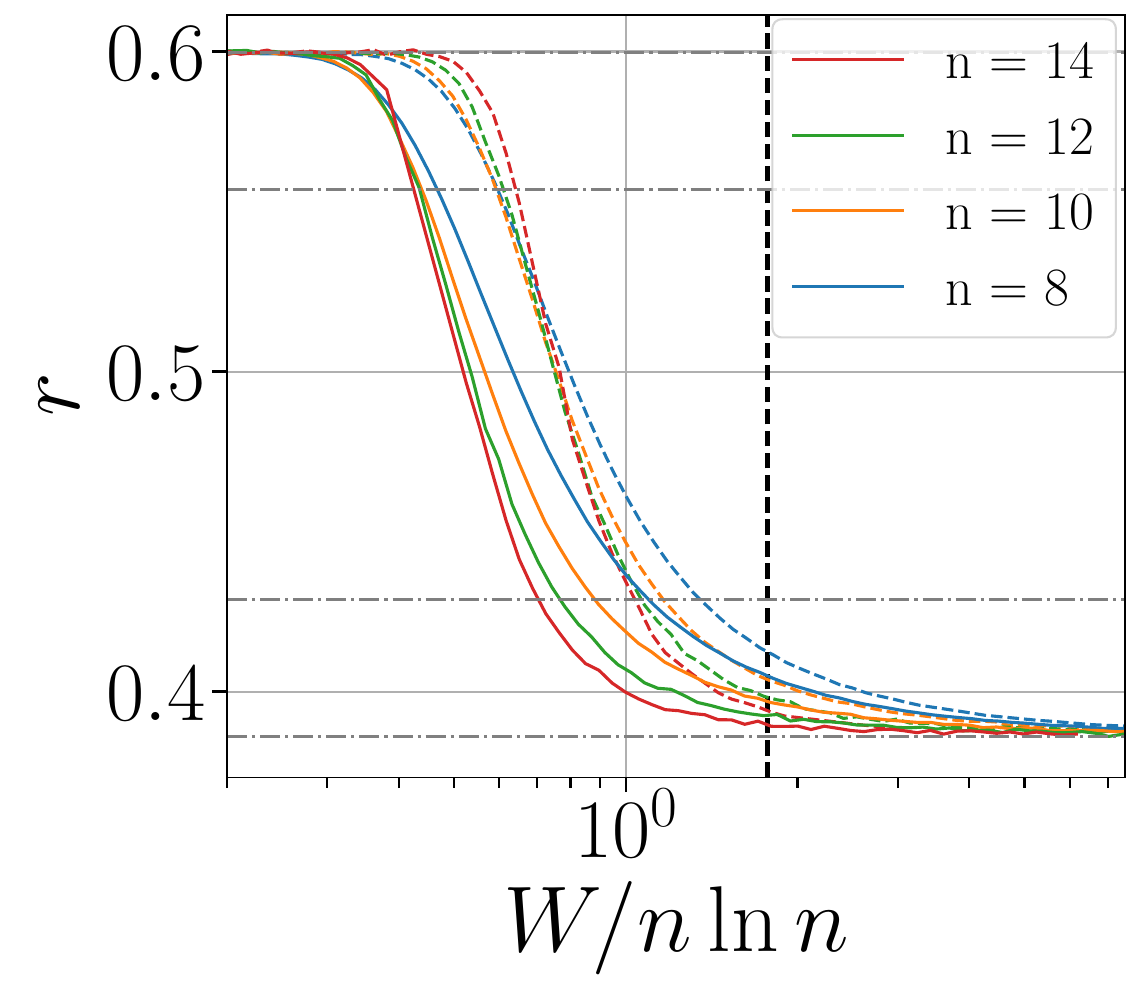}\hfill
  \includegraphics[width=0.33\textwidth]{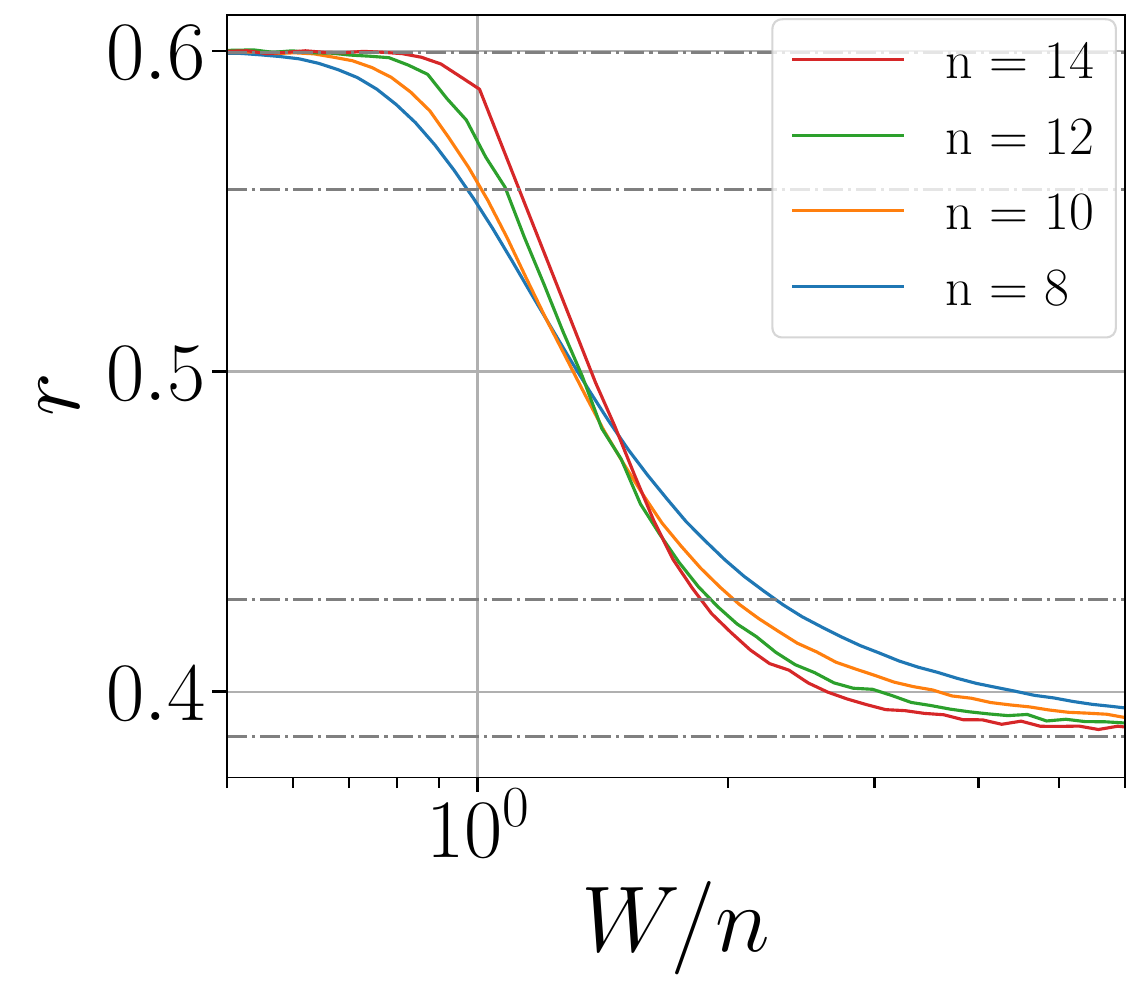}\hfill
  \includegraphics[width=0.33\textwidth]{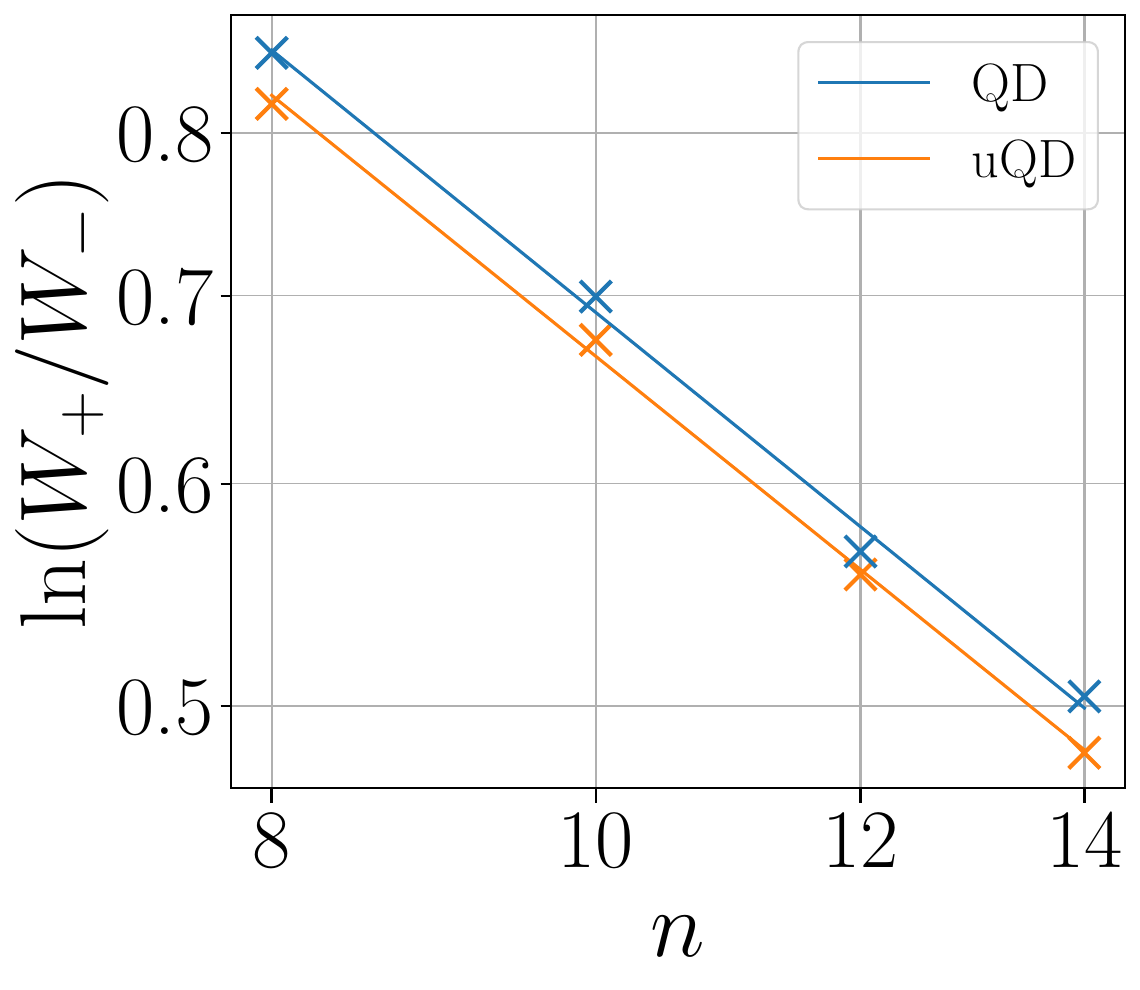}
  \caption{MBL transition in QD and uQD models studied via the level statistics.
  {\it Left panel:} Mean adjacent gap ratio $r$ for the QD (solid lines) and the uQD (dashed lines) models as a function of rescaled disorder
  $W/n \ln n$, with rescaling corresponding to predicted scaling of $W_c$ for the uQD model, 
  (\ref{eq:Wc_RRG_scaling}).
  The data is averaged  over 320000, 20000, 1500, and 350 disorder realizations for $n=8$, 10, 12, and 14, respectively. 
   The vertical dashed line corresponds to the
analytical prediction for $W^{\rm uQD}_c(n)$ in the limit $n\to \infty$, Eq.~\eqref{eq:Wc_RRG_scaling}.
  {\it Middle panel:} Data for the QD model from the left panel plotted as a function of $W/n$. 
  {\it Right panel:} 
  transition width $\ln (W_+/W_-)$ as a function of $n$. The straight lines representing power-law fits
  $\ln (W_+/W_-) \sim n^{-0.96}$ for QD and $\sim n^{-0.95}$ for uQD models are a guide to the eye.
  }
  \label{fig:r_comparison_QD_uQD}
\end{figure*}

In Fig.~\ref{fig:r_comparison_QD_uQD} we show the data for the mean adjacent gap ratio $r$ characterizing the level statistics. In the left panel, the data presented as a function of disorder rescaled as $W/n \ln n$, which corresponds to analytically predicted scaling of $W^{\rm uQD}_c$, see Eq.~(\ref{eq:Wc_RRG_scaling}). We see that this rescaling indeed yields a very good collapse for the uQD data. 
We further observe that the numerical value of the ratio $W_c^{\rm uQD}(n) / n \ln n$ in the considered range of $n$ is smaller by a factor $\approx 2.5$ than the asymptotic $n\to \infty$ value $\sqrt{\pi}$ marked by the vertical dashed line. This is fully analogous to what is observed for the QREM, see Sec.~\ref{subsec:numerics_QREM}. As for the QREM, for increasing $n$, a slow evolution of the above ratio towards its asymptotic value is expected. 

At variance with the uQD data, the QD data in the left panel of Fig.~\eqref{fig:r_comparison_QD_uQD} exhibit a clear (although quite slow) drift to the left. The middle panel shows the QD data with a rescaling of disorder to $W/n$, which leads to a very good collapse. The numerically observed behavior $W_c^{\rm QD}(n) \sim n$ is fully consistent with the analytically derived lower and upper bounds,
Eq.~(\ref{Wc-QD-final}). Obviously, we cannot claim on the basis of the numerical data that 
$W_c^{\rm QD}(n) \sim n$ is an exact large-$n$ asymptotic behavior. 

In the right panel of Fig.~\ref{fig:r_comparison_QD_uQD}, results for the transition width $\ln(W_+/W_-)$ are presented. We see that the behavior is essentially the same for both models: the transition sharpens with increasing $n$, and the effective exponent $\mu(n)$, Eq.~\eqref{eq:transition-width-mu-n}, characterizing this sharpening is $\mu \approx 0.95 - 0.96$.  

\begin{figure*}[t!]
  \includegraphics[width=0.33\textwidth]{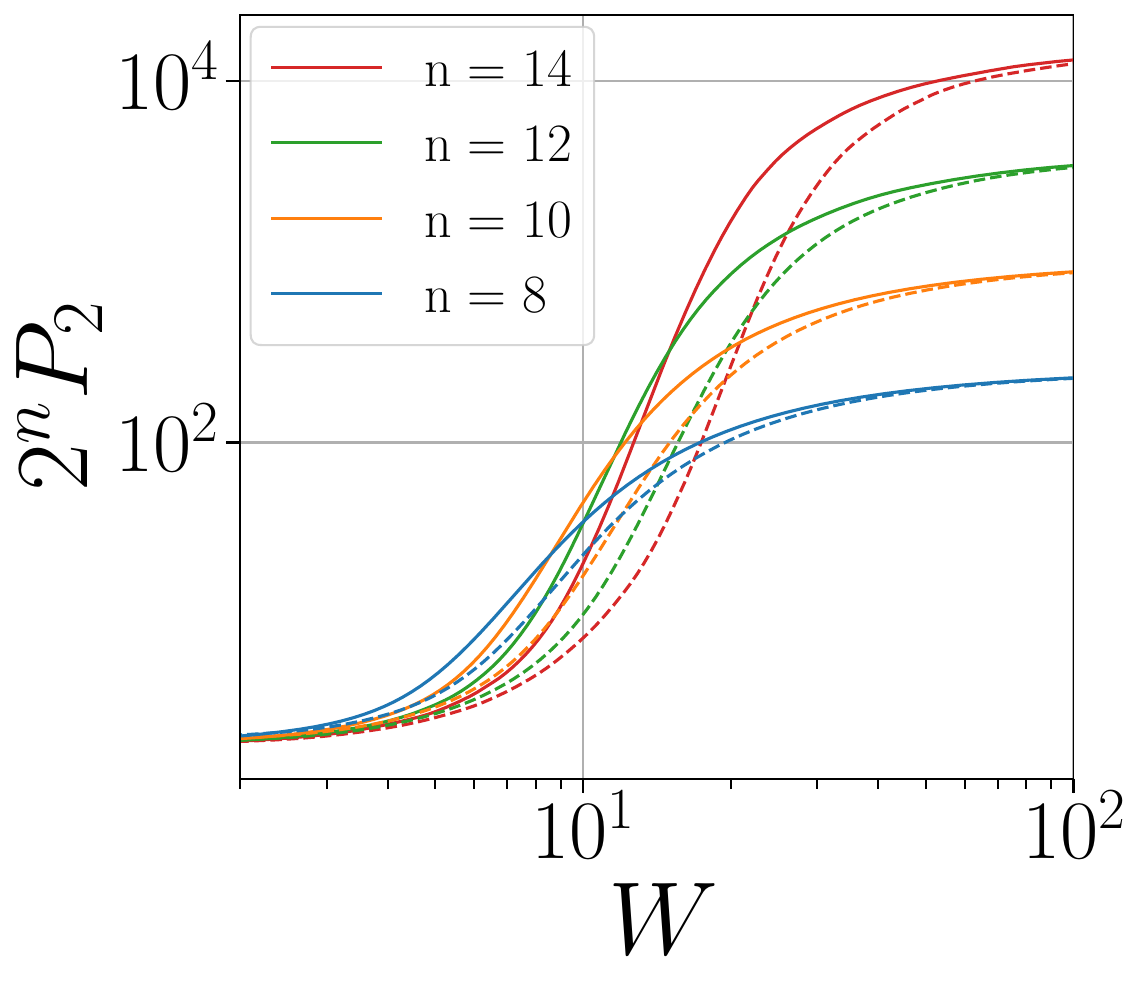}\hfill
  \includegraphics[width=0.33\textwidth]{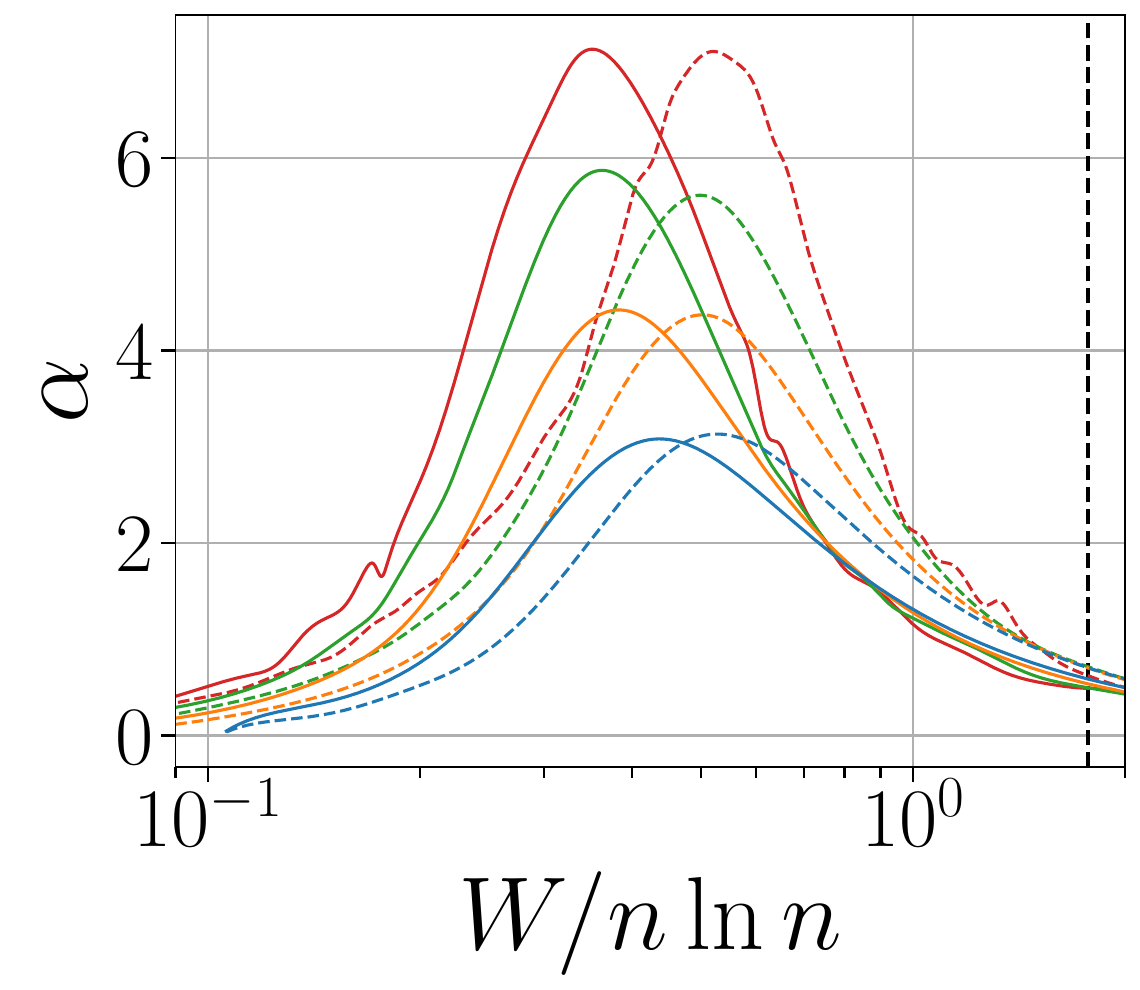}\hfill
  \includegraphics[width=0.33\textwidth]{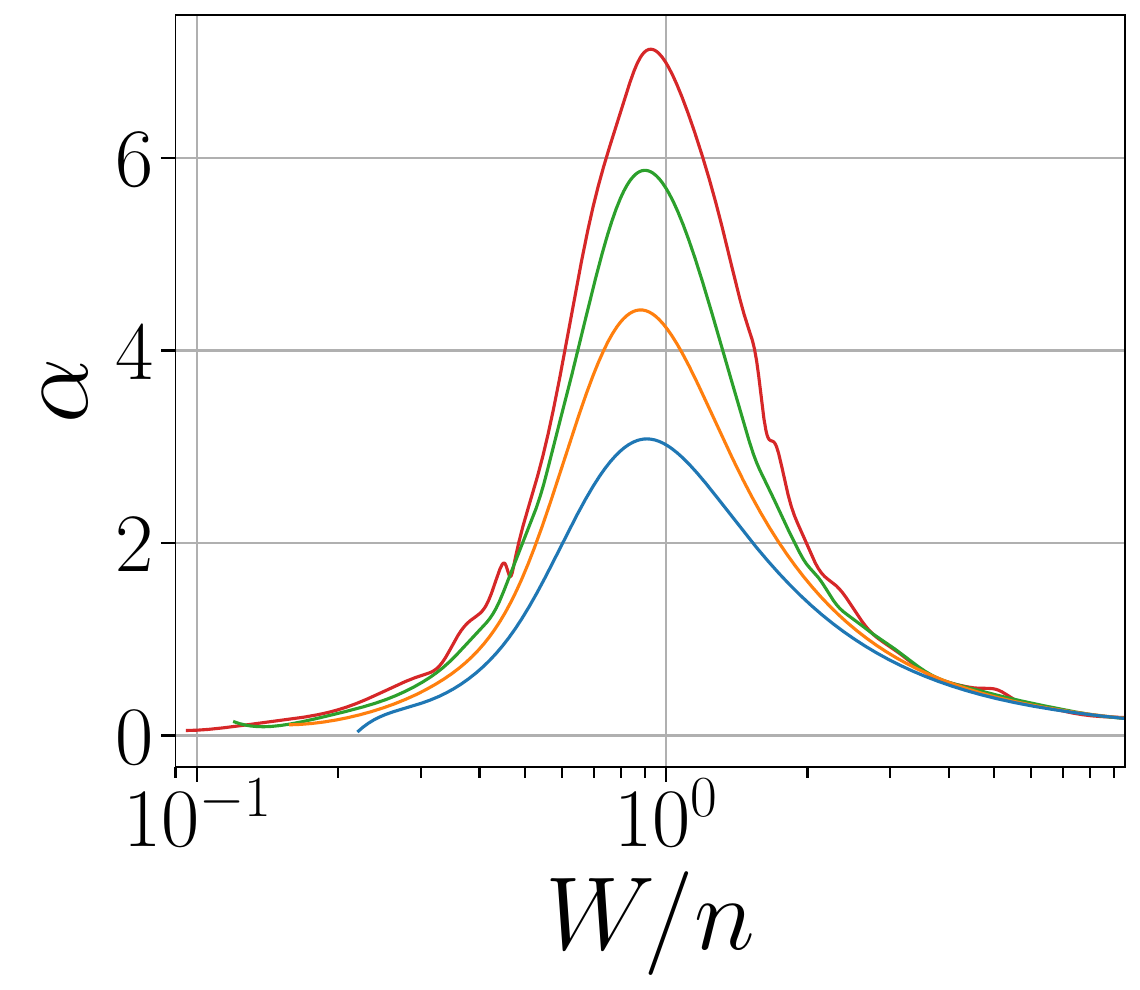}\hfill
  \caption{MBL transition in QD and uQD models studied via the average IPR $P_2$.
  {\it Left panel}: $2^n P_2$ for the QD (solid lines) and the uQD (dashed lines) models as a function of $W$. 
The data is averaged over 320000, 20000, 1500, and 350 disorder realizations for $n=8$, 10, 12, and 14, respectively.
  {\it Middle panel}: logarithmic derivative $\alpha$, Eq.~\eqref{eq:alpha-W-def} as a function of rescaled disorder $W/n \ln n$, with rescaling corresponding to predicted scaling of $W_c$ for the uQD model, (\ref{eq:Wc_RRG_scaling}). The vertical dashed line corresponds to the
analytical prediction for $W^{\rm uQD}_c(n)$ in the limit $n\to \infty$, Eq.~\eqref{eq:Wc_RRG_scaling}.
{\it Right panel:}
Data for the QD model from the middle panel plotted as a function of $W/n$.}
  \label{fig:IPR_comparison_QD_uQD}
\end{figure*}

In Fig.~\ref{fig:IPR_comparison_QD_uQD}, we show numerical results for the IPR (left panel) and for its logarithmic derivative $\alpha$ (middle and right panels). The results confirm the above conclusions made by using the level statistics: a good collapse of the maxima of $\alpha(W)$ is achieved by rescaling of disorder to $W/n\ln n$ for the uQD model and to $W/n$ for the QD model. 

\begin{figure}[t!]
\centering
\includegraphics[width = .85\columnwidth]{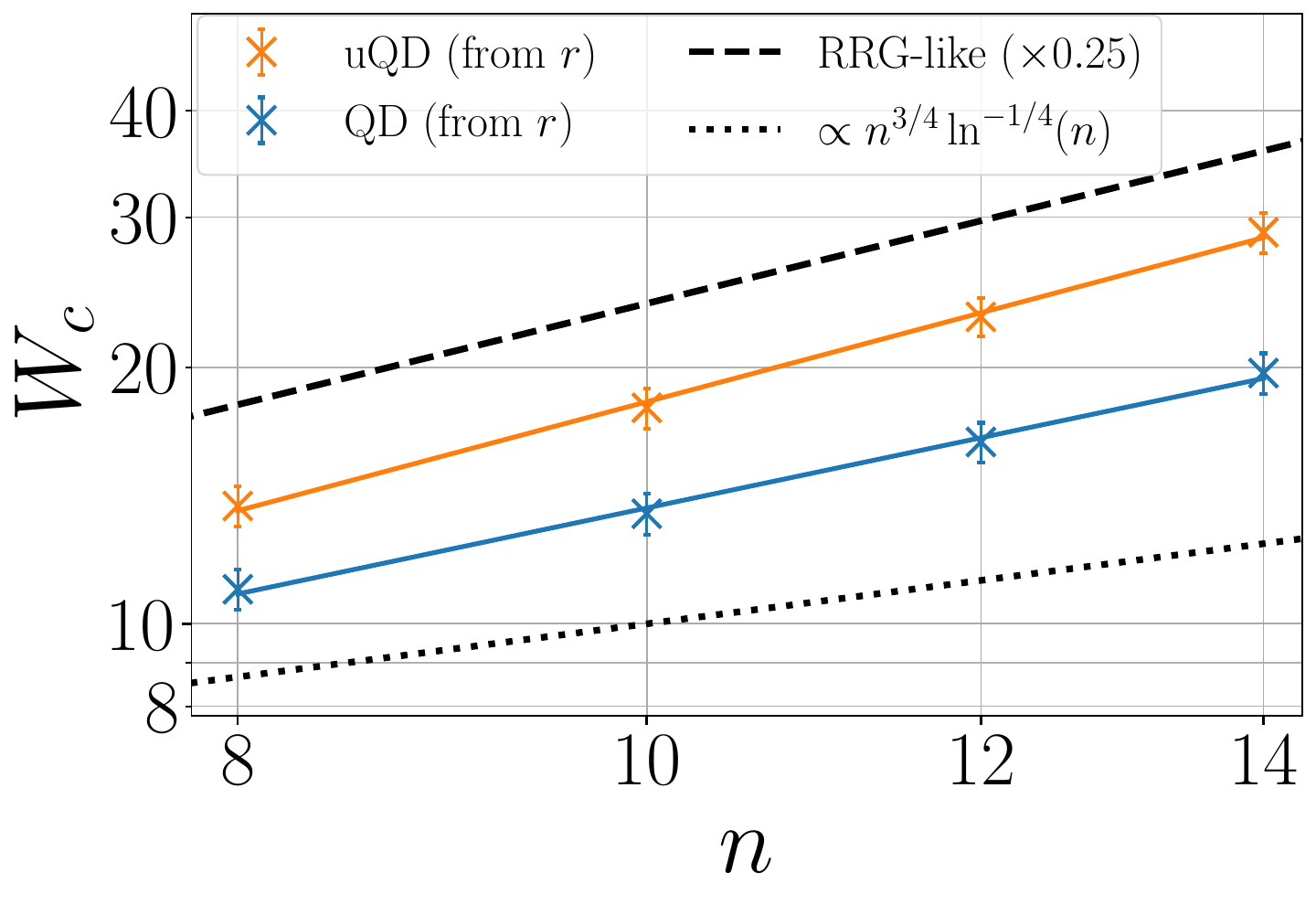}\\
\includegraphics[width = .85\columnwidth]{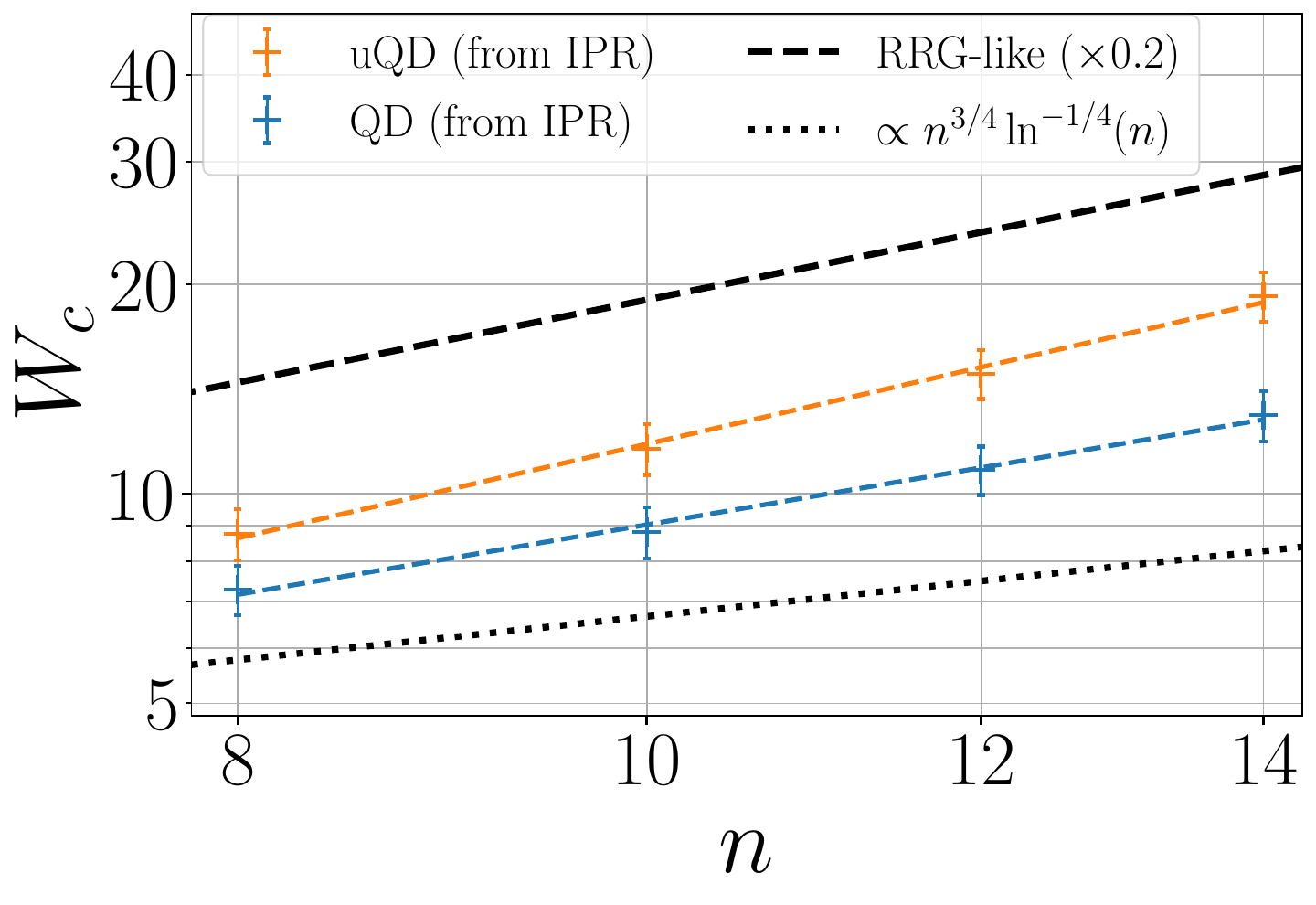}
\caption{
Scaling of critical disorder of the localization transition in QD and uQD models (log-log plots). Symbols are numerical values of $W^{\rm QD}_c(n)$ and $W^{\rm uQD}_c(n)$ based on data for mean gap ratio $r$ (upper panel) and for logarithmic derivative $\alpha$ of IPR $P_2$ (lower panel). Straight lines (corresponding to power-law fits) are guide to the eye. Black dashed lines are the large-$n$ asymptotics of $W_c^{\rm uQD}$ given by the solution of Eq.~\eqref{eq:self_cons_critical_disorder_RRG}
(rescaled by a constant prefactor as shown in the legend for convenience of data presentation). The dotted lines represent the analytical lower bound $\propto n^{3/4} \ln^{-1/4}n$ to $W_c^{\rm QD}$, see Eq.~\eqref{Wc-QD-lower-boundary}.
}
\label{fig:Wc_comparison_QD_n}
\end{figure}

The numerical results for $W_c^{\rm QD}(n)$ and $W_c^{\rm uQD}(n)$ obtained by both approaches are summarized in Fig.~\ref{fig:Wc_comparison_QD_n}. As in the case of QREM, the values of critical disorder obtained from level statistics are somewhat larger than those obtained from IPR but, up to this, they exhibit a nearly identical behavior. Specifically, the uQD data points show the predicted $n \ln n$ scaling and slowly approach the large-$n$ asymptotics given by the solution of Eq.~\eqref{eq:self_cons_critical_disorder_RRG}. The QD data exhibit a slower increase with $n$, which is at the same time faster than the analytically predicted lower bound, $W_c^{\rm QD} > n^{3/4} (\ln n)^{-1/4}$.  
Thus, the observed behavior of $W_c^{\rm QD}(n)$ is in perfect agreement with both the upper and the lower bounds, Eq.~\eqref{Wc-QD-final}. As was pointed out above, $W_c^{\rm QD}(n) \sim n$ turns out to be a good fit to our numerical results.

Thus, our numerical results for the uQD model confirm the validity of the RRG-like approximation in the presence of strong Fock-space energy correlations represented by the matrix $C_E^{\rm uQD}$. Specifically, as discussed in
Sec.~\ref{sec:uQD-u1D-analytics}, strong correlations between energies on nearby sites in the Fock space parametrically enhance the probability of resonances, leading to a faster increase of $W_c(n)$ with $n$ in the uQD model in comparison with the QREM, where these correlations are absent. 
Furthermore, a slower increase of $W_c^{\rm QD}(n)$ in comparison with $W_c^{\rm uQD}(n)$ demonstrates the role of correlations between Fock-space hopping matrix elements encoded in the matrix $C_T$. 
Specifically, as discussed in Sec.~\ref{sec:QD-analytics}, these correlations lead to destructive interference (partial cancelation) of different contributions to hybridization couplings between distant states in the Fock space, thus favoring localization and suppressing $W_c^{\rm QD}(n)$ compared to $W_c^{\rm uQD}(n)$. 

\subsection{Numerical results for the 1D and u1D models}
\label{subsec:comparison_1D_u1D}

We turn now to the numerical analysis of the MBL transition in the 1D and u1D models {\color{black}(see definitions in Secs.~\ref{sec:1D-model-def} and~\ref{sec:u1D-model}).} We recall that these models differ from their quantum-dot counterparts (QD and uQD) by a 1D real-space structure, which is encoded in the Fock-space correlations. Specifically, the energy correlations in the 1D and u1D models depend not only on the Hamming distance $r_{\alpha\beta}$ but also on an additional parameter $q_{\alpha\beta}$ associated with the 1D real-space structure, see Eq.~\eqref{eq:1D-correl_energies_intermediate} and text below it. Importantly, in the 1D model, also Fock-space hopping correlations depend not only on the Hamming distance but rather have a structure that preserves information about the 1D real-space geometry,
see Eq.~\eqref{eq:1D-CT}. As we will see below, this leads to a dramatic change in the behavior of the 1D model in comparison with the QD model.

\begin{figure*}[t!]
  \centering
  \includegraphics[width=0.33\textwidth]{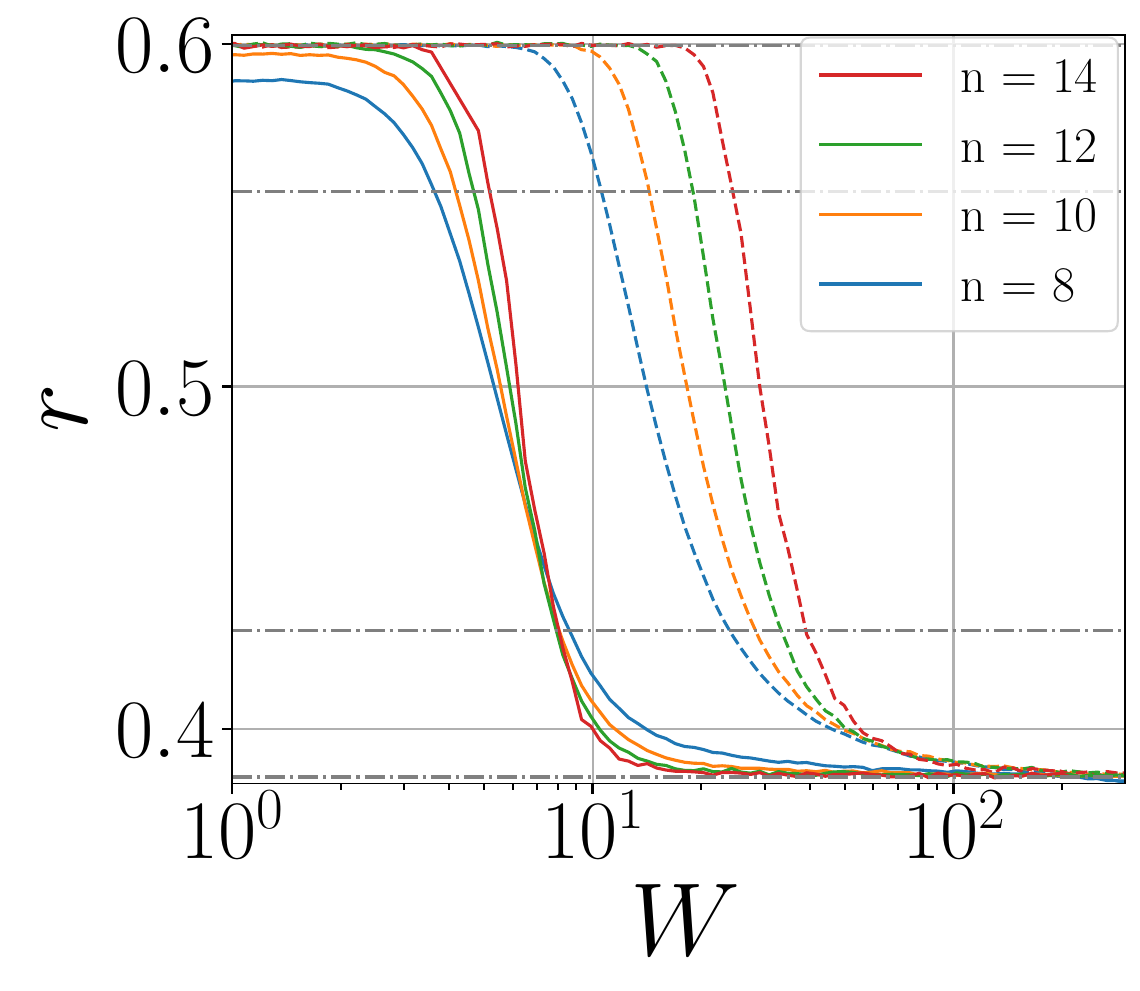}\hfill
  \includegraphics[width=0.33\textwidth]{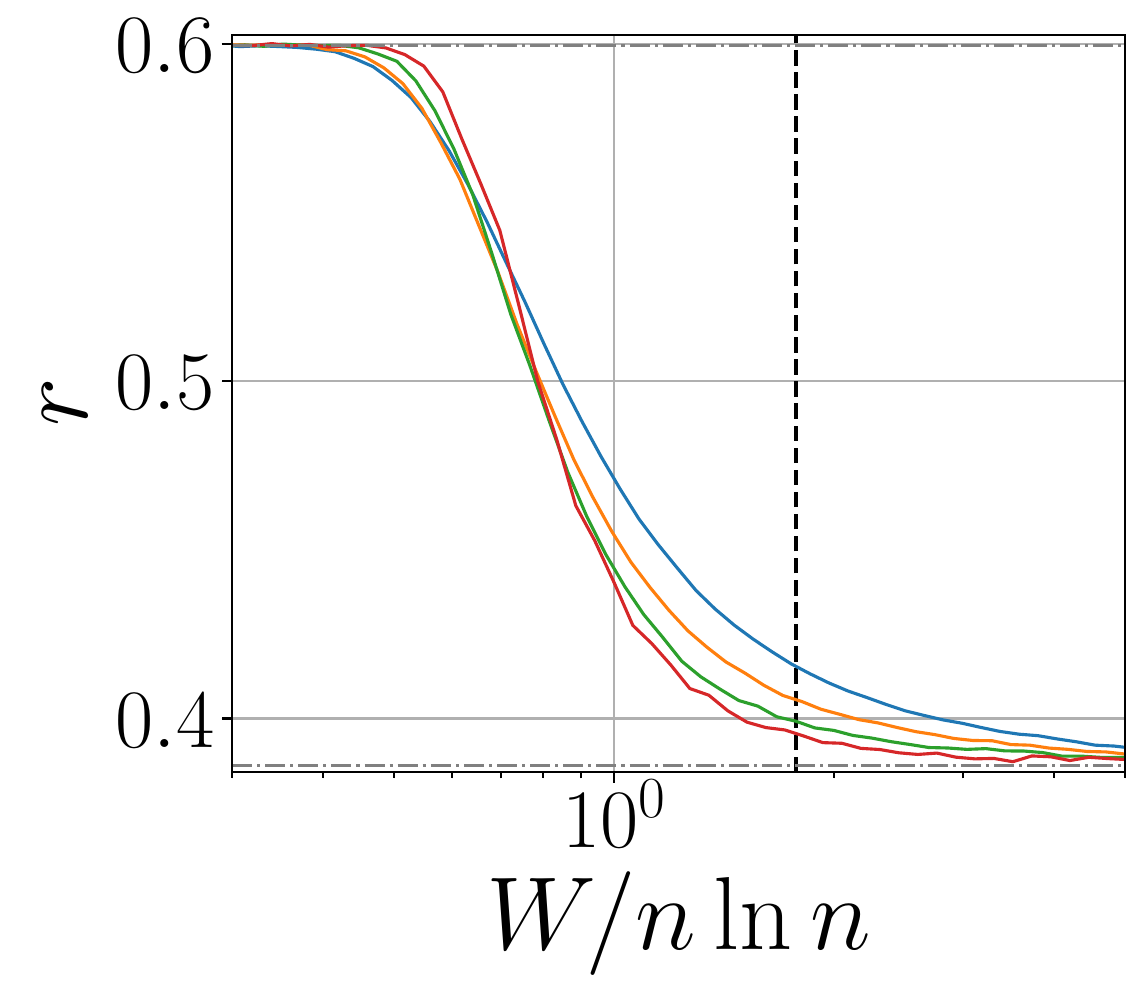}\hfill
  \includegraphics[width=0.33\textwidth]{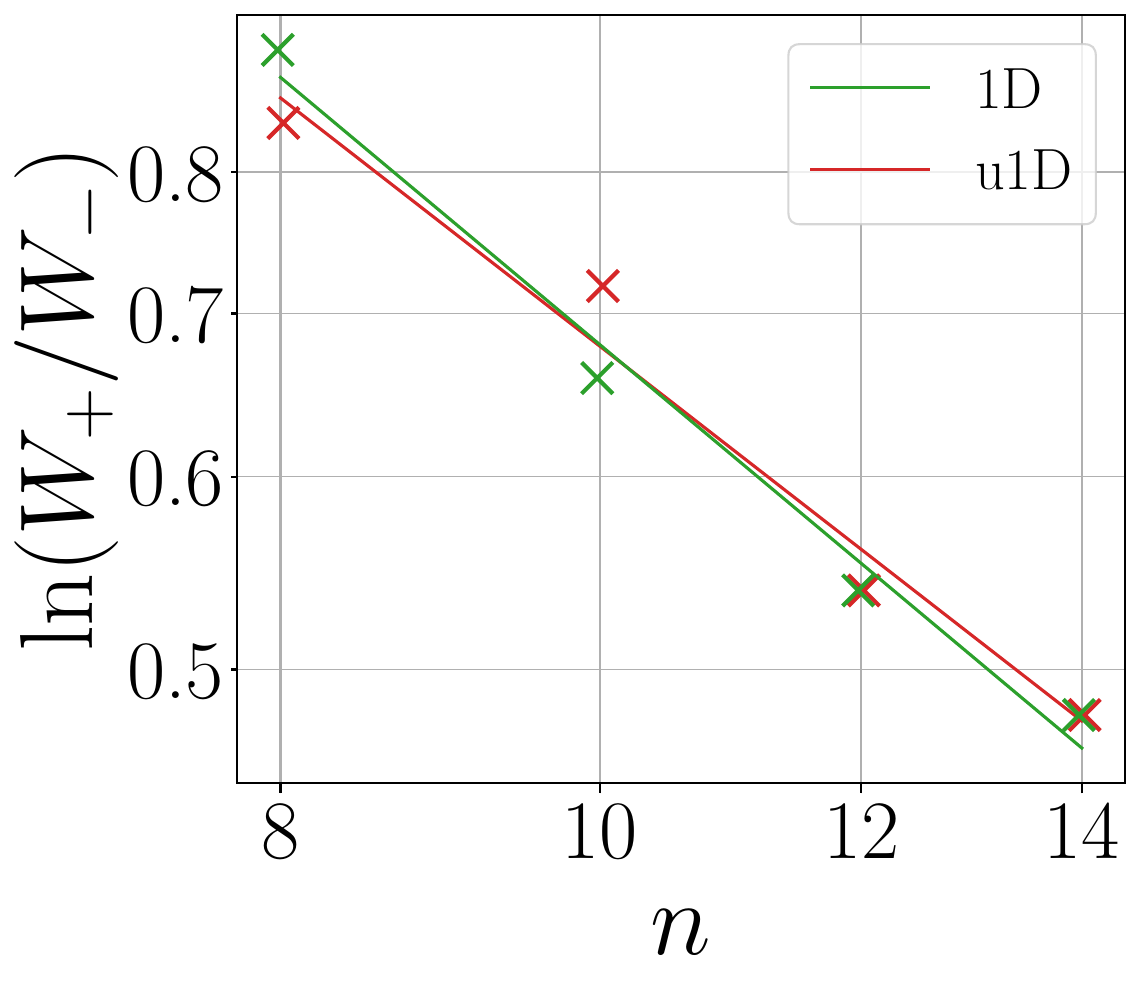}
  \caption{MBL transition in 1D and u1D models studied via the level statistics.
  {\it Left panel:} Mean adjacent gap ratio $r$ for the 1D (solid lines) and the u1D (dashed lines) models as a function of disorder $W$.
  The data is averaged over 160000, 20000, 2500, and 350 disorder realizations for $n=8$, 10, 12, and 14, respectively.
{\it Middle panel:} Data for the u1D model from the left panel plotted as a function of rescaled disorder $W/n \ln n$, with rescaling corresponding to predicted scaling of $W_c$ for the u1D model, (\ref{eq:Wc_RRG_scaling}).  The vertical dashed line corresponds to the
analytical prediction for $W^{\rm u1D}_c(n)$ in the limit $n\to \infty$, Eq.~\eqref{eq:Wc_RRG_scaling}.
 {\it Right panel:} 
  transition width $\ln (W_+/W_-)$ as a function of $n$. The straight lines representing power-law {\color{black} fits $\ln (W_+/W_-) \sim n^{-1.05}$ for 1D and $\sim n^{-1.17}$ for u1D models} are a guide to the eye.
}
  \label{fig:r_comparison_1D_u1D}
\end{figure*}

Figure \ref{fig:r_comparison_1D_u1D} presents the results for the gap ratio $r$ of the level statistics. In the left panel, we show the data for $r(W)$ of both models. It is seen that the data for the 1D model exhibits a good collapse without any rescaling of disorder, in agreement with the analytical expectation of $n$-independent critical disorder $W_c^{\rm 1D} \sim 1$, Eq.~\eqref{eq:1D-Wc-n}. At the same time, the data for the u1D model shows a very different behavior, with a strong drift towards stronger disorder. In the right panel, the u1D data is plotted as a function of disorder rescaled as $W/n\ln n$, in accordance with
Eq.~\eqref{eq:Wc_RRG_scaling}. This yields a good collapse, thus supporting the analytical prediction
$W_c^{\rm u1D} \sim n \ln n$. 

For both models, clear sharpening of the transition with increasing $n$ is observed. The right panel of Fig.~\ref{fig:r_comparison_1D_u1D} quantifies this: we find a power-law shrinking of the transition width, $\ln(W_+/W_-) \sim n^{-\mu}$, {\color{black} with $\mu=1.05$ for the 1D model and $\mu=1.17$ for the u1D model.} These values of the (effective) exponent $\mu$ are remarkably close to those found above for the other three models (QREM, QD, uQD). The fact that u1D and uQD models exhibit, in the same range of $n$, nearly the same exponent $\mu$ as the QREM, is not surprising since these models are described by the RRG-like approximation. The observed values of $\mu$ for these models are not large-$n$ asymptotics but rather flowing exponents $\mu(n)$, as we discussed in Sec.~\ref{subsec:numerics_QREM}. Interestingly, the transition width in the 1D model exhibits essentially the same scaling behavior in the range of $n$ accessible to exact diagonalization.

\begin{figure*}[t!]
  \centering
  \includegraphics[width=0.33\textwidth]{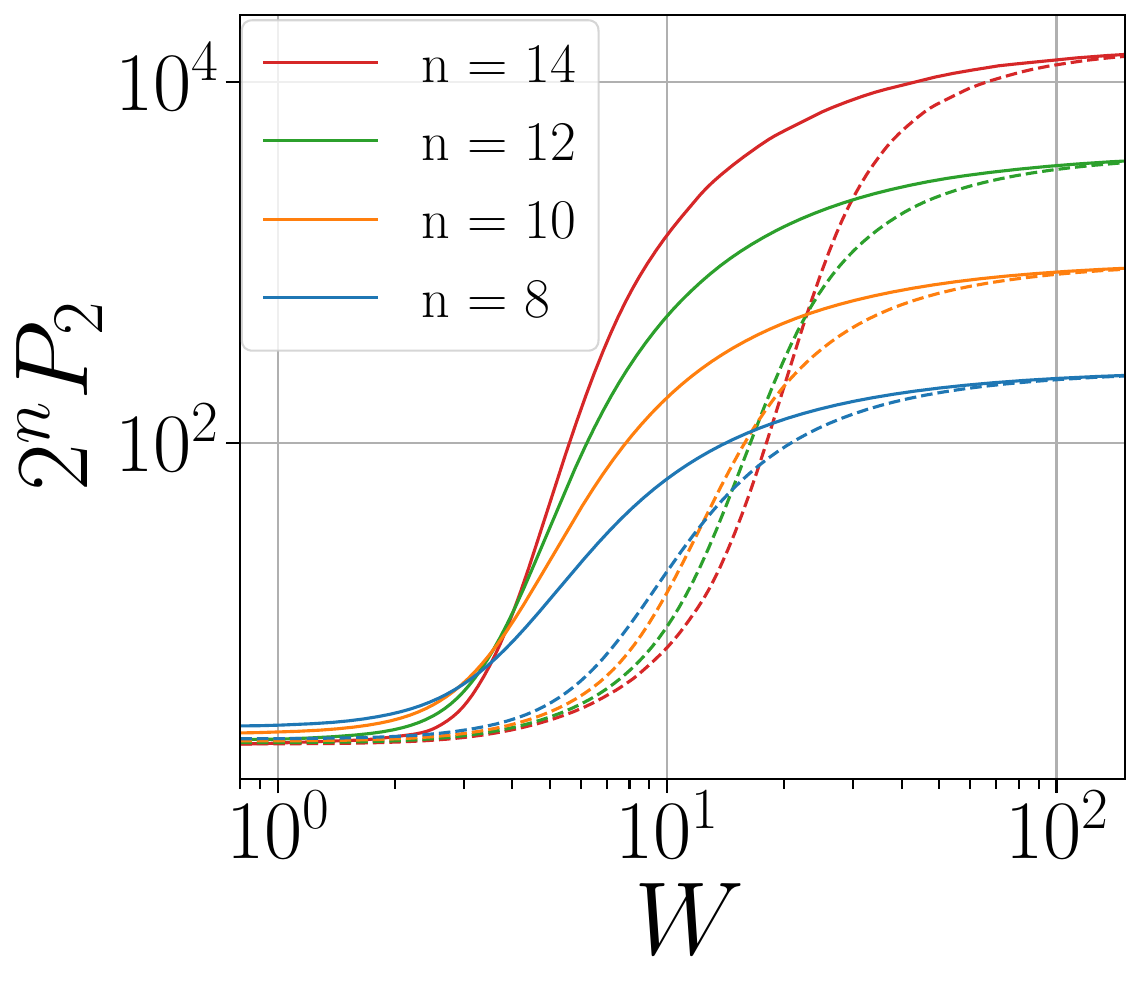}\hfill
  \includegraphics[width=0.33\textwidth]{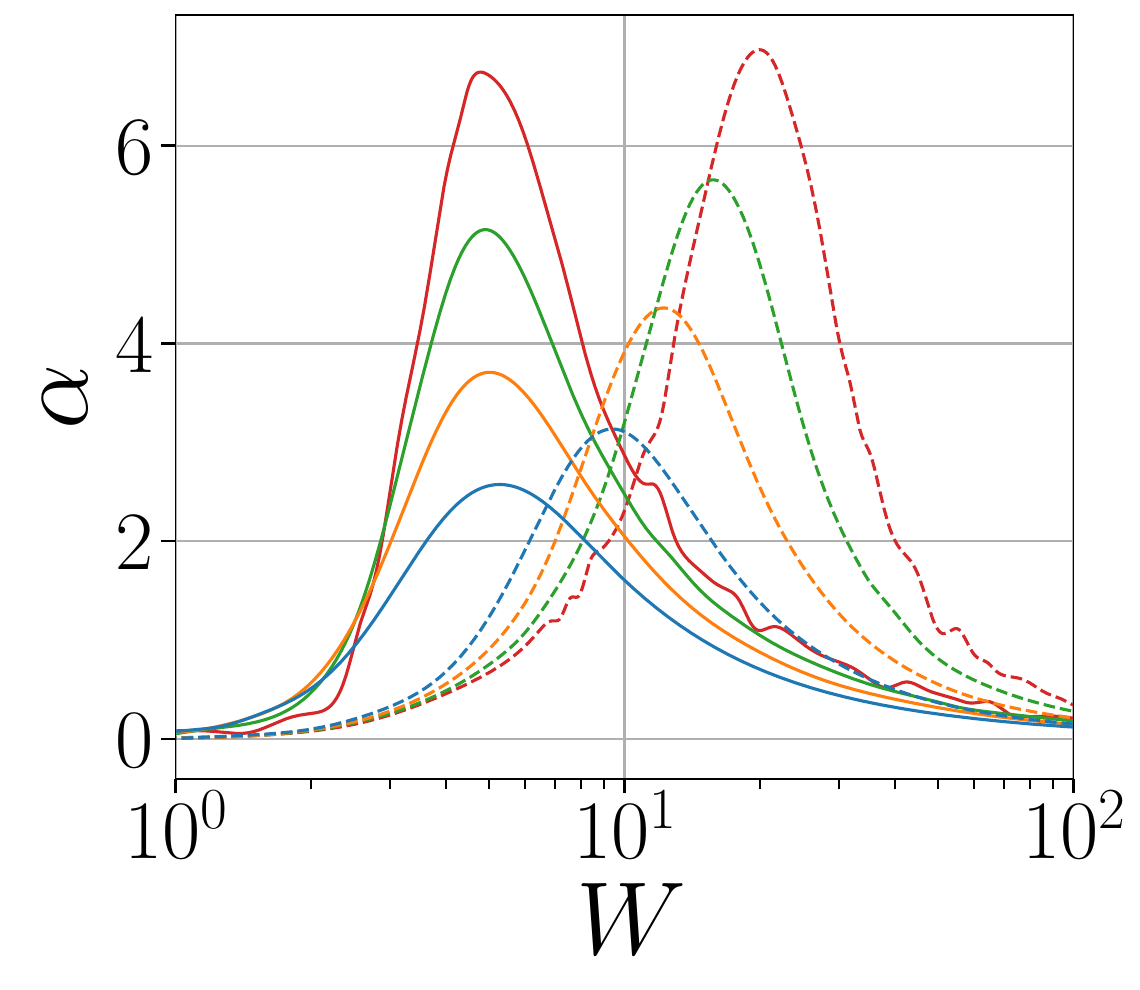}\hfill
  \includegraphics[width=0.33\textwidth]{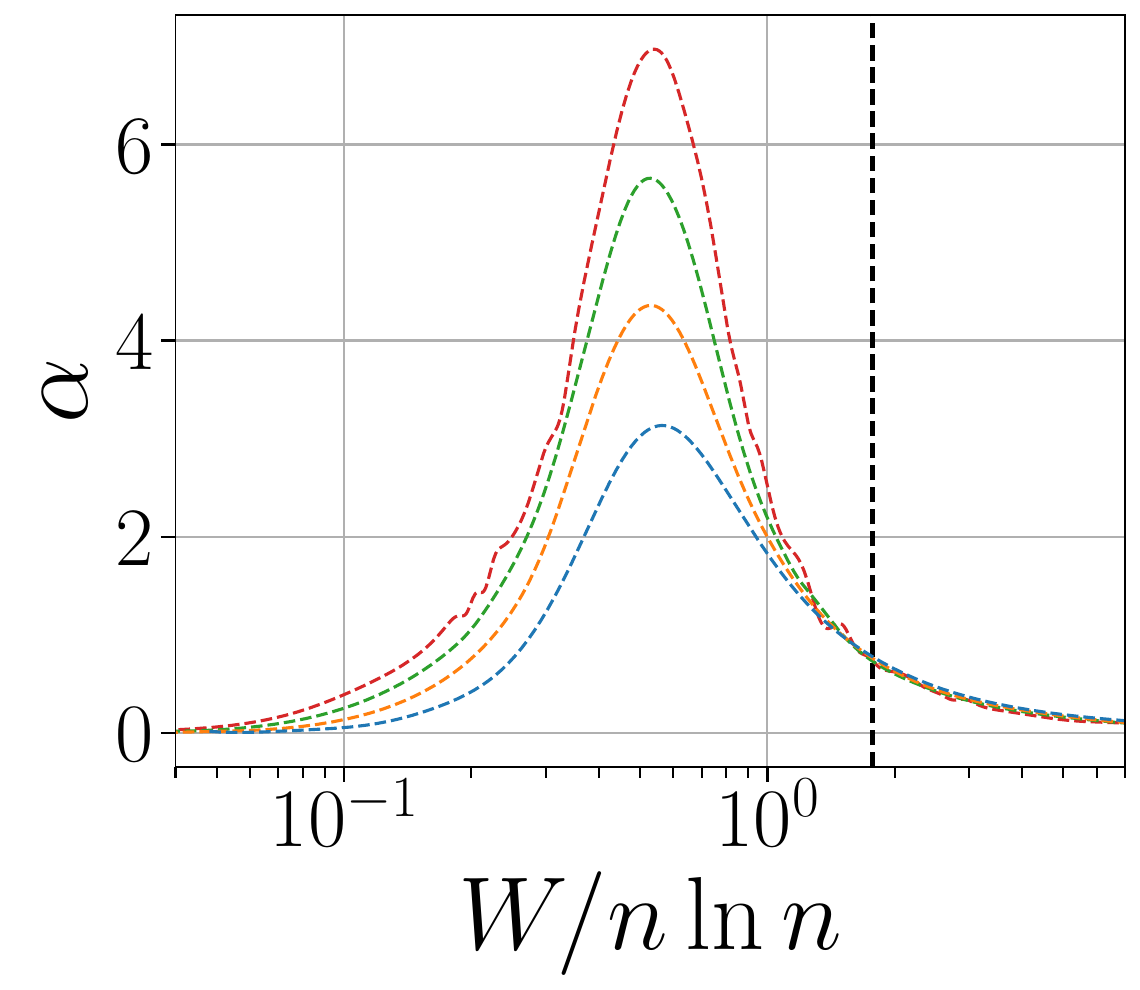}
  \caption{
  MBL transition in 1D and u1D models studied via the average IPR $P_2$.
  {\it Left panel}: $2^n P_2$ for the 1D (solid lines) and the u1D (dashed lines) models as a function of disorder $W$. 
The data is averaged over 320000, 20000, 1500, and 350 disorder realizations for $n=8$, 10, 12, and 14, respectively.
  {\it Middle panel}: logarithmic derivative $\alpha$, Eq.~\eqref{eq:alpha-W-def}, as a function of $W$.
{\it Right panel:}
Data for the u1D model from the middle panel plotted as a function of as a function of rescaled disorder $W/n \ln n$, with rescaling corresponding to predicted scaling of $W_c$ for the u1D model, (\ref{eq:Wc_RRG_scaling}). The vertical dashed line corresponds to the
analytical prediction for $W^{\rm u1D}_c(n)$ in the limit $n\to \infty$, Eq.~\eqref{eq:Wc_RRG_scaling}.
}
  \label{fig:IPR_comparison_1D_u1D}
\end{figure*}

In Fig.~\ref{fig:IPR_comparison_1D_u1D}, we present the results for the IPR (left panel) and its logarithmic derivative $\alpha$ (middle and right panels). The scaling of critical disorder that is inferred from these results is in full agreement with the above findings based on level statistics data and with analytical expectations. Specifically, a collapse of maxima in $\alpha(W)$ is obtained for the 1D model without any rescaling of disorder and for the u1D model with rescaling to $W/n \ln n$ (thus supporting $W_c^{\rm 1D} \sim 1$ and $W_c^{\rm u1D} \sim n \ln n$).

\begin{figure}[t!]
\centering
\includegraphics[width = 0.85\columnwidth]{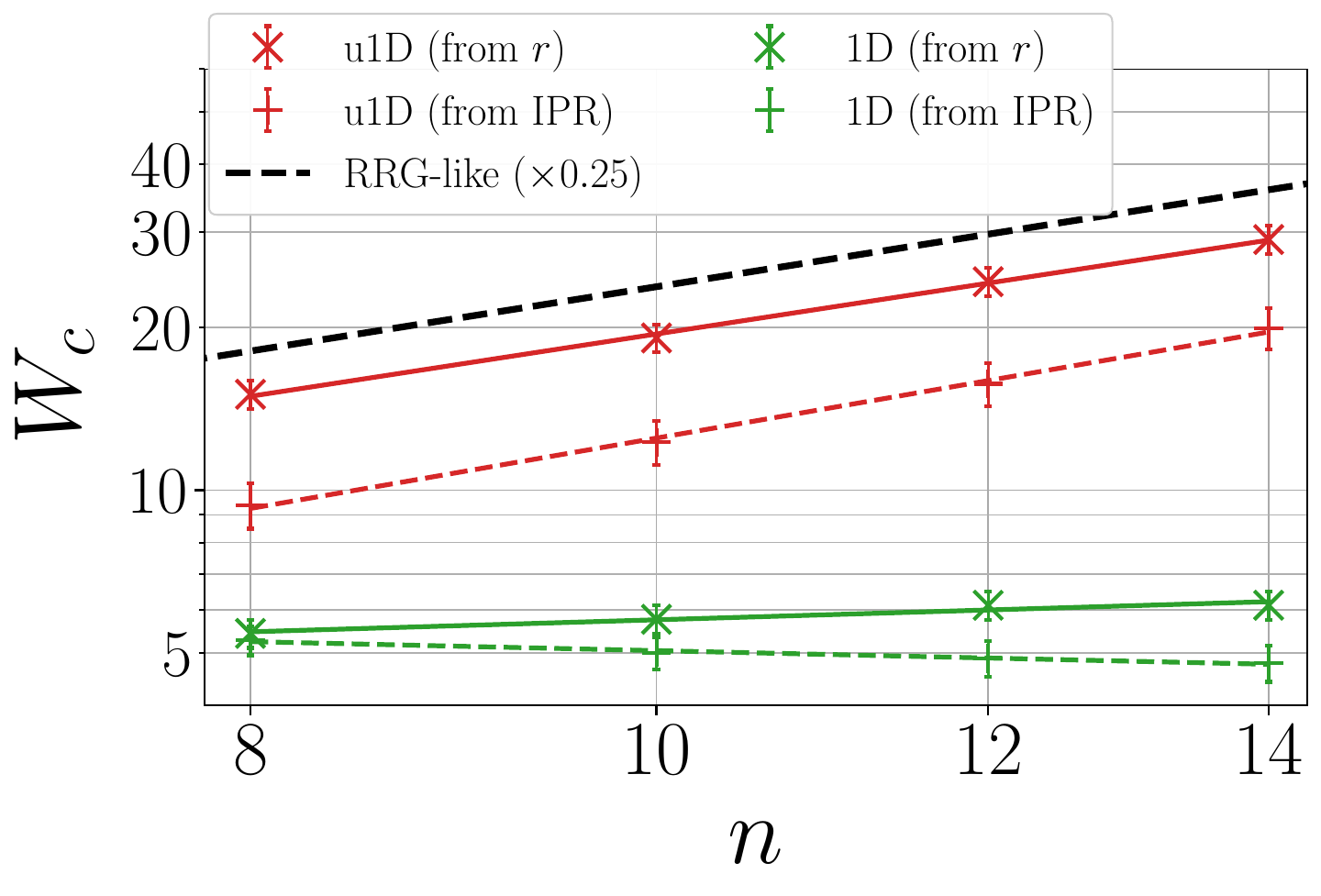}
\caption{
Scaling of critical disorder  of localization transition in 1D and u1D models (log-log plot). Symbols are numerical values of $W^{\rm 1D}_c(n)$ and $W^{\rm u1D}_c(n)$ based on data for mean gap ratio $r$,
 Fig.~\ref{fig:r_comparison_1D_u1D},
and for logarithmic derivative $\alpha$ of IPR $P_2$, Fig.~\ref{fig:IPR_comparison_1D_u1D}. Straight lines (corresponding to power-law fits) are guide to the eye. Black dashed line is the large-$n$ asymptotics of $W_c^{\rm u1D}$ given by the solution of Eq.~\eqref{eq:self_cons_critical_disorder_RRG}
(rescaled by a constant prefactor as shown in the legend for convenience of data presentation). 
}
\label{fig:Wc_comparison_1D_n}
\end{figure}

The results for the critical disorder for 1D and u1D models are summarized in Fig.~\ref{fig:Wc_comparison_1D_n}. A dramatic difference between the scaling $W_c^{\rm 1D}(n) \sim 1 $ and $W_c^{\rm u1D}(n) \sim n \ln n $ is manifest. We recall that both models have exactly the same statistics of Fock-space energies $E_\alpha$ (i.e., the same covariance matrix $C_E$) and exactly the same fluctuations of individual hopping matrix elements $T_{\alpha\beta}$ (i.e., the same diagonal elements of the covariance matrix $C_T$). The only difference between them is in off-diagonal elements of $C_T$, i.e., in correlations between Fock-space transition matrix elements $T_{\alpha\beta}$, which are present in the 1D model and absent in the u1D model. These correlations are thus crucial for the $n$-independent critical disorder of the 1D model. 

To explain such importance of these correlations, we return to the perturbative expansion for hybridization between distant Fock-space basis states,
Sec.~\ref{sec:QD-analytics}. As discussed there, the hybridization amplitude between two distant states $\alpha$ and $\mu$ involves a sum over $r_{\alpha\mu}!$ Fock-space paths, each providing a contribution of the type \eqref{eq:eta-path}. Comparing the 1D and u1D models, we can assume identical sets of energies $E_\alpha$ (since the statistics of energies is the same in both models), so that the denominators of the corresponding terms \eqref{eq:eta-path} in both models will be identical as well. Further, for each individual term, the statistics of the numerator will also be the same in both models. The crucial difference is in correlations between the numerators. In the u1D model, they are uncorrelated, so that there is no interference between the terms. On the other hand, in the 1D model, strong correlations between hopping matrix elements $T_{\alpha\beta}$ lead to major cancellations in the sum of $r_{\alpha\mu}!$ contributions. The $\hat{S}_i^z\hat{S}_j^z$ terms may, in general, counteract these cancellations; however, they are not so efficient in the 1D model since only nearest-neighbor spins interact. As a result of the cancellations, the combinatorial factor $r_{\alpha\mu}!$ gets effectively suppressed down to $\sim p^{r_{\alpha\beta}}$ with $p \sim 1$, thus transforming $W_c^{\rm u1D}(n) \sim n \ln n $ into $W_c^{\rm 1D}(n) \sim 1 $ {\color{black} (see Refs.~\cite{gornyi2016many,gornyi2017spectral} and references therein for technical details on how the interference-induced cancellation of factorials enhances localization)}. {\color{black} A simple way to understand the dramatic difference between $W_c^{\rm u1D}(n)$ and $W_c^{\rm 1D}(n)$ is as follows. For the u1D model, one can choose on each step the closest-in-energy state, which will yield each factor in the denominator of 
Eq.~\eqref{eq:eta-path} of order of $W/n$, thus resulting in $W_c \sim n$. A further optimization can be shown to enhance $W_c$ by an additional logarithmic factor, yielding $W_c^{\rm u1D}(n) \sim n \ln n $. For the 1D model, the cancellation emphasized above leads to an increase of typical values of factors in the denominator up to $\sim W$, leading to $W_c^{\rm 1D}(n) \sim 1$. Our numerical results for $W_c$ contrasting the 1D and u1D models demonstrate that this mechanism is indeed operative.}

\subsection{Comparing QD and 1D models}
\label{subsec:comparison_QD_1D}

\begin{figure*}[t!]
\centering
\includegraphics[width = 0.48\textwidth]{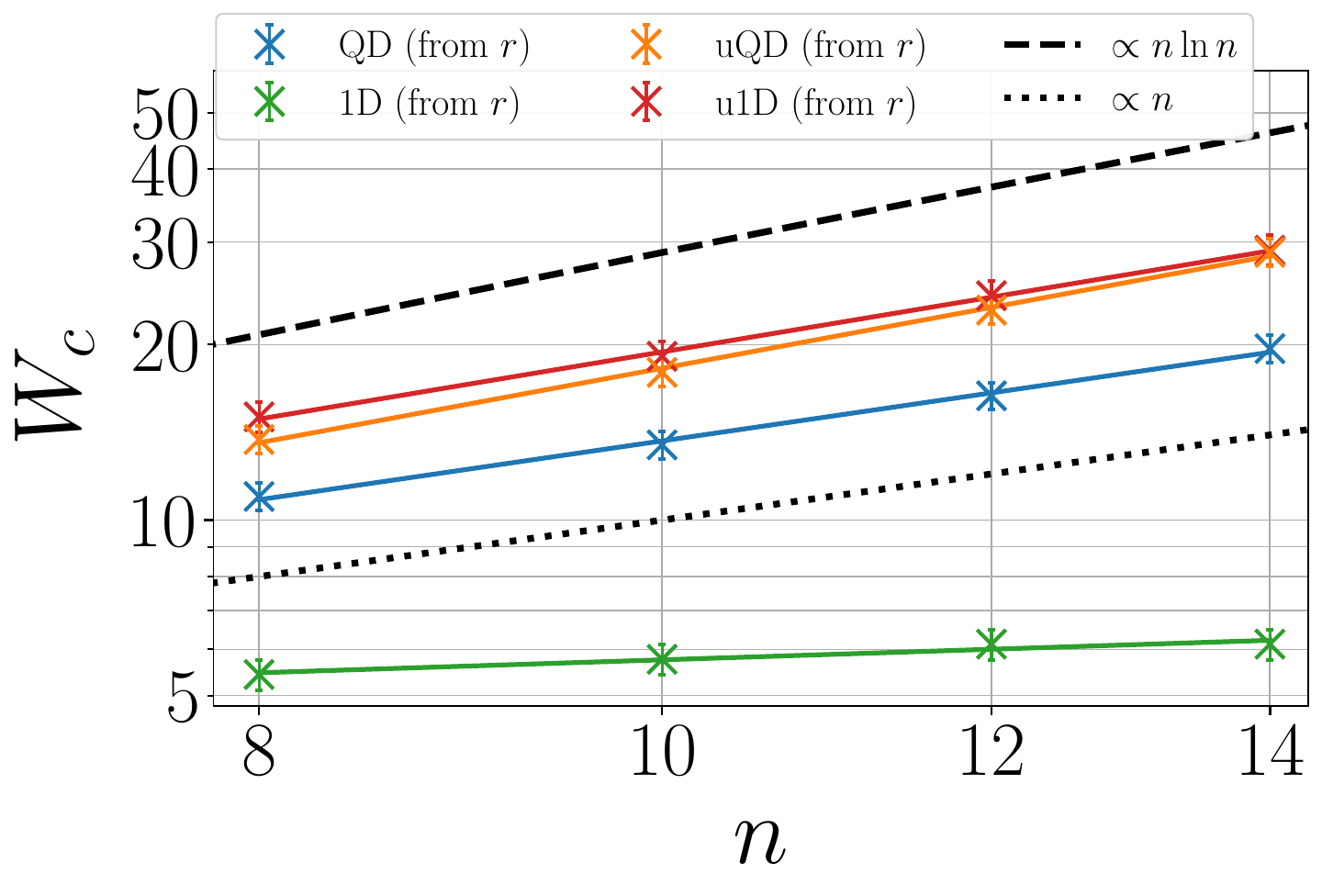}
\hfill
\includegraphics[width = 0.48\textwidth]{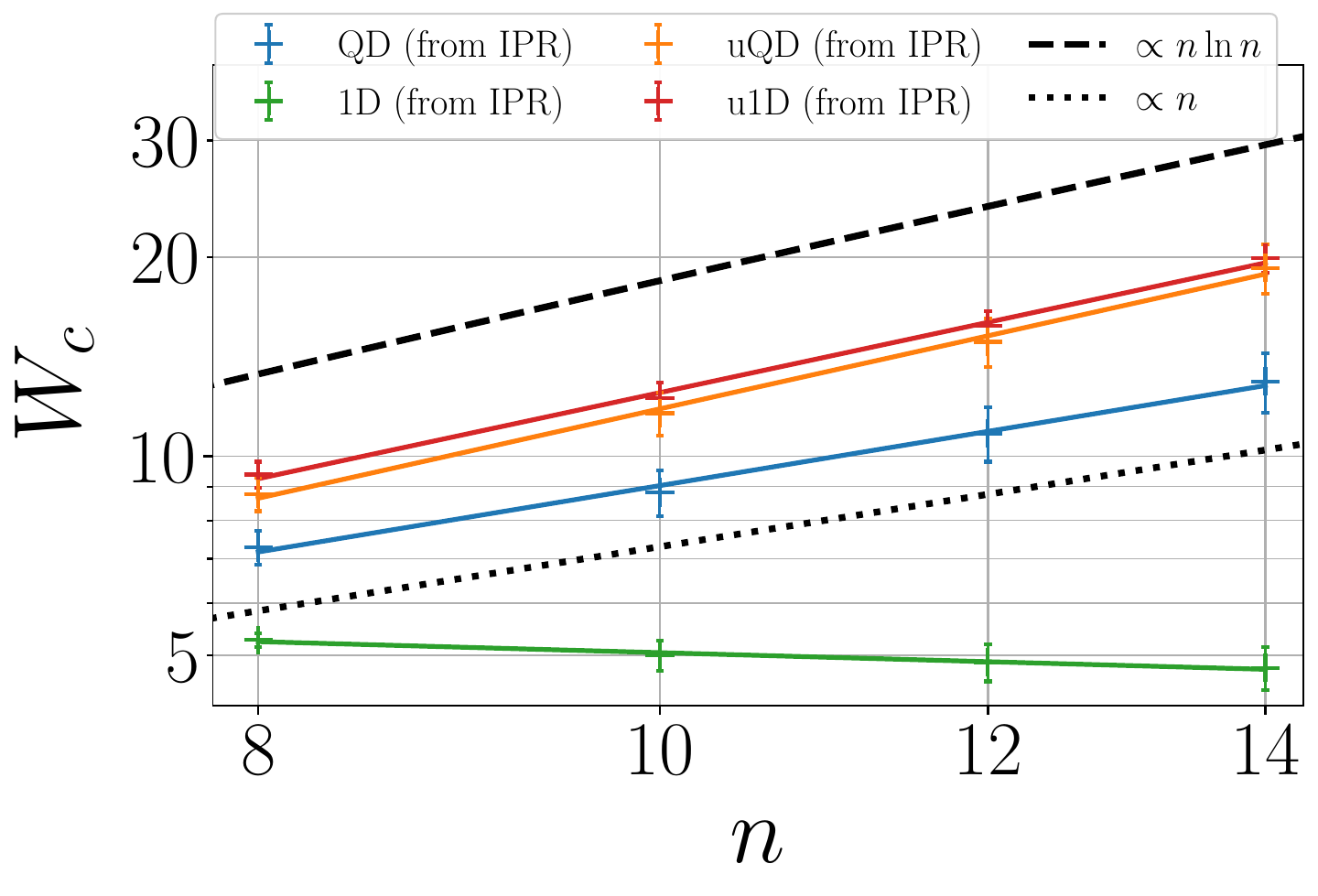}
\caption{
Comparison of scaling of critical disorder of MBL transition in 1D, QD, u1D, and uQD models. Symbols are numerical values of $W_c^{\rm 1D}(n)$, $W_c^{\rm QD}(n)$, $W_c^{\rm u1D}(n)$, and $W_c^{\rm uQD}(n)$ based on data for mean gap ratio $r$ ({\it left  panel}) and for logarithmic derivative $\alpha$ of IPR $P_2$ ({\it right panel}). Straight lines (corresponding to power-law fits) are guide to the eye.   
Black dashed lines are $W \sim n \ln n$
[corresponding to the analytically predicted large-$n$ scaling \eqref{eq:Wc_RRG_scaling} in the uQD and u1D models]; black dotted lines are $W \sim n$.
}
\label{fig:Wc_comparison_all_n}
\end{figure*}

To further compare the QD and 1D models, we now combine the results for $W_c(n)$ in these models as well as in their counterparts with removed Fock-space hopping correlations (uQD and u1D models) in Fig.~\ref{fig:Wc_comparison_all_n}. We recall that from the Fock space point of view, the QD and 1D models are different in that the correlations $C_E$ and $C_T$ of the QD model depend on the Hamming distance only, while these correlations in the 1D model depend on an additional parameter reflecting its 1D real-space structure. It is obvious from Fig.~\ref{fig:Wc_comparison_all_n} that this strongly changes the scaling $W_c(n)$ of the critical disorder of the MBL transition.  At the same time, once the hopping correlations are removed (i.e., off-diagonal matrix elements of $C_T$ are set to zero), we obtain two models---uQD and u1D---with nearly indistinguishable $W_c(n)$, despite a qualitative difference in the form of $C_E$ correlations of their Hamiltonians. The reason for this is clear from a discussion at the end of Sec.~\ref{subsec:comparison_1D_u1D}: 
when numerators of contributions \eqref{eq:eta-path} are uncorrelated (which is the case for both uQD and u1D models), interference between different contributions is absent irrespective of the correlation pattern of energies $E_\alpha$ in the denominators.

\section{Summary and outlook}
\label{sec:summary}

In this paper, we have explored the role of correlations between matrix elements of Hamiltonians in the Fock-space representation in the scaling of MBL transitions. For this purpose, we have investigated five models that all share the same Gaussian distributions of diagonal and off-diagonal Fock-space matrix elements
(energies $E_\alpha$ and hoppings $T_{\alpha\beta}$, respectively) but differ in their correlations, see
Fig.~\ref{fig:Figure_merit}
and Table \ref{tab:correl_summary}. The Hilbert space of all these models is that of a system of $n$ spins 1/2, with a volume $N=2^n$. All considered Hamiltonians, when presented in $\hat{S}_i^z$ basis, involve only single-spin-flip processes. Thus, all of them can be viewed as Anderson models on a graph having the form of an $n$-dimensional hypercube, with hopping matrix elements (i.e., links) associated with the edges of the hypercube. 

Two of the models are ``conventional'' many-body spin models with pair interactions and a random Zeeman field: one of them (1D model) is a spin chain with nearest-neighbor interaction and another one (QD model) involves interactions between all pairs of spins. In the Fock-space representation, both these models are characterized by strong correlations between energies $E_\alpha$ and between hoppings $T_{\alpha\beta}$ (i.e., off-diagonal elements of the corresponding covariance matrices $C_E$ and $C_T$). For the QD model, these correlations depend on the Hamming distance only, while for the 1D model, they have a more complex structure reflecting the 1D real-space geometry. The further two models, uQD and u1D, are obtained from the QD and 1D models, respectively, by removing correlations of Fock-space hoppings, i.e., by setting off-diagonal matrix elements of $C_T$ to zero. Finally, in the fifth model, the QREM, both Fock-space energy and hopping correlations (off-diagonal elements of $C_E$ and $C_T$) are absent. 

We have carried out an exact-diagonalization numerical study of these models, complemented by an analytical treatment. The central question in our study was the scaling of the critical disorder $W_c(n)$ with $n$. Our numerical results are in a very good agreement with analytical predictions of the scaling $W_c \sim n^{1/2} \ln n$ for QREM, $W_c \sim n \ln n$ for the uQD and u1D models, $n^{3/4} (\ln n)^{-1/4} \lesssim W_c \lesssim n \ln n$ for the QD model (with numerical data very well fitted to $W_c \sim n$), and $W_c(n) \sim  \text{const}$ for the 1D model. Comparing these results, we can understand implications of correlations of $E_\alpha$ and $T_{\alpha\beta}$. More specifically:

\begin{itemize}

\item  A comparison of the scaling of $W_c^{\rm QREM}$ with scaling of $W_c^{\rm uQD}$ and $W_c^{\rm u1D}$ reveals that, for a model with uncorrelated random hoppings, strong Fock-space energy correlations parametrically enhance delocalization. The mechanism of this is enhancement of the probability of resonances by energy correlations.

\item  Comparing the scaling of $W_c^{\rm 1D}$  with that of $W_c^{\rm u1D}$, we see that the scaling of the MBL transition in the 1D model crucially depends on a combined effect of correlations of Fock-space energies and hoppings. Making the random hopping matrix elements uncorrelated strongly enhances delocalization. A similar conclusion is made from a comparison of $W_c^{\rm QD}$  scaling with that of $W_c^{\rm uQD}$, although in this case the effect of removing Fock-space hopping correlations is less dramatic. 

\item  Finally, let us compare the scaling of $W_c^{\rm QD}$ with that of $W_c^{\rm 1D}$. Both these models are characterized by strong correlations of Fock-space energies and hoppings. The key difference is that the correlations depend only on the Hamming distance in the QD model, while they have a more complex form in the case of the 1D model. This is responsible for a very different scaling of $W_c$ in both models. Specifically, the behavior $W_c(n) \sim {\rm const}$ in the 1D model crucially depends on the correlations being not simply a function of Hamming distance but rather reflecting the real-space dimensionality.

\end{itemize}

We note that our conclusions differ essentially from those in Refs.~\cite{Roy2020, Roy2020numerics}, {\color{black} which is related to flaws in these papers. Specifically, Ref.~\cite{Roy2020numerics} considers, motivated by Fock-space approach to MBL, a model of RRG with strong correlations of energies, Eq.~\eqref{eq:appendix-correl-RRG-C_E}. As we explain in Appendix \ref{app:RRG-corr}, this model is actually ill-defined (the covariance matrix is not positive definite), in conflict with numerics in Ref.~\cite{Roy2020numerics}. The model can be defined on a hypercube, and the starting point of the analytical study in Ref.~\cite{Roy2020numerics} corresponds to our uQD model. However, the disorder identified by a mean-field argumentation in Ref.~\cite{Roy2020numerics} as a position of the localization transition is, in our notations, $W \sim 1$ and is deeply in the ergodic phase, since the true critical disorder is $W_c \sim n \ln n$, see the discussion below Eq.~(\ref{eq:Wc-RRG-corr-c}).
In Ref.~\cite{Roy2020}, the authors replace the energy correlation $C_E$ by its average $\overline{C_E}(r_{\alpha\beta})$ over Fock-space directions, thus making no distinctions between 1D and QD models (in our notations) and failing to appreciate that they have dramatically different scaling $W_c(n)$, see the discussion at the end of Section~\ref{subsec:1D_model_def_energy_correls}. 
The model addressed analytically in that work is essentially our uQD model but, like in Ref.~\cite{Roy2020numerics}, the authors erroneously identify $W \sim 1$ as the critical disorder, while the correct location of the MBL transition in this model is $W_c \sim n \ln n$.
}

In addition to the scaling of the critical disorder $W_c$, we have analyzed the scaling of the transition width $\ln(W_+/W_-)$ with $n$. For all five models, our numerical results yield $\ln(W_+/W_-) \sim n^{-\mu}$, with $\mu \approx 1.0 - 1.3$, in the range $n= 8 - 14$. These results demonstrate sharpening of the MBL transition with increasing $n$. At the same time, the numerically observed $n^{-\mu}$ scaling of the transition width is different from our analytical prediction $n^{-3} \ln^2 n$ in the large-$n$ limit for the QREM, uQD, and u1D models. This difference is in full consistency with our analytical results that predict a flowing effective exponent $\mu(n)$ in the numerically observed range for $n$ studied in our simulations. We further predict that the asymptotic behavior of the transition width in these models is applicable for $n > n^{\rm crit} \approx 22$, i.e., already for moderately large systems. 

We close the paper with a few comments on our findings in the general context of MBL research.

Rather generally, an MBL transition can be characterized by the dependence $W_c(n)$ of the critical disorder on the system size $n$ (number of spins, atoms, qubits, \ldots) and by the $n$-dependence of the transition width $\ln(W_+(n)/W_-(n)) \approx \Delta W (n) / W_c(n)$. If $\Delta W (n) / W_c(n) \to 0$ at $n \to \infty$, one can speak about a well-defined (sharp) transition in the large-$n$ limit. This notion of the MBL transition applies independently of the large-$n$ behavior of $W_c(n)$; in particular, it does not rely on whether $W_c(n)$ has a finite large-$n$ limit. In fact, for almost all models, $W_c(n)$ grows indefinitely with increasing $n$; one-dimensional systems with a short-range interaction represent a notable exception. While QREM is only a toy-model for the MBL transition, the corresponding analytical results for the $n$-dependence of the critical disorder and transition width, see Fig.~\ref{fig:Wc_QREM_theory}, may serve as a guiding example. 

The sharpening of the transition can be characterized by a flowing exponent $\mu(n)$ defined by Eq.~\eqref{eq:transition-width-mu-n}. 
For every transition, there should be a characteristic system size $n^{\rm crit}$ such that a system of size $n > n^{\rm crit}$ is in the critical regime, implying that the scaling of $\Delta W (n) / W_c(n)$ is close to its asymptotic form. Our results for the QREM (as well as for uQD and u1D models described by the RRG-like approximation) show that one should be cautious when trying to interpret exact-diagonalization data in terms of a large-$n$ asymptotic critical behavior. At the same time, a moderately large value $n^{\rm crit} \approx 22$ estimated for these models gives hope that the critical regime can be achievable in numerical or experimental studies of some genuine MBL models.


{\color{black} 
The choice of particular, single-spin-flip models in this paper was dictated by our wish to have models that have identical Fock-space coordination numbers and identical distributions of Fock-space energies and hoppings, differing only in correlations. This has allowed us to explore the role of Fock-space correlations and to demonstrate differences in scaling of $W_c(n)$ in a particularly 
clear way. We emphasize, however, that our conclusions are expected to have a high degree of universality, with the 1D, QD, u1D, and uQD models addressed in our work being representatives of broad classes of models defined by Eq.~\eqref{eq:Fock_space_rep}. All models of 1D type are characterized by Fock-space correlations reflecting the 1D geometry. Contrary to this, QD models have a structure isotropic in Fock space, leading to a much more efficient many-body delocalization.   For each 1D model (QD model), one can define the corresponding u1D model (respectively, uQD model) by making the non-zero Fock-space hopping matrix elements $T_{\alpha\beta}$ random uncorrelated variables, which promotes delocalization. The scaling of $W_c(n)$ for u1D and uQD models can be determined from the RRG-like approximation. The difference in scaling between a 1D model (e.g., the XXZ model mentioned above in Sec.~\ref{sec:analytics-1D})
and its u1D counterpart is predicted to be particularly dramatic: $W_c^{\rm 1D} \sim 1$ (up to a relatively weak finite-size drift) and $W_c^{\rm u1D} \sim n \ln n$. We note that the Fock-space coordination number is generically $\sim n$ for 1D and u1D models.
For QD models (and their uQD counterparts), the coordination number is frequently a different power of $n$ (specifically, $\sim n^2$ for spin QD models with two-spin-flip interaction and $\sim n^4$ for fermionic QD models), which obviously affects the RRG-like result for $W_c^{\rm uQD}$. The difference in scaling between $W_c^{\rm QD}$ and $W_c^{\rm uQD}$ is much less dramatic than that between $W_c^{\rm 1D}$ and $W_c^{\rm u1D}$. In particular, for fermionic QD model and two-spin-flip spin QD model considered in Refs.~\cite{gornyi2017spectral,Herre2023}, this difference is limited to (at most) a logarthmic factor. We note that the fermionic QD model is essetially equivalent to the deformed Sachdev-Ye-Kitaev model known as $\text{SYK}_4 + \text{SYK}_2$ model \cite{garcia-garcia2018chaotic,micklitz2019nonergodic,monteiro2020minimal,Monteiro2021,nandy2022delayed,larzul2022quenches}.
}

As a final comment, it was argued that MBL is essential for ensuring stability of quantum computers \cite{Berke2022transmon,qian2023mitigating}. Thus, understanding the dependence $W_c(n)$ for systems of coupled qubits may be relevant to scalability of future quantum-information devices of various architectures.

\section*{Acknowledgements}
We acknowledge support from the state of Baden-W{\"u}rttemberg through the Kompetenzzentrum Quantum
Computing (Project QC4BW). We thank {\color{black} Y. Fyodorov}, J.~Herre and K.~Tikhonov for useful communications. ADM also acknowledges a discussion with S.~Roy on Ref.~\cite{Roy2020numerics}.

\appendix

\section{Analytical results for the RRG model and QREM}
\label{app:RRG-QREM}

In this Appendix, we present details of the derivation of analytical results for the QREM (which are closely related to those for the RRG model) presented in Sec.~\ref{sec:RRG_scaling}. Specifically, we are interested here in the behavior of the critical disorder $W_c$ and of the finite-size transition width $\Delta W$. We consider first the RRG model and then ``translate'' the results to the QREM. 

\subsection{RRG model} 
\label{app:RRG}

The RRG model with a large connectivity was studied in Ref.~\cite{Herre2023}. We first outline some results of that work for an Anderson model on an RRG with connectivity $m+1 \gg 1$, uncorrelated random energies $E_\alpha$ sampled from a box distribution on $[-W/2, W/2]$, and with hopping matrix elements $T=1$. In the ``thermodynamic'' limit of a very large number of sites, $N \to \infty$, the critical disorder $W_c$ is a solution of the equation
\begin{equation}
W_c = 4m \ln (W_c/2) \,.
\label{eq:RRG-Wc-equation}
\end{equation}
The number of sites $N$ is related to the linear size $L$ of the system via $N \simeq m^L$. For a finite (but large) $N$, the ``finite-size transition'' $W_c(N)$ is determined by the condition $N_\xi (W) = N$ or, equivalently, $\xi(W) =L$, where $N_\xi$ is the correlation volume and $\xi = \ln N_\xi / \ln m$ is the corresponding correlation length. In these notations, the asymptotic critical disorder, i.e, the solution of Eq.~\eqref{eq:RRG-Wc-equation}, is $W_c(N \to \infty) \equiv W_c(\infty)$. The correlation volume $N_\xi(W)$
is given by 
\begin{equation}
\ln N_\xi = \frac{2\pi \ln(W/2)}{x} \,,
\label{eq:RRG-N-xi-general-1}
\end{equation}
where $x$ is a solution of the equation
\begin{equation}
\frac{\sin x}{x} = \frac{f(W)}{f(W_c(\infty))} \,; \qquad
f(W) = \frac{W}{\ln (W/2)} \,.
\label{eq:RRG-N-xi-general-2}
\end{equation}

When the system volume $N$ is large enough, the critical disorder $W_c(N)$ is close to its asymptotic value $W_c(\infty)$, so that the finite-size transition takes place at disorder $W=W_c(N)$ belonging to the critical regime defined by the condition
\begin{equation}
\frac{W_c(\infty)}{2} < W < W_c(\infty).
\label{eq:RRG-critical regime}
\end{equation}
For disorder strength $W$ within the critical regime,   Eqs.~\eqref{eq:RRG-N-xi-general-1}, \eqref{eq:RRG-N-xi-general-2} yield the following asymptotic behavior of the correlation volume:
\begin{equation}
\ln N_\xi (W) = \pi \ln m \left[\frac{3}{2} \left( 1 - \frac{W}{W_c(\infty)}\right) \right]^{-1/2} .
\label{eq:RRG-N-xi-critical}
\end{equation}
(A more accurate form of Eq.~\eqref{eq:RRG-N-xi-critical}, which includes also subleading corrections that decay slowly with $m$, is presented in Eq. (27) of Ref.~\cite{tikhonov19critical}.)
For a large coordination number, $m \gg 1$, the condition \eqref{eq:RRG-critical regime} of critical regime requires very large system volume $N$, so that the critical regime cannot be studied by exact diagonalization. At the same time, for large $m$ a parametrically broad pre-critical regime emerges, defined by the condition 
\begin{equation}
m < W < \frac{W_c(\infty)}{2} \,.
\end{equation}
In this regime, the correlation volume $N_\xi$ is given by
\begin{equation}
N_\xi (W) \simeq \frac{W^2}{m} \exp\left( \frac{W^2}{2m} \right).
\label{eq:RRG-N-xi-precritical}
\end{equation}
The border $N_\xi^{\rm crit}$ between the pre-critical and critical regimes (which is, of course, not sharp) is obtained by substituting $W=W_c(\infty)/2$ in 
Eqs.~\eqref{eq:RRG-N-xi-general-1}, \eqref{eq:RRG-N-xi-general-2}, which yields
\begin{equation}
N_\xi^{\rm crit} \simeq \frac{1}{m} (2m \ln m)^\gamma \,, \qquad \gamma \approx 3.3 \,.
\label{eq:RRG-N-xi-crit}
\end{equation}
Around this scale, Eq.~\eqref{eq:RRG-N-xi-critical} crosses over into Eq.~\eqref{eq:RRG-N-xi-precritical}.

After this summary of results of Ref.~\cite{Herre2023} relevant to our work, we are ready to use these results for a detailed analysis of the finite-size scaling of the transition. We consider first the critical regime, $N > N_\xi^{\rm crit}$. Combining Eq.~\eqref{eq:RRG-N-xi-critical} with the equation defining the finite-size transition point $W_c(N)$,
\begin{equation}
N_\xi(W_c(N)) = N \,,
\label{eq:RRG-finite-size-Wc}
\end{equation}
and using $N=m^L$, we obtain the finite-size shift of the transition,
\begin{equation}
\frac{W_c(\infty) - W_c(N)}{W_c(\infty)} = \frac{2\pi^2}{3} \: \frac{1}{L^2} \,.
\label{eq:RRG-Wc-finite-size-shift}
\end{equation}
To determine the finite-size width of the transition, we recall that observables that are used to detect the transition (such as the level statistics or the IPR) exhibit for $W < W_c(\infty)$ a ``volumic'' scaling, with $N/N_\xi$ being the relevant scaling parameter \cite{tikhonov2016anderson,garcia-mata17,biroli2018,tikhonov19statistics,tikhonov2021from,garcia-mata2022critical}. Thus, to estimate the width of the transition in a finite system, we should consider $N_\xi(W)$ varying in an interval $[b_+N, b_-N]$, with $b_+, b_- \sim 1 $. The corresponding disorder strengths are determined by the equations following from 
Eq.~\eqref{eq:RRG-finite-size-Wc}:
\begin{equation}
N_\xi(W_-(N)) = b_-N \,, \qquad
N_\xi(W_+(N)) = b_+N \,.
\label{eq:RRG-finite-size-W-plus-minus}
\end{equation}
The transformation $N \longmapsto bN$ corresponds to an additive change of the length $L$:
\begin{equation}
L \: \longmapsto \: L + \frac{\ln b}{ \ln m}\,.
\label{eq:L-b-additive-shift}
\end{equation}
Combining this with Eq.~\eqref{eq:RRG-Wc-finite-size-shift}, we obtain the following result for the width of the transition, $\Delta W(N) = W_+(N)- W_-(N)$, in the critical regime:
\begin{equation}
\frac{\Delta W(N)}{W_c(N)} \simeq 
\frac{\Delta W(N)}{W_c(\infty)} =
\frac{4\pi^2}{3} \: \frac{\ln(b_+/b_-)}{\ln m} \: \frac{1}{L^3}\,.
\label{eq:RRG-Wc-finite-size-width}
\end{equation}

Importantly, the $1/L^3$ scaling of the transition width, Eq.~\eqref{eq:RRG-Wc-finite-size-width}, is different from the $1/L^2$ scaling of the shift \eqref{eq:RRG-Wc-finite-size-shift}. This implies that the transition sharpens much faster than it approaches its asymptotic location. It is instructive to understand the origin of this behavior. Inspecting the derivation, we can trace it back to an exponential relation between the length $L$ and the volume $N=m^L$ (and, correspondingly, between the correlation length $\xi$ and the correlation volume $N_\xi$). For Anderson transition in $d$ dimensions, one would have instead $N=L^d$. Combining this with a power-law divergence of the correlation length, $\xi \sim (W_c-W)^{-\nu}$, and repeating the above analysis, we would get an identical scaling for the finite-size shift and width of the transition,
\begin{equation}
\frac{W_c(\infty) - W_c(N)}{W_c(\infty)} \propto L^{-1/\nu} \,, \ \ \ 
\frac{\Delta W(N)}{W_c(\infty)} \propto L^{-1/\nu} \,.
\label{eq:Anderson-tr-width-shift-scaling}
\end{equation}
At the same time, for the exponentially growing volume, $N=m^L$, we have a distinct scaling behavior,
\begin{equation}
\frac{W_c(\infty) - W_c(N)}{W_c(\infty)} \propto L^{-1/\nu} \,, \ \ \ 
\frac{\Delta W(N)}{W_c(\infty)} \propto L^{-1-1/\nu} \,.
\label{eq:RRG-tr-width-shift-scaling}
\end{equation}
For the RRG model, $\nu=1/2$ on the delocalized side of the transition, and Eq.~\eqref{eq:RRG-tr-width-shift-scaling} reproduces the $1/L^2$ scaling of the shift and $1/L^3$ scaling of the width obtained above.

We turn now to the pre-critical regime, $N < N_\xi^{\rm crit}$. 
Substituting
Eq.~\eqref{eq:RRG-N-xi-precritical}
into Eq.~\eqref{eq:RRG-finite-size-Wc}, we obtain
\begin{equation}
W_c(N) = 2 m \ln m \: (L-1) \,,
\label{eq:RRG-precritical-WcN}
\end{equation}
i.e., a linear drift of the critical disorder with the linear size $L$ of the system. Further, 
using Eqs.~\eqref{eq:RRG-finite-size-W-plus-minus} and
\eqref{eq:L-b-additive-shift} to determine the transition width, we find
\begin{equation}
\frac{\Delta W(N)}{W_c(N)} =
 \frac{\ln(b_+/b_-)}{\ln m} \: \frac{1}{L-1}\,.
\label{eq:RRG-Wc-finite-size-width-precritical}
\end{equation}

These results can be straightforwardly extended to a more general RRG model, with distribution $\gamma(E)$ of (uncorrelated) diagonal energies $E_\alpha$ and with some distribution of (uncorrelated) transition amplitudes $T_{\alpha\beta}$. 
(An underlying assumption is that the distribution $\gamma(E)$ is characterized by a single energy scale $W$.) Then one should perform a substitution \cite{Herre2023}
\begin{equation}
\frac{1}{W} \: \longmapsto \: \gamma(0) \langle | T | \rangle \,,
\label{eq:RRG-substitution-W}
\end{equation}
where $\langle |T| \rangle$ denotes the average value of $|T_{\alpha\beta}|$.
Since the product $\gamma(0) \langle | T | \rangle$ is proportional to $1/W$, this amounts to a rescaling of the disorder $W$ (and correspondingly of $W_c$) in all the formulas. 
 
\subsection{QREM}
\label{app:QREM}

We are now going to apply the results obtained for the RRG model (Sec.~\ref{app:RRG}) to the QREM defined in Sec.~\ref{sec:QREM}. In Fock-space representation, QREM is a tight-binding model defined on a $n$-dimensional hypercube graph (which is a Fock space for $n$ spins 1/2, see Sec.~\ref{sec:Fock-space-rep}). This graph is regular (with coordination number $n$) but not truly random. At the same time, the key property of a random graph is suppression of small-scale loops. For a large $n$, small-scale loops are rare also for paths on an $n$-dimensional hypercube. Therefore, at large $n$, QREM should exhibit properties of the RRG model. Although some deviations may appear at moderately large $n$, we will neglect them in the analysis below.

For our QREM, $\langle |T| \rangle$ is given by Eq.~\eqref{eq:QD_mean_transition_amplitude}
and $\gamma(E)$ is given by Eq.~\eqref{eq:gamma-E}, so that we should perform in formulas of Sec.~\ref{app:RRG}
a substitution \eqref{eq:RRG-substitution-W} with
\begin{equation}
\gamma(0) \langle |T| \rangle = \left( \frac{3}{2n}\right)^{1/2} \frac{1}{W} \,.
\label{eq:RRG-QREM-rescaling}
\end{equation}
Therefore, all $W_c^{\rm RRG}$ below in this subsection are obtained from $W_c$ of Sec.~\ref{app:RRG} by multiplying by $(3/2n)^{1/2}$.
Further, we should make a replacement $m+1 \longmapsto n$ for the coordination number. (Since $n$ is assumed to be large, we will simply replace $m \longmapsto n$.)
There is, however, one more important difference between the RRG model and QREM. Indeed, for the RRG model, the coordination number $m+1$ and the system volume $N = m^L$ are two independent parameters. On the other hand, for the QREM, the coordination number $n$ and the system volume $N=2^n$ are directly related to each other. 
Thus, introducing a notation $W_c^{\rm RRG}(m,N)$ for the critical disorder of the RRG model, we have for the critical disorder of the QREM:
\begin{equation}
W_c^{\rm QREM}(n) \simeq W_c^{\rm RRG}(n, 2^n) \,.
\end{equation}
In the same way, we obtain 
\begin{equation}
\Delta W^{\rm QREM}(n) =  
\Delta W^{\rm RRG}(n, 2^n) \,
\end{equation}
for the transition width.
In the large-$n$ limit, $W_c^{\rm QREM}(n)$ asymptotically approaches $W_c^{\rm RRG}(n,\infty)$,
which is a solution of Eq.~\eqref{eq:RRG-Wc-equation} [with $m \longmapsto n$ and multiplied by $(3/2n)^{1/2}$].

Adjusting RRG formulas to the QREM, we should also take into account that $N=m^L$ for RRG and $N=2^n$ for QREM, so that 
\begin{equation}
L \: \longmapsto \: n \, \frac{\ln 2}{ \ln n}
\end{equation}
(where we used $m \longmapsto n$). 
In the critical regime, Eqs.~\eqref{eq:RRG-Wc-finite-size-shift} and \eqref{eq:RRG-Wc-finite-size-width} of the RRG model now translate into
\begin{equation}
\frac{W_c^{\rm RRG}(n,\infty) - W_c^{\rm QREM}(n)}{W_c^{\rm RRG}(n,\infty)} = \frac{2\pi^2}{3 \ln^2 2} \: \frac{\ln^2 n}{n^2} \,.
\label{eq:QREM-Wc-finite-size-shift}
\end{equation}
and
\begin{equation}
\frac{\Delta W^{\rm QREM}(n)}{W_c^{\rm QREM}(n)} =
\frac{4\pi^2}{3\ln^2 2} \: \ln(b_+/b_-) \: \frac{\ln^2 n}{n^3}\,.
\label{eq:QREM-Wc-finite-size-width}
\end{equation}
In the same way, Eqs.~\eqref{eq:RRG-precritical-WcN} and
\eqref{eq:RRG-Wc-finite-size-width-precritical}
yield for the QREM pre-critical regime 
\begin{equation}
W_c^{\rm QREM}(n) = \sqrt{6} \: \ln 2 \: \cdot n^{3/2}
\label{eq:QREM-Wc-precritical}
\end{equation}
and 
\begin{equation}
\frac{\Delta W^{\rm QREM}(n)}{W_c^{\rm QREM}(n)} = \frac{\ln(b_+/b_-)}{\ln 2} \: \frac{1}{n}.
\label{eq:QREM-DeltaW-precritical}
\end{equation}

The border $n^{\rm crit}$ between the pre-critical and critical regimes for the QREM is determined by the condition 
$W_c^{\rm QREM}(n) = (1/2) W_c^{\rm RRG}(n,\infty)$, which yields $n^{\rm crit} \approx 22$. Let us emphasize that this $n^{\rm crit}$ is a parameterless number ``of order unity'' (even though its value is not so small numerically). This means that the pre-critical regime is not parametrically defined in the case of QREM but rather is restricted to ``small'' systems, $n < 22$. Correspondingly, Eqs.~\eqref{eq:QREM-Wc-precritical} and \eqref{eq:QREM-DeltaW-precritical} are not parametrically controlled.
This should be contrasted to the case of the RRG model where the coordination number $m$ and the system size $N$ are two independent parameters, so that for a large $m$ the pre-critical regime is parametrically broad, 
see Eq.~\eqref{eq:RRG-N-xi-crit}.

Results of this Appendix for the position of the transition $W_c^{\rm QREM}(n)$ and the associated transition width $\Delta W^{\rm QREM}(n)$ for the QREM are presented in Sec.~\ref{sec:RRG_scaling} and, in particular, in Figs.~\ref{fig:Wc_QREM_theory} and
\ref{fig:Wc_QREM_aymptotics}.

\section{Justification of neglect of $V^z$ terms in the RRG-like approximation for uQD and u1D models}
\label{app:RRG_with_energy_correls}

In Sec.~\ref{sec:uQD-u1D-analytics}, we have derived the large-$n$ scaling of the critical disorder $W_c(n)$ for the uQD and u1D models by using the RRG-like approximation with energy correlations. In our derivation there, we have discarded $V_z$ contributions to the energies since they are small compared to the main term at large $n$ (and thus large $W_c$), see a comment below
Eq.~\eqref{Wc-QD-upper-boundary}. In this Appendix, we perform a derivation including these terms and verify that their effect is indeed negligibly small already for our smallest $n$.

With $V^z$ terms taken into account, Eq.~\eqref{Wc-QD-upper-boundary} takes the form
\begin{align}
E_\beta = \pm \epsilon_k \pm \frac{1}{\sqrt{n}}\sum\limits_{\substack{l=1 \\ l \neq k}}^{n} \left(V_{kl}^z + V_{lk}^z \right)
\end{align}
for the QD model  and 
\begin{align}
E_\beta = \pm \epsilon_k \pm \left(V_{k, k+1}^z + V_{k-1, k}^z \right)
\end{align}
for the 1D model. The random energies $\epsilon_k$ are uniformly distributed on $[-W,W]$. Since  $V^z_{kl}$ are normally distributed with unit variance,
the total contribution of the $V_z$ terms
is normally distributed with variance $2$ in both cases.  Thus, the distribution $\gamma_1(E_\beta)$ of energies $E_\beta$ on a site directly coupled to a site with energy $E_\alpha = 0$, which is given by Eq.~\eqref{eq:gamma1-E} when $V^z$ terms are discarded, is modified as follows:
\begin{align}
\gamma_1(E) &= \int{\rm d}s \: \frac{1}{2W}\theta(W-|s|) \: \frac{1}{2\sqrt{\pi}}\exp(-\frac{(E-s)^2}{4})\nonumber\\
&= \frac{1}{4 W}\left[\erf\left(\frac{E+W}{2}\right) - \erf\left(\frac{E-W}{2}\right)\right],
\label{eq:gamma_convolution}
\end{align}
where $\erf(z)$ denotes the error function. 
Using $\langle |T| \rangle = \sqrt{\pi}/2$, we thus get 
$\langle |T| \rangle \gamma_1(\epsilon = 0) = (\sqrt{\pi}/4W)\erf(W/2)$.  Substitution of this result  into Eq.~\eqref{eq:self_cons_critical_disorder_Herre} yields the modified equation for the critical disorder $W_c(n)$ in the uQD and u1D models:
\begin{align}
W_c  = n\sqrt{\pi}\erf\left(\frac{W_c}{2}\right)\ln\left(\frac{4W_c}{\sqrt{\pi}\erf\left(\frac{W_c}{2}\right)}\right).
\label{eq:critical_W_uQD_self_consistent_eq}
\end{align}
For a large $n$, we have $W_c \sim n \ln n \gg 1$, and, Eq.~\eqref{eq:critical_W_uQD_self_consistent_eq} reduces to Eq.~\eqref{eq:self_cons_critical_disorder_RRG} (in view of $\erf(z) \simeq 1$ for $z \gg 1$), 
as expected. Moreover, already for our smallest $n=8$, the numerical difference between $\erf(W_c/2)$ and unity is extremely small, so that the solution of 
Eq.~\eqref{eq:critical_W_uQD_self_consistent_eq} is virtually indistinguishable from that of Eq.~\eqref{eq:self_cons_critical_disorder_RRG}.

\section{Lower bound for the critical disorder in the QD model}
\label{app:QD-lower-bound}

In this Appendix, we present a derivation of the lower bound \eqref{Wc-QD-lower-boundary} on the critical disorder $W_c^{\rm QD}$. The idea of this derivation is analogous to the one developed
in Ref.~\cite{gornyi2017spectral} (see Appendix B there) but it requires a substantial modification since we study here a different QD model. {\color{black} An important difference is a parametrically weaker spectral diffusion in the present QD model as compared to the two-spin-flip quantum-dot model of Ref.~\cite{gornyi2017spectral}. Specifically, in the model of Ref.~\cite{gornyi2017spectral}, the amplitude of a two-spin-flip process and the shift of energy of any spin due to this process are of the same order. At the same time, in our single-spin-flip QD model defined in Sec.~\ref{sec:QD-model-def}, the amplitude of spin flip is of order unity, Eq.~(19), while a shift of energy of any other spin due to such process is $\sim 1/\sqrt{n} \ll 1$. }

As discussed in Sec.~\ref{sec:QD-analytics}, we consider a disorder $W= n /p$, with $1 \ll p \ll n$. In this case, any basis state $\alpha$ has $\sim p$ direct resonance partners; each of them is related to $\alpha$ by flipping a certain spin $\hat{S}_i$. Terms $\sim 1$ in the Hamiltonian of such resonant spin $\hat{S}_i$ are of the form
\begin{equation}
H^{(0)}_i = (\epsilon_i + h_i^z) \hat{S}_i^z
+ h^x_i \hat{S}_i^x + h^y_i \hat{S}_i^y \,,
\label{eq:appendix-QD-H0i}
\end{equation}
where 
$(\epsilon_i + h_i^z), h_i^x h_i^y \sim 1$ are random. Here the terms with couplings $h_i^\alpha$ originate from summation over $n$ random interaction terms with prefactors $\sim n^{-1/2}$ in Eqs.~\eqref{eq:QD-H0} and \eqref{eq:QD-H1}. The Hamiltonian \eqref{eq:appendix-QD-H0i} describes a spin in magnetic field of a strength $\tilde{\epsilon}_i \sim 1$ and random orientation. Performing a unitary rotation of the spin operators $\hat{S}_i^a$, we can orient this field in $z$ direction:
\begin{equation}
H^{(0)}_i = \tilde{\epsilon}_i \hat{R}_i^z \,,
\label{eq:appendix-QD-H0i-rotated}
\end{equation}
where $\hat{R}_i^a$ is the spin operators in the new basis. We perform this for all $p$ resonant spins; the above rotations are random and uncorrelated.

Now we include the terms describing interactions between the resonant spins.  These terms are of the type (the summation here goes over the subset of resonant spins)
\begin{equation}
\hat{H}_{\rm int} = \frac{1}{\sqrt{n}}\smashoperator{\sum}\limits_{\substack{i,j = 1\\ a \in \{x,y,z\}}}^{p}  V_{ij}^{a} \left( \hat{S}_i^z \hat{S}_j^a + \text{H.c.}\right),
\label{eq:appendix-QD-H1}
\end{equation}
with random coefficients $V_{ij}^{a} \sim 1$.
Upon unitary rotations of spins, this takes the form
\begin{equation}
\hat{H}_{\rm int} = \frac{1}{\sqrt{n}}\smashoperator{\sum}\limits_{\substack{i,j = 1\\ a,b \in \{x,y,z\}}}^{p}  \widetilde{V}_{ij}^{ab} \left( \hat{R}_i^a \hat{R}_j^b + \text{H.c.}\right),
\label{eq:appendix-QD-H1-rotated}
\end{equation}
with random coupling $\widetilde{V}_{ij}^{ab} \sim 1$. The total Hamiltonian of the resonant subsystem,
\begin{equation}
\hat{H}_{\rm res} = \sum_{i=1}^p \tilde{\epsilon}_i \hat{R}_i^z + \frac{1}{\sqrt{n}}\smashoperator{\sum}\limits_{\substack{i,j = 1\\ a,b \in \{x,y,z\}}}^{p}  \widetilde{V}_{ij}^{ab} \left( \hat{R}_i^a \hat{R}_j^b + \text{H.c.}\right),
\end{equation}
is a Hamiltonian of a spin quantum dot with random two-spin interactions (including two-spin-flip terms) studied in Ref.~\cite{gornyi2017spectral}. The parameters of the model are the number of spins and the interaction strength, which
were $N$ and $\alpha$ in Ref.~\cite{gornyi2017spectral} and are, respectively, $p$ and $n^{-1/2}$ in our case.
As shown in Ref.~\cite{gornyi2017spectral}, the upper bound for the critical value of $p$ (MBL transition point) in this system is (using notations of the present work, at given $n^{-1/2}$)
\begin{equation}
p_c \lesssim n^{1/4} \ln^{1/4} n \,.
\label{eq:appendix-QD-resonant-MBL}
\end{equation}
The scaling $p \sim n^{-1/4}$ (up to a logarithmic factor) in Eq.~\eqref{eq:appendix-QD-resonant-MBL} can be understood if one compares
the two-spin-flip level spacing $\sim 1/p^2$ to  the interaction matrix element $\sim n^{-1/2}$.

The argument now is that,
for large $n$, the level spacing $\sim 2^{-p}$
in the ergodic resonant subsystem
is much smaller than power-law functions of $n$.
As a consequence, in analogy with Appendix B of Ref.~\cite{gornyi2017spectral}, this ergodic subsystem can serve as a ``bath'' and assist delocalization (spin-flip processes) in the rest of the system. This argument also bears similarity to the avalanche mechanism of many-body delocalization in systems with a structure in real space \cite{roeck17,Thiery2017a}. Substituting Eq.~\eqref{eq:appendix-QD-resonant-MBL} into $W= n/p$, we obtain the lower bound for the critical disorder in our QD model,
\begin{equation}
W_c^{\rm QD}(n) \gtrsim n^{3/4} (\ln n)^{-1/4} \,,
\label{eq:appendix-QD-Wc-lower-bound}
\end{equation}
which is Eq.~\eqref{Wc-QD-lower-boundary} of the main text.

\section{Limitations of implementing Gaussian correlated random fields on RRG}
\label{app:RRG-corr}

As discussed in detail in Sec.~\ref{sec:model_definitions}, all the models studied in this work are characterized by multivariate Gaussian distributions. Let us focus for definiteness on the uQD model, which is characterized by energy correlations that depend only on Hamming distance $r_{\alpha\beta}$. The corresponding covariance matrix $C_E^\text{QD}$ is given by Eq.~(\ref{eq:correl_energies_final}), with states $\alpha$ being vertices of an $n$-dimensional hypercube. This multivariate Gaussian distribution on a hypercube graph emerges, by virtue of the central limit theorem, since energies $E_\alpha$ of the uQD model are given by Eq.~\eqref{eq:QD_on_site_energies}.

\begin{figure*}[t!]
  \centering
  \includegraphics[width=0.33\textwidth]{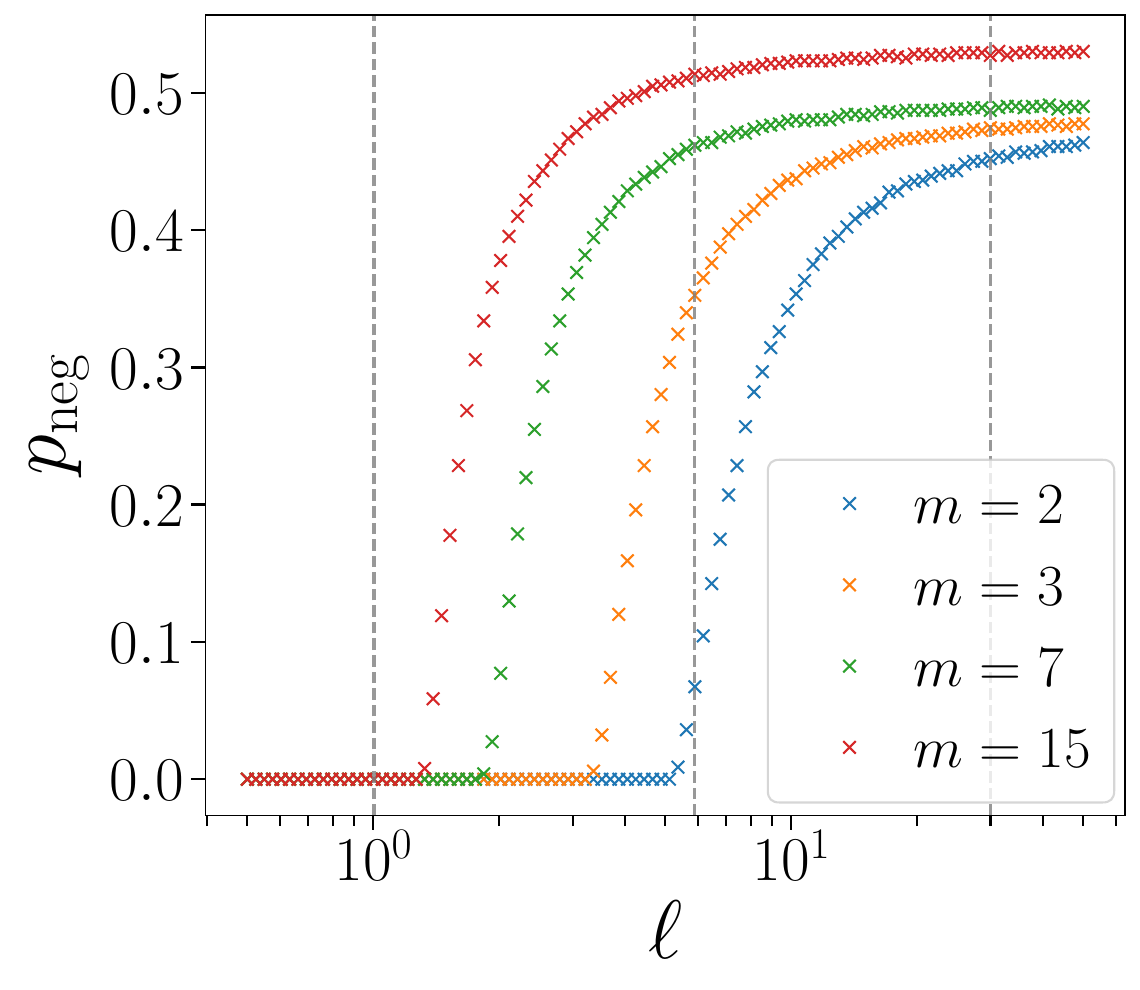}\hfill
  \includegraphics[width=0.33\textwidth]{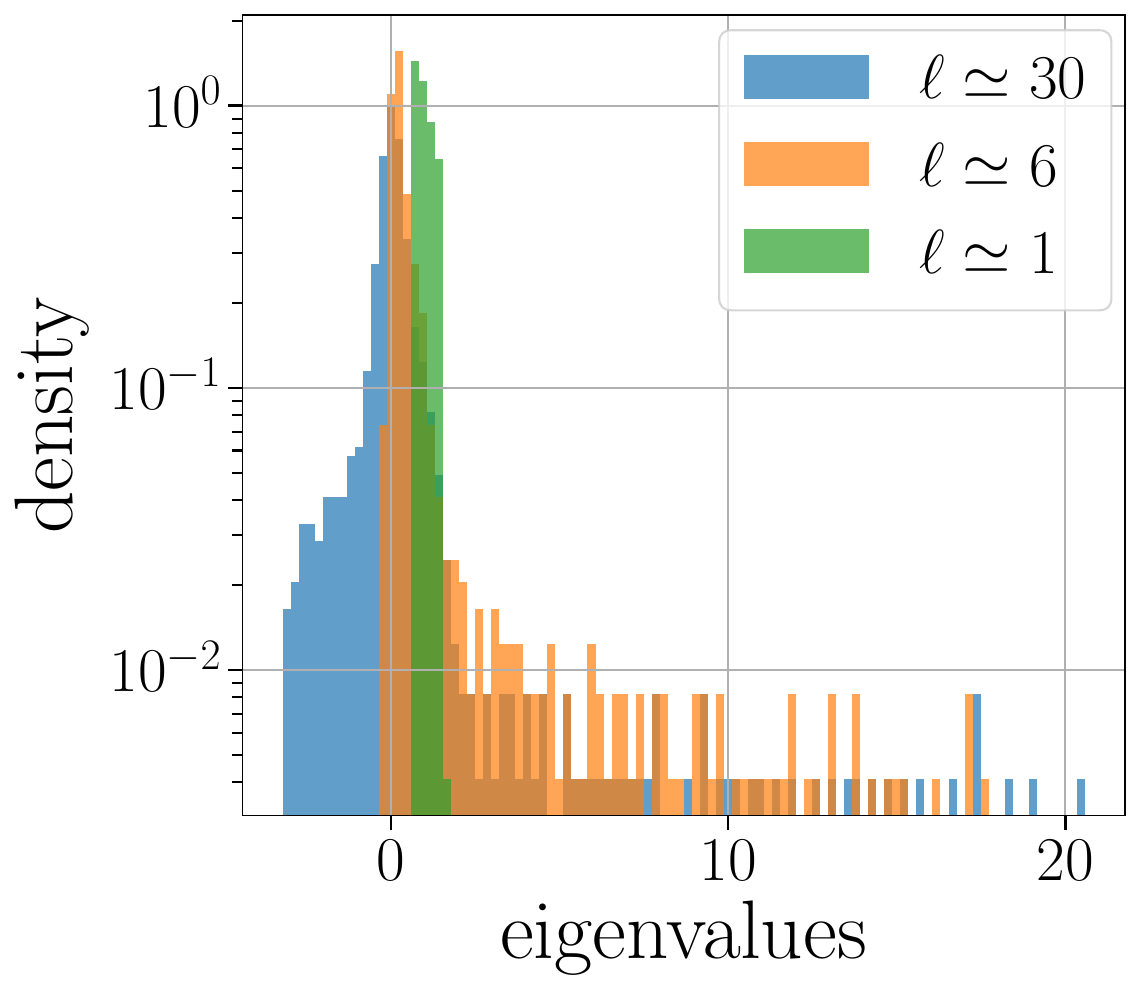}\hfill
  \includegraphics[width=0.33\textwidth]{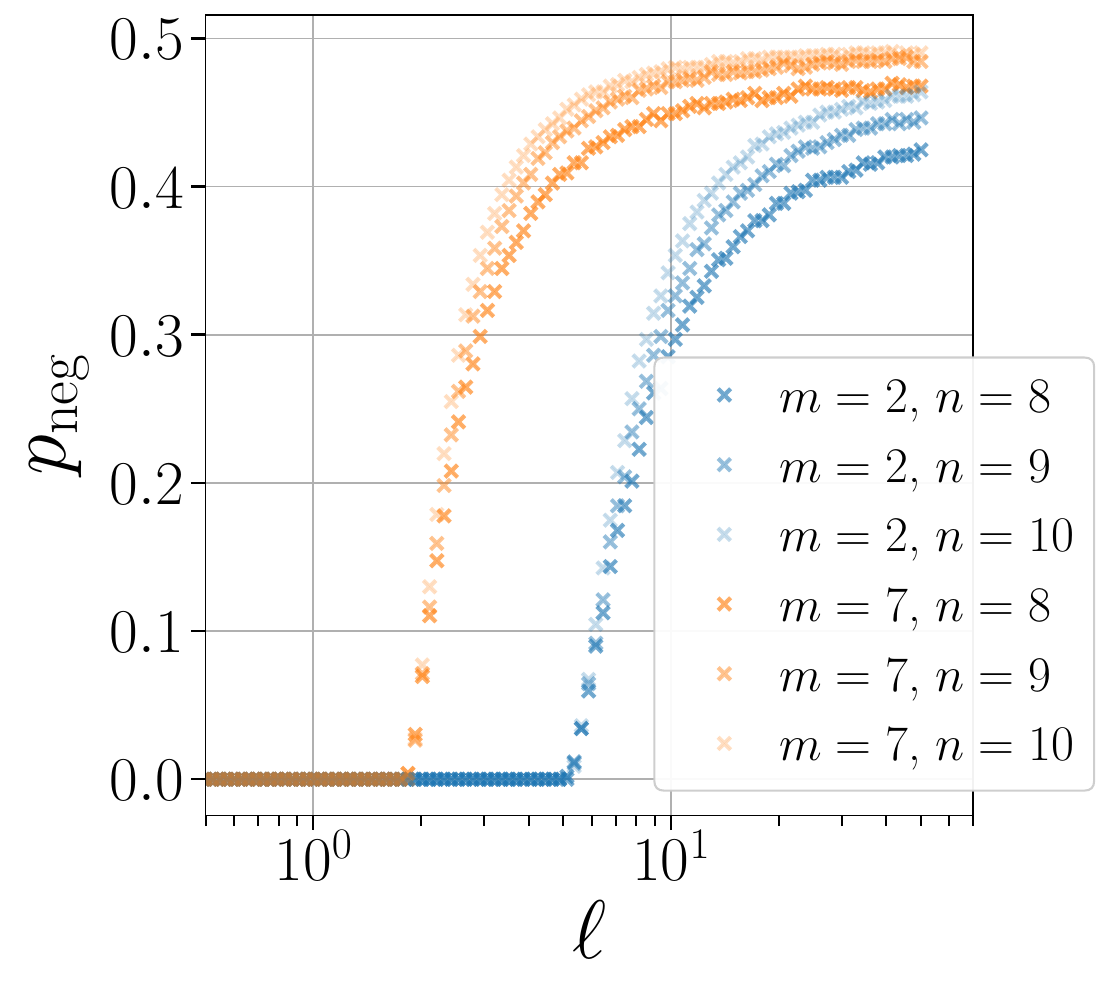}
  \caption{Emergence of negative eigenvalues of the matrix \eqref{eq:appendix-correl-RRG-C_E} on RRG with coordination number $m+1$ and number of vertices $N=2^n$.
  {\it Left panel:} Fraction of negative eigenvalues $p_{\rm neg}$ as a function of correlation length $\ell$ on RRG with $n=10$ and $m=2$, 3, 7, and 15. 
  {\it Middle panel:} Distribution of eigenvalues for $n=10$, $m=2$, and three values of correlation length (shown by vertical dashed lines in the left panel). For $\ell = 1$ there are no negative eigenvalues; for $\ell = 6$ many eigenvalues are negative though this is still a relatively small fraction of all eigenvalues; for $\ell = 30$ approximately a half of all eigenvalues are negative.
  {\it Right panel:} $p_{\rm neg}$  for $m=2$ and 7, and various values of system size: $n=8$, 9, and 10. It is seen that the threshold $\ell_*$ does not depend on $n$. 
}
  \label{fig:RRG-corr-p-neg}
\end{figure*}

One may be interested in generalizing the problem by considering a model with a multivariate Gaussian distribution $\mathcal{N}(0, C_E)$ of energies defined by a covariance matrix $C_E = f(r_{\alpha\beta})$ on a certain graph. Importantly, this distribution is defined only if the matrix $C_E$ is positive definite, i.e.,  all its eigenvalues are positive. In the case of uQD model, this condition is fulfilled, as the hypercube graph is the Fock space of the model of $n$ spins, and the covariance matrix $C_E^\text{QD}$ given by Eq.~(\ref{eq:correl_energies_final}) follows from Eq.~\eqref{eq:QD_on_site_energies}. Similarly, multivariate Gaussian distributions characterizing Fock-space representations of QD, 1D, and u1D models are well defined. However, in general, the positive definiteness of $C_E$ is by no means guaranteed if one picks up a certain structure of the graph and a certain correlation function $f(r_{\alpha\beta})$.

In view of connections to the RRG model, it may be tempting to define a model of the uQD type not on the hypercube but rather on RRG with $2^n$ nodes and coordination number $m+1$. This would require implementing a multivariate Gaussian energy distribution on RRG, with energy correlations similar to $C_E^\text{QD}$, i.e., given by $f(r_{\alpha\beta})$ analogous to Eq.~(\ref{eq:correl_energies_final}). 
If possible, this would provide an additional link between the uncorrelated RRG model (where $m$ and $n$ are independent parameters) and the uQD and u1D models for which we developed the RRG-like approximation with energy correlations, Sec.~\ref{sec:uQD-u1D-analytics}, and for which the coordination number is strictly bound to $m+1 = n$.

We show, however, in this Appendix that an attempt to implement energy correlations on RRG, with a correlation function of the type (\ref{eq:correl_energies_final}), i.e., with strong correlations extending up to the largest Hamming distance on the graph in general fails. Specifically, we show that, for a typical covariance function $f(r_{\alpha\beta})$ of this class, the requirement of positive definiteness of the matrix is strongly violated, with approximately half of its eigenvalues being negative. Only when correlations are strongly suppressed (e.g, by reduction of the correlation length down to a value $\sim 1$), one finds a positive definite covariance matrix and thus a well-defined model. 

We choose a correlation of the form  
\begin{align}
(C_E)_{\alpha\beta} = f(r_{\alpha\beta}) = 
C_0 \, e^{- r_{\alpha\beta}/\ell} \,,
\label{eq:appendix-correl-RRG-C_E}
\end{align}
as was proposed in Ref.~\cite{Roy2020numerics}. Here $\ell > 0$ is a correlation length; the limit $\ell \to 0$ corresponds to an uncorrelated model. The overall constant prefactor $C_0$ is of no importance for the discussion in the present Appendix. It is easy to check that, if considered on a hypercube (the Fock-space graph of the models studied in this paper), the covariance matrix 
\eqref{eq:appendix-correl-RRG-C_E} is positive definite for any $\ell$. (The corresponding eigenvalues can be straightforwardly calculated analytically
by diagonalizing the matrix \eqref{eq:appendix-correl-RRG-C_E} on a hypercube by Fourier transformation.) Taking in this case $\ell \sim n$, one gets strong correlations analogous to the case of uQD model, Eq.~(\ref{eq:correl_energies_final}). 
We show now numerically that the situation changes dramatically if one tries to define such a model on RRG, since the matrix \eqref{eq:appendix-correl-RRG-C_E} is then positive definite only for a short correlation length $\ell$.

In the left panel of Fig.~\ref{fig:RRG-corr-p-neg}, the fraction $p_{\rm neg}$ of negative eigenvalues of the matrix \eqref{eq:appendix-correl-RRG-C_E} on RRG is shown as a function of correlation length $\ell$, for the system size $N= 2^n$ with $n=10$ and for several values of the coordination number $m+1$. It is seen that
$p_{\rm neg}=0$ for small $\ell$, so that 
Eq.~\eqref{eq:appendix-correl-RRG-C_E} defines a valid covariance matrix. On the other hand, for $\ell > \ell_*$, a macroscopically large number of eigenvalues is negative, so that $C_E$ is not a valid covariance matrix any longer. Emergence of negative eigenvalues and increase of $p_{\rm neg}$ with increasing $\ell$ is illustrated in the histograms shown in the middle panel. We find that the threshold value of the correlation length is $\ell_* \simeq c_* / \ln m$ with $c_* \approx 3.56$.  As demonstrated in the right panel, at given $m$, the threshold $\ell_*$ (and thus the constant $c_*$) does not depend on $n$ (i.e., on the number of graph vertices $N=2^n$). Equivalently, one can define the ``correlation volume'' for $C_E$ via $N_\ell = m^\ell$. Once $N_\ell$ becomes larger than $N_{\ell_*} = e^{c_*} \approx 35$, negative eigenvalues arise and $C_E$ ceases to be a valid covariance matrix. 

Our results show, in particular, that one cannot define on RRG an ensemble with energies obeying Gaussian statistics with covariance 
\eqref{eq:appendix-correl-RRG-C_E}
and correlations extending over the whole system (i.e., with $\ell \sim L = \ln N / \ln m$). Indeed, such large $\ell$ is deeply in the regime where nearly half of all eigenvalues are negative. Our conclusion is in conflict with a numerical study in Ref.~\cite{Roy2020numerics}.

\bibliography{rrg}

\begin{thebibliography}{115}%
\makeatletter
\providecommand \@ifxundefined [1]{%
 \@ifx{#1\undefined}
}%
\providecommand \@ifnum [1]{%
 \ifnum #1\expandafter \@firstoftwo
 \else \expandafter \@secondoftwo
 \fi
}%
\providecommand \@ifx [1]{%
 \ifx #1\expandafter \@firstoftwo
 \else \expandafter \@secondoftwo
 \fi
}%
\providecommand \natexlab [1]{#1}%
\providecommand \enquote  [1]{``#1''}%
\providecommand \bibnamefont  [1]{#1}%
\providecommand \bibfnamefont [1]{#1}%
\providecommand \citenamefont [1]{#1}%
\providecommand \href@noop [0]{\@secondoftwo}%
\providecommand \href [0]{\begingroup \@sanitize@url \@href}%
\providecommand \@href[1]{\@@startlink{#1}\@@href}%
\providecommand \@@href[1]{\endgroup#1\@@endlink}%
\providecommand \@sanitize@url [0]{\catcode `\\12\catcode `\$12\catcode
  `\&12\catcode `\#12\catcode `\^12\catcode `\_12\catcode `\%12\relax}%
\providecommand \@@startlink[1]{}%
\providecommand \@@endlink[0]{}%
\providecommand \url  [0]{\begingroup\@sanitize@url \@url }%
\providecommand \@url [1]{\endgroup\@href {#1}{\urlprefix }}%
\providecommand \urlprefix  [0]{URL }%
\providecommand \Eprint [0]{\href }%
\providecommand \doibase [0]{https://doi.org/}%
\providecommand \selectlanguage [0]{\@gobble}%
\providecommand \bibinfo  [0]{\@secondoftwo}%
\providecommand \bibfield  [0]{\@secondoftwo}%
\providecommand \translation [1]{[#1]}%
\providecommand \BibitemOpen [0]{}%
\providecommand \bibitemStop [0]{}%
\providecommand \bibitemNoStop [0]{.\EOS\space}%
\providecommand \EOS [0]{\spacefactor3000\relax}%
\providecommand \BibitemShut  [1]{\csname bibitem#1\endcsname}%
\let\auto@bib@innerbib\@empty
\bibitem [{\citenamefont {Gornyi}\ \emph {et~al.}(2005)\citenamefont {Gornyi},
  \citenamefont {Mirlin},\ and\ \citenamefont
  {Polyakov}}]{gornyi2005interacting}%
  \BibitemOpen
  \bibfield  {author} {\bibinfo {author} {\bibfnamefont {I.}~\bibnamefont
  {Gornyi}}, \bibinfo {author} {\bibfnamefont {A.}~\bibnamefont {Mirlin}},\
  and\ \bibinfo {author} {\bibfnamefont {D.}~\bibnamefont {Polyakov}},\
  }\bibfield  {title} {\bibinfo {title} {Interacting electrons in disordered
  wires: Anderson localization and low-{T} transport},\ }\href@noop {}
  {\bibfield  {journal} {\bibinfo  {journal} {Physical Review Letters}\
  }\textbf {\bibinfo {volume} {95}},\ \bibinfo {pages} {206603} (\bibinfo
  {year} {2005})}\BibitemShut {NoStop}%
\bibitem [{\citenamefont {Basko}\ \emph {et~al.}(2006)\citenamefont {Basko},
  \citenamefont {Aleiner},\ and\ \citenamefont {Altshuler}}]{basko2006metal}%
  \BibitemOpen
  \bibfield  {author} {\bibinfo {author} {\bibfnamefont {D.}~\bibnamefont
  {Basko}}, \bibinfo {author} {\bibfnamefont {I.}~\bibnamefont {Aleiner}},\
  and\ \bibinfo {author} {\bibfnamefont {B.}~\bibnamefont {Altshuler}},\
  }\bibfield  {title} {\bibinfo {title} {Metal--insulator transition in a
  weakly interacting many-electron system with localized single-particle
  states},\ }\href@noop {} {\bibfield  {journal} {\bibinfo  {journal} {Annals
  of Physics}\ }\textbf {\bibinfo {volume} {321}},\ \bibinfo {pages} {1126}
  (\bibinfo {year} {2006})}\BibitemShut {NoStop}%
\bibitem [{\citenamefont {Anderson}(1958)}]{anderson58}%
  \BibitemOpen
  \bibfield  {author} {\bibinfo {author} {\bibfnamefont {P.~W.}\ \bibnamefont
  {Anderson}},\ }\bibfield  {title} {\bibinfo {title} {Absence of diffusion in
  certain random lattices},\ }\href@noop {} {\bibfield  {journal} {\bibinfo
  {journal} {Phys. Rev.}\ }\textbf {\bibinfo {volume} {109}},\ \bibinfo {pages}
  {1492} (\bibinfo {year} {1958})}\BibitemShut {NoStop}%
\bibitem [{\citenamefont {Evers}\ and\ \citenamefont {Mirlin}(2008)}]{evers08}%
  \BibitemOpen
  \bibfield  {author} {\bibinfo {author} {\bibfnamefont {F.}~\bibnamefont
  {Evers}}\ and\ \bibinfo {author} {\bibfnamefont {A.~D.}\ \bibnamefont
  {Mirlin}},\ }\bibfield  {title} {\bibinfo {title} {Anderson transitions},\
  }\href@noop {} {\bibfield  {journal} {\bibinfo  {journal} {Reviews of Modern
  Physics}\ }\textbf {\bibinfo {volume} {80}},\ \bibinfo {pages} {1355}
  (\bibinfo {year} {2008})}\BibitemShut {NoStop}%
\bibitem [{\citenamefont {Nandkishore}\ and\ \citenamefont
  {Huse}(2015)}]{nandkishore15}%
  \BibitemOpen
  \bibfield  {author} {\bibinfo {author} {\bibfnamefont {R.}~\bibnamefont
  {Nandkishore}}\ and\ \bibinfo {author} {\bibfnamefont {D.~A.}\ \bibnamefont
  {Huse}},\ }\bibfield  {title} {\bibinfo {title} {Many-body localization and
  thermalization in quantum statistical mechanics},\ }\href@noop {} {\bibfield
  {journal} {\bibinfo  {journal} {Annu. Rev. Condens. Matter Phys.}\ }\textbf
  {\bibinfo {volume} {6}},\ \bibinfo {pages} {15} (\bibinfo {year}
  {2015})}\BibitemShut {NoStop}%
\bibitem [{\citenamefont {Alet}\ and\ \citenamefont
  {Laflorencie}(2018)}]{Alet2018a}%
  \BibitemOpen
  \bibfield  {author} {\bibinfo {author} {\bibfnamefont {F.}~\bibnamefont
  {Alet}}\ and\ \bibinfo {author} {\bibfnamefont {N.}~\bibnamefont
  {Laflorencie}},\ }\bibfield  {title} {\bibinfo {title} {Many-body
  localization: An introduction and selected topics},\ }\href@noop {}
  {\bibfield  {journal} {\bibinfo  {journal} {Comptes Rendus Physique}\
  }\textbf {\bibinfo {volume} {19}},\ \bibinfo {pages} {498} (\bibinfo {year}
  {2018})}\BibitemShut {NoStop}%
\bibitem [{\citenamefont {Abanin}\ \emph {et~al.}(2019)\citenamefont {Abanin},
  \citenamefont {Altman}, \citenamefont {Bloch},\ and\ \citenamefont
  {Serbyn}}]{abanin2019colloquium}%
  \BibitemOpen
  \bibfield  {author} {\bibinfo {author} {\bibfnamefont {D.~A.}\ \bibnamefont
  {Abanin}}, \bibinfo {author} {\bibfnamefont {E.}~\bibnamefont {Altman}},
  \bibinfo {author} {\bibfnamefont {I.}~\bibnamefont {Bloch}},\ and\ \bibinfo
  {author} {\bibfnamefont {M.}~\bibnamefont {Serbyn}},\ }\bibfield  {title}
  {\bibinfo {title} {Colloquium: Many-body localization, thermalization, and
  entanglement},\ }\href@noop {} {\bibfield  {journal} {\bibinfo  {journal}
  {Reviews of Modern Physics}\ }\textbf {\bibinfo {volume} {91}},\ \bibinfo
  {pages} {021001} (\bibinfo {year} {2019})}\BibitemShut {NoStop}%
\bibitem [{\citenamefont {Gopalakrishnan}\ and\ \citenamefont
  {Parameswaran}(2020)}]{gopalakrishnan2020dynamics}%
  \BibitemOpen
  \bibfield  {author} {\bibinfo {author} {\bibfnamefont {S.}~\bibnamefont
  {Gopalakrishnan}}\ and\ \bibinfo {author} {\bibfnamefont {S.}~\bibnamefont
  {Parameswaran}},\ }\bibfield  {title} {\bibinfo {title} {Dynamics and
  transport at the threshold of many-body localization},\ }\href@noop {}
  {\bibfield  {journal} {\bibinfo  {journal} {Physics Reports}\ }\textbf
  {\bibinfo {volume} {862}},\ \bibinfo {pages} {1} (\bibinfo {year}
  {2020})}\BibitemShut {NoStop}%
\bibitem [{\citenamefont {Tikhonov}\ and\ \citenamefont
  {Mirlin}(2021{\natexlab{a}})}]{tikhonov2021from}%
  \BibitemOpen
  \bibfield  {author} {\bibinfo {author} {\bibfnamefont {K.~S.}\ \bibnamefont
  {Tikhonov}}\ and\ \bibinfo {author} {\bibfnamefont {A.~D.}\ \bibnamefont
  {Mirlin}},\ }\bibfield  {title} {\bibinfo {title} {From {Anderson}
  localization on random regular graphs to many-body localization},\
  }\href@noop {} {\bibfield  {journal} {\bibinfo  {journal} {Annals of
  Physics}\ }\textbf {\bibinfo {volume} {435}},\ \bibinfo {pages} {168525}
  (\bibinfo {year} {2021}{\natexlab{a}})}\BibitemShut {NoStop}%
\bibitem [{\citenamefont {Doggen}\ \emph {et~al.}(2021)\citenamefont {Doggen},
  \citenamefont {Gornyi}, \citenamefont {Mirlin},\ and\ \citenamefont
  {Polyakov}}]{doggen2021many}%
  \BibitemOpen
  \bibfield  {author} {\bibinfo {author} {\bibfnamefont {E.~V.}\ \bibnamefont
  {Doggen}}, \bibinfo {author} {\bibfnamefont {I.~V.}\ \bibnamefont {Gornyi}},
  \bibinfo {author} {\bibfnamefont {A.~D.}\ \bibnamefont {Mirlin}},\ and\
  \bibinfo {author} {\bibfnamefont {D.~G.}\ \bibnamefont {Polyakov}},\
  }\bibfield  {title} {\bibinfo {title} {Many-body localization in large
  systems: {M}atrix-product-state approach},\ }\href
  {https://doi.org/https://doi.org/10.1016/j.aop.2021.168437} {\bibfield
  {journal} {\bibinfo  {journal} {Annals of Physics}\ }\textbf {\bibinfo
  {volume} {435}},\ \bibinfo {pages} {168437} (\bibinfo {year} {2021})},\
  \bibinfo {note} {special Issue on Localisation 2020}\BibitemShut {NoStop}%
\bibitem [{\citenamefont {Sierant}\ \emph {et~al.}(2024)\citenamefont
  {Sierant}, \citenamefont {Lewenstein}, \citenamefont {Scardicchio},
  \citenamefont {Vidmar},\ and\ \citenamefont {Zakrzewski}}]{Sierant2024}%
  \BibitemOpen
  \bibfield  {author} {\bibinfo {author} {\bibfnamefont {P.}~\bibnamefont
  {Sierant}}, \bibinfo {author} {\bibfnamefont {M.}~\bibnamefont {Lewenstein}},
  \bibinfo {author} {\bibfnamefont {A.}~\bibnamefont {Scardicchio}}, \bibinfo
  {author} {\bibfnamefont {L.}~\bibnamefont {Vidmar}},\ and\ \bibinfo {author}
  {\bibfnamefont {J.}~\bibnamefont {Zakrzewski}},\ }\href@noop {} {\bibinfo
  {title} {Many-body localization in the age of classical computing}} (\bibinfo
  {year} {2024}),\ \Eprint {https://arxiv.org/abs/2403.07111} {arXiv:2403.07111
  [cond-mat.dis-nn]} \BibitemShut {NoStop}%
\bibitem [{\citenamefont {Monteiro}\ \emph {et~al.}(2021)\citenamefont
  {Monteiro}, \citenamefont {Tezuka}, \citenamefont {Altland}, \citenamefont
  {Huse},\ and\ \citenamefont {Micklitz}}]{Monteiro2021}%
  \BibitemOpen
  \bibfield  {author} {\bibinfo {author} {\bibfnamefont {F.}~\bibnamefont
  {Monteiro}}, \bibinfo {author} {\bibfnamefont {M.}~\bibnamefont {Tezuka}},
  \bibinfo {author} {\bibfnamefont {A.}~\bibnamefont {Altland}}, \bibinfo
  {author} {\bibfnamefont {D.~A.}\ \bibnamefont {Huse}},\ and\ \bibinfo
  {author} {\bibfnamefont {T.}~\bibnamefont {Micklitz}},\ }\bibfield  {title}
  {\bibinfo {title} {Quantum ergodicity in the many-body localization
  problem},\ }\href {https://doi.org/10.1103/PhysRevLett.127.030601} {\bibfield
   {journal} {\bibinfo  {journal} {Phys. Rev. Lett.}\ }\textbf {\bibinfo
  {volume} {127}},\ \bibinfo {pages} {030601} (\bibinfo {year}
  {2021})}\BibitemShut {NoStop}%
\bibitem [{Note1()}]{Note1}%
  \BibitemOpen
  \bibinfo {note} {{\protect \color {black} It is worth noting that there are
  different definitions of ergodicity in various fields of classical and
  quantum physics. In particular, the definition of quantum (non-)ergodicity in
  the context of MBL (see a recent discussion in Ref.~\cite {Monteiro2021})
  differs from that traditionally adopted in the field of spin glasses, where a
  system can be called non-ergodic once full equilibration takes a time
  exponentially long in system size $n$. Indeed, this time can still be much
  shorter than the Heisenberg time (also exponential in $n$), in which case the
  system will satisfy the condition of quantum ergodicity that we adopt. What
  we call ``ergodicity'' reflects the full mixture of many-body states within
  an energy shell containing many states (cf. microcanonical ensemble of
  statistical mechanics and the eigenstate thermalization hypothesis). For a
  recent review of connections between quantum spin glasses and MBL, see
  Ref.~\cite {Cugliandolo}}}\BibitemShut {NoStop}%
\bibitem [{\citenamefont {Altshuler}\ \emph {et~al.}(1997)\citenamefont
  {Altshuler}, \citenamefont {Gefen}, \citenamefont {Kamenev},\ and\
  \citenamefont {Levitov}}]{altshuler1997quasiparticle}%
  \BibitemOpen
  \bibfield  {author} {\bibinfo {author} {\bibfnamefont {B.~L.}\ \bibnamefont
  {Altshuler}}, \bibinfo {author} {\bibfnamefont {Y.}~\bibnamefont {Gefen}},
  \bibinfo {author} {\bibfnamefont {A.}~\bibnamefont {Kamenev}},\ and\ \bibinfo
  {author} {\bibfnamefont {L.~S.}\ \bibnamefont {Levitov}},\ }\bibfield
  {title} {\bibinfo {title} {Quasiparticle lifetime in a finite system: A
  nonperturbative approach},\ }\href@noop {} {\bibfield  {journal} {\bibinfo
  {journal} {Physical Review Letters}\ }\textbf {\bibinfo {volume} {78}},\
  \bibinfo {pages} {2803} (\bibinfo {year} {1997})}\BibitemShut {NoStop}%
\bibitem [{\citenamefont {Jacquod}\ and\ \citenamefont
  {Shepelyansky}(1997)}]{jacquod1997emergence}%
  \BibitemOpen
  \bibfield  {author} {\bibinfo {author} {\bibfnamefont {P.}~\bibnamefont
  {Jacquod}}\ and\ \bibinfo {author} {\bibfnamefont {D.}~\bibnamefont
  {Shepelyansky}},\ }\bibfield  {title} {\bibinfo {title} {Emergence of quantum
  chaos in finite interacting {F}ermi systems},\ }\href@noop {} {\bibfield
  {journal} {\bibinfo  {journal} {Physical Review Letters}\ }\textbf {\bibinfo
  {volume} {79}},\ \bibinfo {pages} {1837} (\bibinfo {year}
  {1997})}\BibitemShut {NoStop}%
\bibitem [{\citenamefont {Mirlin}\ and\ \citenamefont
  {Fyodorov}(1997)}]{mirlin1997localization}%
  \BibitemOpen
  \bibfield  {author} {\bibinfo {author} {\bibfnamefont {A.~D.}\ \bibnamefont
  {Mirlin}}\ and\ \bibinfo {author} {\bibfnamefont {Y.~V.}\ \bibnamefont
  {Fyodorov}},\ }\bibfield  {title} {\bibinfo {title} {Localization and
  fluctuations of local spectral density on treelike structures with large
  connectivity: Application to the quasiparticle line shape in quantum dots},\
  }\href@noop {} {\bibfield  {journal} {\bibinfo  {journal} {Phys. Rev. B}\
  }\textbf {\bibinfo {volume} {56}},\ \bibinfo {pages} {13393} (\bibinfo {year}
  {1997})}\BibitemShut {NoStop}%
\bibitem [{\citenamefont {Silvestrov}(1997)}]{silvestrov1997decay}%
  \BibitemOpen
  \bibfield  {author} {\bibinfo {author} {\bibfnamefont {P.}~\bibnamefont
  {Silvestrov}},\ }\bibfield  {title} {\bibinfo {title} {Decay of a
  quasiparticle in a quantum dot: The role of energy resolution},\ }\href@noop
  {} {\bibfield  {journal} {\bibinfo  {journal} {Physical Review Letters}\
  }\textbf {\bibinfo {volume} {79}},\ \bibinfo {pages} {3994} (\bibinfo {year}
  {1997})}\BibitemShut {NoStop}%
\bibitem [{\citenamefont {Silvestrov}(1998)}]{silvestrov1998chaos}%
  \BibitemOpen
  \bibfield  {author} {\bibinfo {author} {\bibfnamefont {P.}~\bibnamefont
  {Silvestrov}},\ }\bibfield  {title} {\bibinfo {title} {Chaos thresholds in
  finite {F}ermi systems},\ }\href@noop {} {\bibfield  {journal} {\bibinfo
  {journal} {Physical Review E}\ }\textbf {\bibinfo {volume} {58}},\ \bibinfo
  {pages} {5629} (\bibinfo {year} {1998})}\BibitemShut {NoStop}%
\bibitem [{\citenamefont {Gornyi}\ \emph {et~al.}(2016)\citenamefont {Gornyi},
  \citenamefont {Mirlin},\ and\ \citenamefont {Polyakov}}]{gornyi2016many}%
  \BibitemOpen
  \bibfield  {author} {\bibinfo {author} {\bibfnamefont {I.}~\bibnamefont
  {Gornyi}}, \bibinfo {author} {\bibfnamefont {A.}~\bibnamefont {Mirlin}},\
  and\ \bibinfo {author} {\bibfnamefont {D.}~\bibnamefont {Polyakov}},\
  }\bibfield  {title} {\bibinfo {title} {Many-body delocalization transition
  and relaxation in a quantum dot},\ }\href@noop {} {\bibfield  {journal}
  {\bibinfo  {journal} {Physical Review B}\ }\textbf {\bibinfo {volume} {93}},\
  \bibinfo {pages} {125419} (\bibinfo {year} {2016})}\BibitemShut {NoStop}%
\bibitem [{\citenamefont {Gornyi}\ \emph {et~al.}(2017)\citenamefont {Gornyi},
  \citenamefont {Mirlin}, \citenamefont {Polyakov},\ and\ \citenamefont
  {Burin}}]{gornyi2017spectral}%
  \BibitemOpen
  \bibfield  {author} {\bibinfo {author} {\bibfnamefont {I.}~\bibnamefont
  {Gornyi}}, \bibinfo {author} {\bibfnamefont {A.}~\bibnamefont {Mirlin}},
  \bibinfo {author} {\bibfnamefont {D.}~\bibnamefont {Polyakov}},\ and\
  \bibinfo {author} {\bibfnamefont {A.}~\bibnamefont {Burin}},\ }\bibfield
  {title} {\bibinfo {title} {Spectral diffusion and scaling of many-body
  delocalization transitions},\ }\href@noop {} {\bibfield  {journal} {\bibinfo
  {journal} {Annalen der Physik}\ }\textbf {\bibinfo {volume} {529}},\ \bibinfo
  {pages} {1600360} (\bibinfo {year} {2017})}\BibitemShut {NoStop}%
\bibitem [{\citenamefont {Georgeot}\ and\ \citenamefont
  {Shepelyansky}(1997)}]{georgeot1997breit}%
  \BibitemOpen
  \bibfield  {author} {\bibinfo {author} {\bibfnamefont {B.}~\bibnamefont
  {Georgeot}}\ and\ \bibinfo {author} {\bibfnamefont {D.~L.}\ \bibnamefont
  {Shepelyansky}},\ }\bibfield  {title} {\bibinfo {title} {Breit-{W}igner width
  and inverse participation ratio in finite interacting {F}ermi systems},\
  }\href@noop {} {\bibfield  {journal} {\bibinfo  {journal} {Physical Review
  Letters}\ }\textbf {\bibinfo {volume} {79}},\ \bibinfo {pages} {4365}
  (\bibinfo {year} {1997})}\BibitemShut {NoStop}%
\bibitem [{\citenamefont {Leyronas}\ \emph {et~al.}(2000)\citenamefont
  {Leyronas}, \citenamefont {Silvestrov},\ and\ \citenamefont
  {Beenakker}}]{leyronas2000scaling}%
  \BibitemOpen
  \bibfield  {author} {\bibinfo {author} {\bibfnamefont {X.}~\bibnamefont
  {Leyronas}}, \bibinfo {author} {\bibfnamefont {P.}~\bibnamefont
  {Silvestrov}},\ and\ \bibinfo {author} {\bibfnamefont {C.}~\bibnamefont
  {Beenakker}},\ }\bibfield  {title} {\bibinfo {title} {Scaling at the chaos
  threshold for interacting electrons in a quantum dot},\ }\href@noop {}
  {\bibfield  {journal} {\bibinfo  {journal} {Physical Review Letters}\
  }\textbf {\bibinfo {volume} {84}},\ \bibinfo {pages} {3414} (\bibinfo {year}
  {2000})}\BibitemShut {NoStop}%
\bibitem [{\citenamefont {Shepelyansky}(2001)}]{shepelyansky2001quantum}%
  \BibitemOpen
  \bibfield  {author} {\bibinfo {author} {\bibfnamefont {D.}~\bibnamefont
  {Shepelyansky}},\ }\bibfield  {title} {\bibinfo {title} {Quantum chaos and
  quantum computers},\ }\href@noop {} {\bibfield  {journal} {\bibinfo
  {journal} {Physica Scripta}\ }\textbf {\bibinfo {volume} {2001}},\ \bibinfo
  {pages} {112} (\bibinfo {year} {2001})}\BibitemShut {NoStop}%
\bibitem [{\citenamefont {Song}(2000)}]{PhysRevE.62.R7575}%
  \BibitemOpen
  \bibfield  {author} {\bibinfo {author} {\bibfnamefont {P.~H.}\ \bibnamefont
  {Song}},\ }\bibfield  {title} {\bibinfo {title} {Scaling near the quantum
  chaos border in interacting {F}ermi systems},\ }\href@noop {} {\bibfield
  {journal} {\bibinfo  {journal} {Phys. Rev. E}\ }\textbf {\bibinfo {volume}
  {62}},\ \bibinfo {pages} {R7575} (\bibinfo {year} {2000})}\BibitemShut
  {NoStop}%
\bibitem [{\citenamefont {Jacquod}\ and\ \citenamefont
  {Varga}(2001)}]{jacquod2001duality}%
  \BibitemOpen
  \bibfield  {author} {\bibinfo {author} {\bibfnamefont {P.}~\bibnamefont
  {Jacquod}}\ and\ \bibinfo {author} {\bibfnamefont {I.}~\bibnamefont
  {Varga}},\ }\bibfield  {title} {\bibinfo {title} {Duality between the weak
  and strong interaction limits of deformed fermionic two-body random
  ensembles},\ }\href@noop {} {\bibfield  {journal} {\bibinfo  {journal} {Phys.
  Rev. Lett.}\ }\textbf {\bibinfo {volume} {89}},\ \bibinfo {pages} {134101}
  (\bibinfo {year} {2001})}\BibitemShut {NoStop}%
\bibitem [{\citenamefont {Rivas}\ \emph {et~al.}(2002)\citenamefont {Rivas},
  \citenamefont {Mucciolo},\ and\ \citenamefont
  {Kamenev}}]{rivas2002numerical}%
  \BibitemOpen
  \bibfield  {author} {\bibinfo {author} {\bibfnamefont {A.~M.}\ \bibnamefont
  {Rivas}}, \bibinfo {author} {\bibfnamefont {E.~R.}\ \bibnamefont
  {Mucciolo}},\ and\ \bibinfo {author} {\bibfnamefont {A.}~\bibnamefont
  {Kamenev}},\ }\bibfield  {title} {\bibinfo {title} {Numerical study of
  quasiparticle lifetime in quantum dots},\ }\href@noop {} {\bibfield
  {journal} {\bibinfo  {journal} {Physical Review B}\ }\textbf {\bibinfo
  {volume} {65}},\ \bibinfo {pages} {155309} (\bibinfo {year}
  {2002})}\BibitemShut {NoStop}%
\bibitem [{\citenamefont {Bulchandani}\ \emph {et~al.}(2022)\citenamefont
  {Bulchandani}, \citenamefont {Huse},\ and\ \citenamefont
  {Gopalakrishnan}}]{bulchandani2022onset}%
  \BibitemOpen
  \bibfield  {author} {\bibinfo {author} {\bibfnamefont {V.~B.}\ \bibnamefont
  {Bulchandani}}, \bibinfo {author} {\bibfnamefont {D.~A.}\ \bibnamefont
  {Huse}},\ and\ \bibinfo {author} {\bibfnamefont {S.}~\bibnamefont
  {Gopalakrishnan}},\ }\bibfield  {title} {\bibinfo {title} {Onset of many-body
  quantum chaos due to breaking integrability},\ }\href@noop {} {\bibfield
  {journal} {\bibinfo  {journal} {Phys. Rev. B}\ }\textbf {\bibinfo {volume}
  {105}},\ \bibinfo {pages} {214308} (\bibinfo {year} {2022})}\BibitemShut
  {NoStop}%
\bibitem [{\citenamefont {Garc\'{\i}a-Garc\'{\i}a}\ \emph
  {et~al.}(2018)\citenamefont {Garc\'{\i}a-Garc\'{\i}a}, \citenamefont
  {Loureiro}, \citenamefont {Romero-Berm\'udez},\ and\ \citenamefont
  {Tezuka}}]{garcia-garcia2018chaotic}%
  \BibitemOpen
  \bibfield  {author} {\bibinfo {author} {\bibfnamefont {A.~M.}\ \bibnamefont
  {Garc\'{\i}a-Garc\'{\i}a}}, \bibinfo {author} {\bibfnamefont
  {B.}~\bibnamefont {Loureiro}}, \bibinfo {author} {\bibfnamefont
  {A.}~\bibnamefont {Romero-Berm\'udez}},\ and\ \bibinfo {author}
  {\bibfnamefont {M.}~\bibnamefont {Tezuka}},\ }\bibfield  {title} {\bibinfo
  {title} {Chaotic-integrable transition in the {Sachdev-Ye-Kitaev} model},\
  }\href {https://doi.org/10.1103/PhysRevLett.120.241603} {\bibfield  {journal}
  {\bibinfo  {journal} {Phys. Rev. Lett.}\ }\textbf {\bibinfo {volume} {120}},\
  \bibinfo {pages} {241603} (\bibinfo {year} {2018})}\BibitemShut {NoStop}%
\bibitem [{\citenamefont {Micklitz}\ \emph {et~al.}(2019)\citenamefont
  {Micklitz}, \citenamefont {Monteiro},\ and\ \citenamefont
  {Altland}}]{micklitz2019nonergodic}%
  \BibitemOpen
  \bibfield  {author} {\bibinfo {author} {\bibfnamefont {T.}~\bibnamefont
  {Micklitz}}, \bibinfo {author} {\bibfnamefont {F.}~\bibnamefont {Monteiro}},\
  and\ \bibinfo {author} {\bibfnamefont {A.}~\bibnamefont {Altland}},\
  }\bibfield  {title} {\bibinfo {title} {Nonergodic extended states in the
  {S}achdev-{Y}e-{K}itaev model},\ }\href@noop {} {\bibfield  {journal}
  {\bibinfo  {journal} {Physical Review Letters}\ }\textbf {\bibinfo {volume}
  {123}},\ \bibinfo {pages} {125701} (\bibinfo {year} {2019})}\BibitemShut
  {NoStop}%
\bibitem [{\citenamefont {Monteiro}\ \emph {et~al.}(2020)\citenamefont
  {Monteiro}, \citenamefont {Micklitz}, \citenamefont {Tezuka},\ and\
  \citenamefont {Altland}}]{monteiro2020minimal}%
  \BibitemOpen
  \bibfield  {author} {\bibinfo {author} {\bibfnamefont {F.}~\bibnamefont
  {Monteiro}}, \bibinfo {author} {\bibfnamefont {T.}~\bibnamefont {Micklitz}},
  \bibinfo {author} {\bibfnamefont {M.}~\bibnamefont {Tezuka}},\ and\ \bibinfo
  {author} {\bibfnamefont {A.}~\bibnamefont {Altland}},\ }\bibfield  {title}
  {\bibinfo {title} {Minimal model of many-body localization},\ }\href@noop {}
  {\bibfield  {journal} {\bibinfo  {journal} {Physical Review Research}\
  }\textbf {\bibinfo {volume} {3}},\ \bibinfo {pages} {013023} (\bibinfo {year}
  {2020})}\BibitemShut {NoStop}%
\bibitem [{\citenamefont {Nandy}\ \emph {et~al.}(2022)\citenamefont {Nandy},
  \citenamefont {\ifmmode \check{C}\else \v{C}\fi{}ade\ifmmode~\check{z}\else
  \v{z}\fi{}}, \citenamefont {Dietz}, \citenamefont {Andreanov},\ and\
  \citenamefont {Rosa}}]{nandy2022delayed}%
  \BibitemOpen
  \bibfield  {author} {\bibinfo {author} {\bibfnamefont {D.~K.}\ \bibnamefont
  {Nandy}}, \bibinfo {author} {\bibfnamefont {T.}~\bibnamefont {\ifmmode
  \check{C}\else \v{C}\fi{}ade\ifmmode~\check{z}\else \v{z}\fi{}}}, \bibinfo
  {author} {\bibfnamefont {B.}~\bibnamefont {Dietz}}, \bibinfo {author}
  {\bibfnamefont {A.}~\bibnamefont {Andreanov}},\ and\ \bibinfo {author}
  {\bibfnamefont {D.}~\bibnamefont {Rosa}},\ }\bibfield  {title} {\bibinfo
  {title} {Delayed thermalization in the mass-deformed {Sachdev-Ye-Kitaev}
  model},\ }\href {https://doi.org/10.1103/PhysRevB.106.245147} {\bibfield
  {journal} {\bibinfo  {journal} {Phys. Rev. B}\ }\textbf {\bibinfo {volume}
  {106}},\ \bibinfo {pages} {245147} (\bibinfo {year} {2022})}\BibitemShut
  {NoStop}%
\bibitem [{\citenamefont {Larzul}\ and\ \citenamefont
  {Schir\'o}(2022)}]{larzul2022quenches}%
  \BibitemOpen
  \bibfield  {author} {\bibinfo {author} {\bibfnamefont {A.}~\bibnamefont
  {Larzul}}\ and\ \bibinfo {author} {\bibfnamefont {M.}~\bibnamefont
  {Schir\'o}},\ }\bibfield  {title} {\bibinfo {title} {Quenches and
  (pre)thermalization in a mixed {Sachdev-Ye-Kitaev} model},\ }\href
  {https://doi.org/10.1103/PhysRevB.105.045105} {\bibfield  {journal} {\bibinfo
   {journal} {Phys. Rev. B}\ }\textbf {\bibinfo {volume} {105}},\ \bibinfo
  {pages} {045105} (\bibinfo {year} {2022})}\BibitemShut {NoStop}%
\bibitem [{\citenamefont {Herre}\ \emph {et~al.}(2023)\citenamefont {Herre},
  \citenamefont {Karcher}, \citenamefont {Tikhonov},\ and\ \citenamefont
  {Mirlin}}]{Herre2023}%
  \BibitemOpen
  \bibfield  {author} {\bibinfo {author} {\bibfnamefont {J.-N.}\ \bibnamefont
  {Herre}}, \bibinfo {author} {\bibfnamefont {J.~F.}\ \bibnamefont {Karcher}},
  \bibinfo {author} {\bibfnamefont {K.~S.}\ \bibnamefont {Tikhonov}},\ and\
  \bibinfo {author} {\bibfnamefont {A.~D.}\ \bibnamefont {Mirlin}},\ }\bibfield
   {title} {\bibinfo {title} {Ergodicity-to-localization transition on random
  regular graphs with large connectivity and in many-body quantum dots},\
  }\href {https://doi.org/10.1103/PhysRevB.108.014203} {\bibfield  {journal}
  {\bibinfo  {journal} {Phys. Rev. B}\ }\textbf {\bibinfo {volume} {108}},\
  \bibinfo {pages} {014203} (\bibinfo {year} {2023})}\BibitemShut {NoStop}%
\bibitem [{\citenamefont {Serbyn}\ \emph {et~al.}(2017)\citenamefont {Serbyn},
  \citenamefont {Papi{\'c}},\ and\ \citenamefont
  {Abanin}}]{serbyn2017thouless}%
  \BibitemOpen
  \bibfield  {author} {\bibinfo {author} {\bibfnamefont {M.}~\bibnamefont
  {Serbyn}}, \bibinfo {author} {\bibfnamefont {Z.}~\bibnamefont {Papi{\'c}}},\
  and\ \bibinfo {author} {\bibfnamefont {D.~A.}\ \bibnamefont {Abanin}},\
  }\bibfield  {title} {\bibinfo {title} {Thouless energy and multifractality
  across the many-body localization transition},\ }\href@noop {} {\bibfield
  {journal} {\bibinfo  {journal} {Physical Review B}\ }\textbf {\bibinfo
  {volume} {96}},\ \bibinfo {pages} {104201} (\bibinfo {year}
  {2017})}\BibitemShut {NoStop}%
\bibitem [{\citenamefont {Tikhonov}\ and\ \citenamefont
  {Mirlin}(2018)}]{tikhonov18}%
  \BibitemOpen
  \bibfield  {author} {\bibinfo {author} {\bibfnamefont {K.~S.}\ \bibnamefont
  {Tikhonov}}\ and\ \bibinfo {author} {\bibfnamefont {A.~D.}\ \bibnamefont
  {Mirlin}},\ }\bibfield  {title} {\bibinfo {title} {Many-body localization
  transition with power-law interactions: Statistics of eigenstates},\
  }\href@noop {} {\bibfield  {journal} {\bibinfo  {journal} {Phys. Rev. B}\
  }\textbf {\bibinfo {volume} {97}},\ \bibinfo {pages} {214205} (\bibinfo
  {year} {2018})}\BibitemShut {NoStop}%
\bibitem [{\citenamefont {Mac\'e}\ \emph {et~al.}(2019)\citenamefont {Mac\'e},
  \citenamefont {Alet},\ and\ \citenamefont
  {Laflorencie}}]{mace19multifractal}%
  \BibitemOpen
  \bibfield  {author} {\bibinfo {author} {\bibfnamefont {N.}~\bibnamefont
  {Mac\'e}}, \bibinfo {author} {\bibfnamefont {F.}~\bibnamefont {Alet}},\ and\
  \bibinfo {author} {\bibfnamefont {N.}~\bibnamefont {Laflorencie}},\
  }\bibfield  {title} {\bibinfo {title} {Multifractal scalings across the
  many-body localization transition},\ }\href@noop {} {\bibfield  {journal}
  {\bibinfo  {journal} {Phys. Rev. Lett.}\ }\textbf {\bibinfo {volume} {123}},\
  \bibinfo {pages} {180601} (\bibinfo {year} {2019})}\BibitemShut {NoStop}%
\bibitem [{\citenamefont {Tikhonov}\ and\ \citenamefont
  {Mirlin}(2021{\natexlab{b}})}]{tikhonov2021eigenstate}%
  \BibitemOpen
  \bibfield  {author} {\bibinfo {author} {\bibfnamefont {K.~S.}\ \bibnamefont
  {Tikhonov}}\ and\ \bibinfo {author} {\bibfnamefont {A.~D.}\ \bibnamefont
  {Mirlin}},\ }\bibfield  {title} {\bibinfo {title} {Eigenstate correlations
  around the many-body localization transition},\ }\href
  {https://doi.org/10.1103/PhysRevB.103.064204} {\bibfield  {journal} {\bibinfo
   {journal} {Phys. Rev. B}\ }\textbf {\bibinfo {volume} {103}},\ \bibinfo
  {pages} {064204} (\bibinfo {year} {2021}{\natexlab{b}})}\BibitemShut
  {NoStop}%
\bibitem [{\citenamefont {Nag}\ and\ \citenamefont
  {Garg}(2019)}]{nag2019many-body}%
  \BibitemOpen
  \bibfield  {author} {\bibinfo {author} {\bibfnamefont {S.}~\bibnamefont
  {Nag}}\ and\ \bibinfo {author} {\bibfnamefont {A.}~\bibnamefont {Garg}},\
  }\bibfield  {title} {\bibinfo {title} {Many-body localization in the presence
  of long-range interactions and long-range hopping},\ }\href
  {https://doi.org/10.1103/PhysRevB.99.224203} {\bibfield  {journal} {\bibinfo
  {journal} {Phys. Rev. B}\ }\textbf {\bibinfo {volume} {99}},\ \bibinfo
  {pages} {224203} (\bibinfo {year} {2019})}\BibitemShut {NoStop}%
\bibitem [{\citenamefont {Tarzia}(2020)}]{tarzia2020many}%
  \BibitemOpen
  \bibfield  {author} {\bibinfo {author} {\bibfnamefont {M.}~\bibnamefont
  {Tarzia}},\ }\bibfield  {title} {\bibinfo {title} {Many-body localization
  transition in {Hilbert} space},\ }\href@noop {} {\bibfield  {journal}
  {\bibinfo  {journal} {Physical Review B}\ }\textbf {\bibinfo {volume}
  {102}},\ \bibinfo {pages} {014208} (\bibinfo {year} {2020})}\BibitemShut
  {NoStop}%
\bibitem [{\citenamefont {Roy}\ and\ \citenamefont
  {Logan}(2021)}]{roy2021fock}%
  \BibitemOpen
  \bibfield  {author} {\bibinfo {author} {\bibfnamefont {S.}~\bibnamefont
  {Roy}}\ and\ \bibinfo {author} {\bibfnamefont {D.~E.}\ \bibnamefont
  {Logan}},\ }\bibfield  {title} {\bibinfo {title} {Fock-space anatomy of
  eigenstates across the many-body localization transition},\ }\href
  {https://doi.org/10.1103/PhysRevB.104.174201} {\bibfield  {journal} {\bibinfo
   {journal} {Phys. Rev. B}\ }\textbf {\bibinfo {volume} {104}},\ \bibinfo
  {pages} {174201} (\bibinfo {year} {2021})}\BibitemShut {NoStop}%
\bibitem [{\citenamefont {Crowley}\ and\ \citenamefont
  {Chandran}(2022)}]{crowley2022constructive}%
  \BibitemOpen
  \bibfield  {author} {\bibinfo {author} {\bibfnamefont {P.~J.~D.}\
  \bibnamefont {Crowley}}\ and\ \bibinfo {author} {\bibfnamefont
  {A.}~\bibnamefont {Chandran}},\ }\bibfield  {title} {\bibinfo {title} {{A
  constructive theory of the numerically accessible many-body localized to
  thermal crossover}},\ }\href {https://doi.org/10.21468/SciPostPhys.12.6.201}
  {\bibfield  {journal} {\bibinfo  {journal} {SciPost Phys.}\ }\textbf
  {\bibinfo {volume} {12}},\ \bibinfo {pages} {201} (\bibinfo {year}
  {2022})}\BibitemShut {NoStop}%
\bibitem [{\citenamefont {Creed}\ \emph {et~al.}(2023)\citenamefont {Creed},
  \citenamefont {Logan},\ and\ \citenamefont {Roy}}]{creed2023probability}%
  \BibitemOpen
  \bibfield  {author} {\bibinfo {author} {\bibfnamefont {I.}~\bibnamefont
  {Creed}}, \bibinfo {author} {\bibfnamefont {D.~E.}\ \bibnamefont {Logan}},\
  and\ \bibinfo {author} {\bibfnamefont {S.}~\bibnamefont {Roy}},\ }\bibfield
  {title} {\bibinfo {title} {Probability transport on the {Fock} space of a
  disordered quantum spin chain},\ }\href
  {https://doi.org/10.1103/PhysRevB.107.094206} {\bibfield  {journal} {\bibinfo
   {journal} {Phys. Rev. B}\ }\textbf {\bibinfo {volume} {107}},\ \bibinfo
  {pages} {094206} (\bibinfo {year} {2023})}\BibitemShut {NoStop}%
\bibitem [{\citenamefont {Ghosh}\ \emph {et~al.}(2024)\citenamefont {Ghosh},
  \citenamefont {Sutradhar}, \citenamefont {Mukerjee},\ and\ \citenamefont
  {Banerjee}}]{Ghosh2024}%
  \BibitemOpen
  \bibfield  {author} {\bibinfo {author} {\bibfnamefont {S.}~\bibnamefont
  {Ghosh}}, \bibinfo {author} {\bibfnamefont {J.}~\bibnamefont {Sutradhar}},
  \bibinfo {author} {\bibfnamefont {S.}~\bibnamefont {Mukerjee}},\ and\
  \bibinfo {author} {\bibfnamefont {S.}~\bibnamefont {Banerjee}},\ }\href@noop
  {} {\bibinfo {title} {Scaling of {Fock} space propagator in quasiperiodic
  many-body localizing systems}} (\bibinfo {year} {2024}),\ \Eprint
  {https://arxiv.org/abs/2401.03027} {arXiv:2401.03027 [cond-mat.dis-nn]}
  \BibitemShut {NoStop}%
\bibitem [{\citenamefont {Sutradhar}\ \emph {et~al.}(2022)\citenamefont
  {Sutradhar}, \citenamefont {Ghosh}, \citenamefont {Roy}, \citenamefont
  {Logan}, \citenamefont {Mukerjee},\ and\ \citenamefont
  {Banerjee}}]{Sutradhar2022}%
  \BibitemOpen
  \bibfield  {author} {\bibinfo {author} {\bibfnamefont {J.}~\bibnamefont
  {Sutradhar}}, \bibinfo {author} {\bibfnamefont {S.}~\bibnamefont {Ghosh}},
  \bibinfo {author} {\bibfnamefont {S.}~\bibnamefont {Roy}}, \bibinfo {author}
  {\bibfnamefont {D.~E.}\ \bibnamefont {Logan}}, \bibinfo {author}
  {\bibfnamefont {S.}~\bibnamefont {Mukerjee}},\ and\ \bibinfo {author}
  {\bibfnamefont {S.}~\bibnamefont {Banerjee}},\ }\bibfield  {title} {\bibinfo
  {title} {Scaling of the {Fock}-space propagator and multifractality across
  the many-body localization transition},\ }\href
  {https://doi.org/10.1103/PhysRevB.106.054203} {\bibfield  {journal} {\bibinfo
   {journal} {Phys. Rev. B}\ }\textbf {\bibinfo {volume} {106}},\ \bibinfo
  {pages} {054203} (\bibinfo {year} {2022})}\BibitemShut {NoStop}%
\bibitem [{\citenamefont {Roy}\ \emph {et~al.}(2023)\citenamefont {Roy},
  \citenamefont {Sutradhar},\ and\ \citenamefont {Banerjee}}]{Roy2023}%
  \BibitemOpen
  \bibfield  {author} {\bibinfo {author} {\bibfnamefont {N.}~\bibnamefont
  {Roy}}, \bibinfo {author} {\bibfnamefont {J.}~\bibnamefont {Sutradhar}},\
  and\ \bibinfo {author} {\bibfnamefont {S.}~\bibnamefont {Banerjee}},\
  }\bibfield  {title} {\bibinfo {title} {Diagnostics of nonergodic extended
  states and many body localization proximity effect through real-space and
  {Fock}-space excitations},\ }\href
  {https://doi.org/10.1103/PhysRevB.107.115155} {\bibfield  {journal} {\bibinfo
   {journal} {Phys. Rev. B}\ }\textbf {\bibinfo {volume} {107}},\ \bibinfo
  {pages} {115155} (\bibinfo {year} {2023})}\BibitemShut {NoStop}%
\bibitem [{\citenamefont {De~Tomasi}\ \emph {et~al.}(2021)\citenamefont
  {De~Tomasi}, \citenamefont {Khaymovich}, \citenamefont {Pollmann},\ and\
  \citenamefont {Warzel}}]{DeTomasi2021}%
  \BibitemOpen
  \bibfield  {author} {\bibinfo {author} {\bibfnamefont {G.}~\bibnamefont
  {De~Tomasi}}, \bibinfo {author} {\bibfnamefont {I.~M.}\ \bibnamefont
  {Khaymovich}}, \bibinfo {author} {\bibfnamefont {F.}~\bibnamefont
  {Pollmann}},\ and\ \bibinfo {author} {\bibfnamefont {S.}~\bibnamefont
  {Warzel}},\ }\bibfield  {title} {\bibinfo {title} {Rare thermal bubbles at
  the many-body localization transition from the {Fock} space point of view},\
  }\href {https://doi.org/10.1103/PhysRevB.104.024202} {\bibfield  {journal}
  {\bibinfo  {journal} {Phys. Rev. B}\ }\textbf {\bibinfo {volume} {104}},\
  \bibinfo {pages} {024202} (\bibinfo {year} {2021})}\BibitemShut {NoStop}%
\bibitem [{\citenamefont {Smith}\ \emph {et~al.}(2016)\citenamefont {Smith},
  \citenamefont {Lee}, \citenamefont {Richerme}, \citenamefont {Neyenhuis},
  \citenamefont {Hess}, \citenamefont {Hauke}, \citenamefont {Heyl},
  \citenamefont {Huse},\ and\ \citenamefont {Monroe}}]{smith2016many-body}%
  \BibitemOpen
  \bibfield  {author} {\bibinfo {author} {\bibfnamefont {J.}~\bibnamefont
  {Smith}}, \bibinfo {author} {\bibfnamefont {A.}~\bibnamefont {Lee}}, \bibinfo
  {author} {\bibfnamefont {P.}~\bibnamefont {Richerme}}, \bibinfo {author}
  {\bibfnamefont {B.}~\bibnamefont {Neyenhuis}}, \bibinfo {author}
  {\bibfnamefont {P.~W.}\ \bibnamefont {Hess}}, \bibinfo {author}
  {\bibfnamefont {P.}~\bibnamefont {Hauke}}, \bibinfo {author} {\bibfnamefont
  {M.}~\bibnamefont {Heyl}}, \bibinfo {author} {\bibfnamefont {D.~A.}\
  \bibnamefont {Huse}},\ and\ \bibinfo {author} {\bibfnamefont
  {C.}~\bibnamefont {Monroe}},\ }\bibfield  {title} {\bibinfo {title}
  {Many-body localization in a quantum simulator with programmable random
  disorder},\ }\href {https://doi.org/10.1038/nphys3783} {\bibfield  {journal}
  {\bibinfo  {journal} {Nature Physics}\ }\textbf {\bibinfo {volume} {12}},\
  \bibinfo {pages} {907} (\bibinfo {year} {2016})}\BibitemShut {NoStop}%
\bibitem [{\citenamefont {Roushan}\ \emph {et~al.}(2017)\citenamefont
  {Roushan}, \citenamefont {Neill}, \citenamefont {Tangpanitanon},
  \citenamefont {Bastidas}, \citenamefont {Megrant}, \citenamefont {Barends},
  \citenamefont {Chen}, \citenamefont {Chen}, \citenamefont {Chiaro},
  \citenamefont {Dunsworth}, \citenamefont {Fowler}, \citenamefont {Foxen},
  \citenamefont {Giustina}, \citenamefont {Jeffrey}, \citenamefont {Kelly},
  \citenamefont {Lucero}, \citenamefont {Mutus}, \citenamefont {Neeley},
  \citenamefont {Quintana}, \citenamefont {Sank}, \citenamefont {Vainsencher},
  \citenamefont {Wenner}, \citenamefont {White}, \citenamefont {Neven},
  \citenamefont {Angelakis},\ and\ \citenamefont
  {Martinis}}]{roushan2017spectroscopic}%
  \BibitemOpen
  \bibfield  {author} {\bibinfo {author} {\bibfnamefont {P.}~\bibnamefont
  {Roushan}}, \bibinfo {author} {\bibfnamefont {C.}~\bibnamefont {Neill}},
  \bibinfo {author} {\bibfnamefont {J.}~\bibnamefont {Tangpanitanon}}, \bibinfo
  {author} {\bibfnamefont {V.~M.}\ \bibnamefont {Bastidas}}, \bibinfo {author}
  {\bibfnamefont {A.}~\bibnamefont {Megrant}}, \bibinfo {author} {\bibfnamefont
  {R.}~\bibnamefont {Barends}}, \bibinfo {author} {\bibfnamefont
  {Y.}~\bibnamefont {Chen}}, \bibinfo {author} {\bibfnamefont {Z.}~\bibnamefont
  {Chen}}, \bibinfo {author} {\bibfnamefont {B.}~\bibnamefont {Chiaro}},
  \bibinfo {author} {\bibfnamefont {A.}~\bibnamefont {Dunsworth}}, \bibinfo
  {author} {\bibfnamefont {A.}~\bibnamefont {Fowler}}, \bibinfo {author}
  {\bibfnamefont {B.}~\bibnamefont {Foxen}}, \bibinfo {author} {\bibfnamefont
  {M.}~\bibnamefont {Giustina}}, \bibinfo {author} {\bibfnamefont
  {E.}~\bibnamefont {Jeffrey}}, \bibinfo {author} {\bibfnamefont
  {J.}~\bibnamefont {Kelly}}, \bibinfo {author} {\bibfnamefont
  {E.}~\bibnamefont {Lucero}}, \bibinfo {author} {\bibfnamefont
  {J.}~\bibnamefont {Mutus}}, \bibinfo {author} {\bibfnamefont
  {M.}~\bibnamefont {Neeley}}, \bibinfo {author} {\bibfnamefont
  {C.}~\bibnamefont {Quintana}}, \bibinfo {author} {\bibfnamefont
  {D.}~\bibnamefont {Sank}}, \bibinfo {author} {\bibfnamefont {A.}~\bibnamefont
  {Vainsencher}}, \bibinfo {author} {\bibfnamefont {J.}~\bibnamefont {Wenner}},
  \bibinfo {author} {\bibfnamefont {T.}~\bibnamefont {White}}, \bibinfo
  {author} {\bibfnamefont {H.}~\bibnamefont {Neven}}, \bibinfo {author}
  {\bibfnamefont {D.~G.}\ \bibnamefont {Angelakis}},\ and\ \bibinfo {author}
  {\bibfnamefont {J.}~\bibnamefont {Martinis}},\ }\bibfield  {title} {\bibinfo
  {title} {Spectroscopic signatures of localization with interacting photons in
  superconducting qubits},\ }\href@noop {} {\bibfield  {journal} {\bibinfo
  {journal} {Science}\ }\textbf {\bibinfo {volume} {358}},\ \bibinfo {pages}
  {1175} (\bibinfo {year} {2017})}\BibitemShut {NoStop}%
\bibitem [{\citenamefont {Xu}\ \emph {et~al.}(2018)\citenamefont {Xu},
  \citenamefont {Chen}, \citenamefont {Zeng}, \citenamefont {Zhang},
  \citenamefont {Song}, \citenamefont {Liu}, \citenamefont {Guo}, \citenamefont
  {Zhang}, \citenamefont {Xu}, \citenamefont {Deng}, \citenamefont {Huang},
  \citenamefont {Wang}, \citenamefont {Zhu}, \citenamefont {Zheng},\ and\
  \citenamefont {Fan}}]{xu2018emulating}%
  \BibitemOpen
  \bibfield  {author} {\bibinfo {author} {\bibfnamefont {K.}~\bibnamefont
  {Xu}}, \bibinfo {author} {\bibfnamefont {J.-J.}\ \bibnamefont {Chen}},
  \bibinfo {author} {\bibfnamefont {Y.}~\bibnamefont {Zeng}}, \bibinfo {author}
  {\bibfnamefont {Y.-R.}\ \bibnamefont {Zhang}}, \bibinfo {author}
  {\bibfnamefont {C.}~\bibnamefont {Song}}, \bibinfo {author} {\bibfnamefont
  {W.}~\bibnamefont {Liu}}, \bibinfo {author} {\bibfnamefont {Q.}~\bibnamefont
  {Guo}}, \bibinfo {author} {\bibfnamefont {P.}~\bibnamefont {Zhang}}, \bibinfo
  {author} {\bibfnamefont {D.}~\bibnamefont {Xu}}, \bibinfo {author}
  {\bibfnamefont {H.}~\bibnamefont {Deng}}, \bibinfo {author} {\bibfnamefont
  {K.}~\bibnamefont {Huang}}, \bibinfo {author} {\bibfnamefont
  {H.}~\bibnamefont {Wang}}, \bibinfo {author} {\bibfnamefont {X.}~\bibnamefont
  {Zhu}}, \bibinfo {author} {\bibfnamefont {D.}~\bibnamefont {Zheng}},\ and\
  \bibinfo {author} {\bibfnamefont {H.}~\bibnamefont {Fan}},\ }\bibfield
  {title} {\bibinfo {title} {Emulating many-body localization with a
  superconducting quantum processor},\ }\href
  {https://doi.org/10.1103/PhysRevLett.120.050507} {\bibfield  {journal}
  {\bibinfo  {journal} {Phys. Rev. Lett.}\ }\textbf {\bibinfo {volume} {120}},\
  \bibinfo {pages} {050507} (\bibinfo {year} {2018})}\BibitemShut {NoStop}%
\bibitem [{\citenamefont {Lukin}\ \emph {et~al.}(2019)\citenamefont {Lukin},
  \citenamefont {Rispoli}, \citenamefont {Schittko}, \citenamefont {Tai},
  \citenamefont {Kaufman}, \citenamefont {Choi}, \citenamefont {Khemani},
  \citenamefont {L{\'e}onard},\ and\ \citenamefont
  {Greiner}}]{lukin2019probing}%
  \BibitemOpen
  \bibfield  {author} {\bibinfo {author} {\bibfnamefont {A.}~\bibnamefont
  {Lukin}}, \bibinfo {author} {\bibfnamefont {M.}~\bibnamefont {Rispoli}},
  \bibinfo {author} {\bibfnamefont {R.}~\bibnamefont {Schittko}}, \bibinfo
  {author} {\bibfnamefont {M.~E.}\ \bibnamefont {Tai}}, \bibinfo {author}
  {\bibfnamefont {A.~M.}\ \bibnamefont {Kaufman}}, \bibinfo {author}
  {\bibfnamefont {S.}~\bibnamefont {Choi}}, \bibinfo {author} {\bibfnamefont
  {V.}~\bibnamefont {Khemani}}, \bibinfo {author} {\bibfnamefont
  {J.}~\bibnamefont {L{\'e}onard}},\ and\ \bibinfo {author} {\bibfnamefont
  {M.}~\bibnamefont {Greiner}},\ }\bibfield  {title} {\bibinfo {title} {Probing
  entanglement in a many-body-localized system},\ }\href@noop {} {\bibfield
  {journal} {\bibinfo  {journal} {Science}\ }\textbf {\bibinfo {volume}
  {364}},\ \bibinfo {pages} {256} (\bibinfo {year} {2019})}\BibitemShut
  {NoStop}%
\bibitem [{\citenamefont {Yao}\ \emph {et~al.}(2023)\citenamefont {Yao},
  \citenamefont {Xiang}, \citenamefont {Guo}, \citenamefont {Bao},
  \citenamefont {Yang}, \citenamefont {Song}, \citenamefont {Shi},
  \citenamefont {Zhu}, \citenamefont {Jin}, \citenamefont {Chen}, \citenamefont
  {Xu}, \citenamefont {Zhu}, \citenamefont {Shen}, \citenamefont {Wang},
  \citenamefont {Zhang}, \citenamefont {Wu}, \citenamefont {Zou}, \citenamefont
  {Zhang}, \citenamefont {Li}, \citenamefont {Wang}, \citenamefont {Song},
  \citenamefont {Cheng}, \citenamefont {Mondaini}, \citenamefont {Wang},
  \citenamefont {You}, \citenamefont {Zhu}, \citenamefont {Ying},\ and\
  \citenamefont {Guo}}]{Yao2023}%
  \BibitemOpen
  \bibfield  {author} {\bibinfo {author} {\bibfnamefont {Y.}~\bibnamefont
  {Yao}}, \bibinfo {author} {\bibfnamefont {L.}~\bibnamefont {Xiang}}, \bibinfo
  {author} {\bibfnamefont {Z.}~\bibnamefont {Guo}}, \bibinfo {author}
  {\bibfnamefont {Z.}~\bibnamefont {Bao}}, \bibinfo {author} {\bibfnamefont
  {Y.-F.}\ \bibnamefont {Yang}}, \bibinfo {author} {\bibfnamefont
  {Z.}~\bibnamefont {Song}}, \bibinfo {author} {\bibfnamefont {H.}~\bibnamefont
  {Shi}}, \bibinfo {author} {\bibfnamefont {X.}~\bibnamefont {Zhu}}, \bibinfo
  {author} {\bibfnamefont {F.}~\bibnamefont {Jin}}, \bibinfo {author}
  {\bibfnamefont {J.}~\bibnamefont {Chen}}, \bibinfo {author} {\bibfnamefont
  {S.}~\bibnamefont {Xu}}, \bibinfo {author} {\bibfnamefont {Z.}~\bibnamefont
  {Zhu}}, \bibinfo {author} {\bibfnamefont {F.}~\bibnamefont {Shen}}, \bibinfo
  {author} {\bibfnamefont {N.}~\bibnamefont {Wang}}, \bibinfo {author}
  {\bibfnamefont {C.}~\bibnamefont {Zhang}}, \bibinfo {author} {\bibfnamefont
  {Y.}~\bibnamefont {Wu}}, \bibinfo {author} {\bibfnamefont {Y.}~\bibnamefont
  {Zou}}, \bibinfo {author} {\bibfnamefont {P.}~\bibnamefont {Zhang}}, \bibinfo
  {author} {\bibfnamefont {H.}~\bibnamefont {Li}}, \bibinfo {author}
  {\bibfnamefont {Z.}~\bibnamefont {Wang}}, \bibinfo {author} {\bibfnamefont
  {C.}~\bibnamefont {Song}}, \bibinfo {author} {\bibfnamefont {C.}~\bibnamefont
  {Cheng}}, \bibinfo {author} {\bibfnamefont {R.}~\bibnamefont {Mondaini}},
  \bibinfo {author} {\bibfnamefont {H.}~\bibnamefont {Wang}}, \bibinfo {author}
  {\bibfnamefont {J.~Q.}\ \bibnamefont {You}}, \bibinfo {author} {\bibfnamefont
  {S.-Y.}\ \bibnamefont {Zhu}}, \bibinfo {author} {\bibfnamefont
  {L.}~\bibnamefont {Ying}},\ and\ \bibinfo {author} {\bibfnamefont
  {Q.}~\bibnamefont {Guo}},\ }\bibfield  {title} {\bibinfo {title} {Observation
  of many-body {Fock} space dynamics in two dimensions},\ }\href
  {https://doi.org/10.1038/s41567-023-02133-0} {\bibfield  {journal} {\bibinfo
  {journal} {Nature Physics}\ }\textbf {\bibinfo {volume} {19}},\ \bibinfo
  {pages} {1459} (\bibinfo {year} {2023})}\BibitemShut {NoStop}%
\bibitem [{\citenamefont {Biroli}\ \emph {et~al.}(2012)\citenamefont {Biroli},
  \citenamefont {Ribeiro-Teixeira},\ and\ \citenamefont
  {Tarzia}}]{biroli2012difference}%
  \BibitemOpen
  \bibfield  {author} {\bibinfo {author} {\bibfnamefont {G.}~\bibnamefont
  {Biroli}}, \bibinfo {author} {\bibfnamefont {A.}~\bibnamefont
  {Ribeiro-Teixeira}},\ and\ \bibinfo {author} {\bibfnamefont {M.}~\bibnamefont
  {Tarzia}},\ }\bibfield  {title} {\bibinfo {title} {Difference between level
  statistics, ergodicity and localization transitions on the bethe lattice},\
  }\href@noop {} {\bibfield  {journal} {\bibinfo  {journal} {arXiv:1211.7334}\
  } (\bibinfo {year} {2012})}\BibitemShut {NoStop}%
\bibitem [{\citenamefont {De~Luca}\ \emph {et~al.}(2014)\citenamefont
  {De~Luca}, \citenamefont {Altshuler}, \citenamefont {Kravtsov},\ and\
  \citenamefont {Scardicchio}}]{de2014anderson}%
  \BibitemOpen
  \bibfield  {author} {\bibinfo {author} {\bibfnamefont {A.}~\bibnamefont
  {De~Luca}}, \bibinfo {author} {\bibfnamefont {B.}~\bibnamefont {Altshuler}},
  \bibinfo {author} {\bibfnamefont {V.}~\bibnamefont {Kravtsov}},\ and\
  \bibinfo {author} {\bibfnamefont {A.}~\bibnamefont {Scardicchio}},\
  }\bibfield  {title} {\bibinfo {title} {Anderson localization on the {B}ethe
  lattice: nonergodicity of extended states},\ }\href@noop {} {\bibfield
  {journal} {\bibinfo  {journal} {Physical Review Letters}\ }\textbf {\bibinfo
  {volume} {113}},\ \bibinfo {pages} {046806} (\bibinfo {year}
  {2014})}\BibitemShut {NoStop}%
\bibitem [{\citenamefont {Tikhonov}\ \emph {et~al.}(2016)\citenamefont
  {Tikhonov}, \citenamefont {Mirlin},\ and\ \citenamefont
  {Skvortsov}}]{tikhonov2016anderson}%
  \BibitemOpen
  \bibfield  {author} {\bibinfo {author} {\bibfnamefont {K.}~\bibnamefont
  {Tikhonov}}, \bibinfo {author} {\bibfnamefont {A.}~\bibnamefont {Mirlin}},\
  and\ \bibinfo {author} {\bibfnamefont {M.}~\bibnamefont {Skvortsov}},\
  }\bibfield  {title} {\bibinfo {title} {Anderson localization and ergodicity
  on random regular graphs},\ }\href@noop {} {\bibfield  {journal} {\bibinfo
  {journal} {Phys. Rev. B}\ }\textbf {\bibinfo {volume} {94}},\ \bibinfo
  {pages} {220203} (\bibinfo {year} {2016})}\BibitemShut {NoStop}%
\bibitem [{\citenamefont {Garc{\'\i}a-Mata}\ \emph {et~al.}(2017)\citenamefont
  {Garc{\'\i}a-Mata}, \citenamefont {Giraud}, \citenamefont {Georgeot},
  \citenamefont {Martin}, \citenamefont {Dubertrand},\ and\ \citenamefont
  {Lemari{\'e}}}]{garcia-mata17}%
  \BibitemOpen
  \bibfield  {author} {\bibinfo {author} {\bibfnamefont {I.}~\bibnamefont
  {Garc{\'\i}a-Mata}}, \bibinfo {author} {\bibfnamefont {O.}~\bibnamefont
  {Giraud}}, \bibinfo {author} {\bibfnamefont {B.}~\bibnamefont {Georgeot}},
  \bibinfo {author} {\bibfnamefont {J.}~\bibnamefont {Martin}}, \bibinfo
  {author} {\bibfnamefont {R.}~\bibnamefont {Dubertrand}},\ and\ \bibinfo
  {author} {\bibfnamefont {G.}~\bibnamefont {Lemari{\'e}}},\ }\bibfield
  {title} {\bibinfo {title} {Scaling theory of the {A}nderson transition in
  random graphs: ergodicity and universality},\ }\href@noop {} {\bibfield
  {journal} {\bibinfo  {journal} {Physical Review Letters}\ }\textbf {\bibinfo
  {volume} {118}},\ \bibinfo {pages} {166801} (\bibinfo {year}
  {2017})}\BibitemShut {NoStop}%
\bibitem [{\citenamefont {Metz}\ and\ \citenamefont
  {Castillo}(2017)}]{metz2017level}%
  \BibitemOpen
  \bibfield  {author} {\bibinfo {author} {\bibfnamefont {F.~L.}\ \bibnamefont
  {Metz}}\ and\ \bibinfo {author} {\bibfnamefont {I.~P.}\ \bibnamefont
  {Castillo}},\ }\bibfield  {title} {\bibinfo {title} {Level compressibility
  for the {A}nderson model on regular random graphs and the eigenvalue
  statistics in the extended phase},\ }\href@noop {} {\bibfield  {journal}
  {\bibinfo  {journal} {Phys. Rev. B}\ }\textbf {\bibinfo {volume} {96}},\
  \bibinfo {pages} {064202} (\bibinfo {year} {2017})}\BibitemShut {NoStop}%
\bibitem [{\citenamefont {Biroli}\ and\ \citenamefont
  {Tarzia}(2017)}]{Biroli2017}%
  \BibitemOpen
  \bibfield  {author} {\bibinfo {author} {\bibfnamefont {G.}~\bibnamefont
  {Biroli}}\ and\ \bibinfo {author} {\bibfnamefont {M.}~\bibnamefont
  {Tarzia}},\ }\bibfield  {title} {\bibinfo {title} {Delocalized glassy
  dynamics and many-body localization},\ }\href@noop {} {\bibfield  {journal}
  {\bibinfo  {journal} {Phys. Rev. B}\ }\textbf {\bibinfo {volume} {96}},\
  \bibinfo {pages} {201114} (\bibinfo {year} {2017})}\BibitemShut {NoStop}%
\bibitem [{\citenamefont {Kravtsov}\ \emph {et~al.}(2018)\citenamefont
  {Kravtsov}, \citenamefont {Altshuler},\ and\ \citenamefont
  {Ioffe}}]{kravtsov2018non}%
  \BibitemOpen
  \bibfield  {author} {\bibinfo {author} {\bibfnamefont {V.}~\bibnamefont
  {Kravtsov}}, \bibinfo {author} {\bibfnamefont {B.}~\bibnamefont
  {Altshuler}},\ and\ \bibinfo {author} {\bibfnamefont {L.}~\bibnamefont
  {Ioffe}},\ }\bibfield  {title} {\bibinfo {title} {Non-ergodic delocalized
  phase in {A}nderson model on {B}ethe lattice and regular graph},\ }\href@noop
  {} {\bibfield  {journal} {\bibinfo  {journal} {Annals of Physics}\ }\textbf
  {\bibinfo {volume} {389}},\ \bibinfo {pages} {148} (\bibinfo {year}
  {2018})}\BibitemShut {NoStop}%
\bibitem [{\citenamefont {Biroli}\ and\ \citenamefont
  {Tarzia}(2018)}]{biroli2018}%
  \BibitemOpen
  \bibfield  {author} {\bibinfo {author} {\bibfnamefont {G.}~\bibnamefont
  {Biroli}}\ and\ \bibinfo {author} {\bibfnamefont {M.}~\bibnamefont
  {Tarzia}},\ }\bibfield  {title} {\bibinfo {title} {Delocalization and
  ergodicity of the {A}nderson model on {B}ethe lattices},\ }\href@noop {}
  {\bibfield  {journal} {\bibinfo  {journal} {arXiv:1810.07545}\ } (\bibinfo
  {year} {2018})}\BibitemShut {NoStop}%
\bibitem [{\citenamefont {Bera}\ \emph {et~al.}(2018)\citenamefont {Bera},
  \citenamefont {De~Tomasi}, \citenamefont {Khaymovich},\ and\ \citenamefont
  {Scardicchio}}]{PhysRevB.98.134205}%
  \BibitemOpen
  \bibfield  {author} {\bibinfo {author} {\bibfnamefont {S.}~\bibnamefont
  {Bera}}, \bibinfo {author} {\bibfnamefont {G.}~\bibnamefont {De~Tomasi}},
  \bibinfo {author} {\bibfnamefont {I.~M.}\ \bibnamefont {Khaymovich}},\ and\
  \bibinfo {author} {\bibfnamefont {A.}~\bibnamefont {Scardicchio}},\
  }\bibfield  {title} {\bibinfo {title} {Return probability for the {A}nderson
  model on the random regular graph},\ }\href@noop {} {\bibfield  {journal}
  {\bibinfo  {journal} {Phys. Rev. B}\ }\textbf {\bibinfo {volume} {98}},\
  \bibinfo {pages} {134205} (\bibinfo {year} {2018})}\BibitemShut {NoStop}%
\bibitem [{\citenamefont {Tikhonov}\ and\ \citenamefont
  {Mirlin}(2019{\natexlab{a}})}]{tikhonov19statistics}%
  \BibitemOpen
  \bibfield  {author} {\bibinfo {author} {\bibfnamefont {K.~S.}\ \bibnamefont
  {Tikhonov}}\ and\ \bibinfo {author} {\bibfnamefont {A.~D.}\ \bibnamefont
  {Mirlin}},\ }\bibfield  {title} {\bibinfo {title} {Statistics of eigenstates
  near the localization transition on random regular graphs},\ }\href@noop {}
  {\bibfield  {journal} {\bibinfo  {journal} {Phys. Rev. B}\ }\textbf {\bibinfo
  {volume} {99}},\ \bibinfo {pages} {024202} (\bibinfo {year}
  {2019}{\natexlab{a}})}\BibitemShut {NoStop}%
\bibitem [{\citenamefont {Tikhonov}\ and\ \citenamefont
  {Mirlin}(2019{\natexlab{b}})}]{tikhonov19critical}%
  \BibitemOpen
  \bibfield  {author} {\bibinfo {author} {\bibfnamefont {K.~S.}\ \bibnamefont
  {Tikhonov}}\ and\ \bibinfo {author} {\bibfnamefont {A.~D.}\ \bibnamefont
  {Mirlin}},\ }\bibfield  {title} {\bibinfo {title} {Critical behavior at the
  localization transition on random regular graphs},\ }\href@noop {} {\bibfield
   {journal} {\bibinfo  {journal} {Phys. Rev. B}\ }\textbf {\bibinfo {volume}
  {99}},\ \bibinfo {pages} {214202} (\bibinfo {year}
  {2019}{\natexlab{b}})}\BibitemShut {NoStop}%
\bibitem [{\citenamefont {Garc\'{\i}a-Mata}\ \emph {et~al.}(2020)\citenamefont
  {Garc\'{\i}a-Mata}, \citenamefont {Martin}, \citenamefont {Dubertrand},
  \citenamefont {Giraud}, \citenamefont {Georgeot},\ and\ \citenamefont
  {Lemari\'e}}]{PhysRevResearch.2.012020}%
  \BibitemOpen
  \bibfield  {author} {\bibinfo {author} {\bibfnamefont {I.}~\bibnamefont
  {Garc\'{\i}a-Mata}}, \bibinfo {author} {\bibfnamefont {J.}~\bibnamefont
  {Martin}}, \bibinfo {author} {\bibfnamefont {R.}~\bibnamefont {Dubertrand}},
  \bibinfo {author} {\bibfnamefont {O.}~\bibnamefont {Giraud}}, \bibinfo
  {author} {\bibfnamefont {B.}~\bibnamefont {Georgeot}},\ and\ \bibinfo
  {author} {\bibfnamefont {G.}~\bibnamefont {Lemari\'e}},\ }\bibfield  {title}
  {\bibinfo {title} {Two critical localization lengths in the {A}nderson
  transition on random graphs},\ }\href@noop {} {\bibfield  {journal} {\bibinfo
   {journal} {Phys. Rev. Research}\ }\textbf {\bibinfo {volume} {2}},\ \bibinfo
  {pages} {012020} (\bibinfo {year} {2020})}\BibitemShut {NoStop}%
\bibitem [{\citenamefont {Roy}\ and\ \citenamefont
  {Logan}(2020{\natexlab{a}})}]{Roy2020numerics}%
  \BibitemOpen
  \bibfield  {author} {\bibinfo {author} {\bibfnamefont {S.}~\bibnamefont
  {Roy}}\ and\ \bibinfo {author} {\bibfnamefont {D.~E.}\ \bibnamefont
  {Logan}},\ }\bibfield  {title} {\bibinfo {title} {Localization on certain
  graphs with strongly correlated disorder},\ }\href@noop {} {\bibfield
  {journal} {\bibinfo  {journal} {Phys. Rev. Lett.}\ }\textbf {\bibinfo
  {volume} {125}},\ \bibinfo {pages} {250402} (\bibinfo {year}
  {2020}{\natexlab{a}})}\BibitemShut {NoStop}%
\bibitem [{\citenamefont {Garc\'{\i}a-Mata}\ \emph {et~al.}(2022)\citenamefont
  {Garc\'{\i}a-Mata}, \citenamefont {Martin}, \citenamefont {Giraud},
  \citenamefont {Georgeot}, \citenamefont {Dubertrand},\ and\ \citenamefont
  {Lemari\'e}}]{garcia-mata2022critical}%
  \BibitemOpen
  \bibfield  {author} {\bibinfo {author} {\bibfnamefont {I.}~\bibnamefont
  {Garc\'{\i}a-Mata}}, \bibinfo {author} {\bibfnamefont {J.}~\bibnamefont
  {Martin}}, \bibinfo {author} {\bibfnamefont {O.}~\bibnamefont {Giraud}},
  \bibinfo {author} {\bibfnamefont {B.}~\bibnamefont {Georgeot}}, \bibinfo
  {author} {\bibfnamefont {R.}~\bibnamefont {Dubertrand}},\ and\ \bibinfo
  {author} {\bibfnamefont {G.}~\bibnamefont {Lemari\'e}},\ }\bibfield  {title}
  {\bibinfo {title} {Critical properties of the {A}nderson transition on random
  graphs: {T}wo-parameter scaling theory, {Kosterlitz-Thouless} type flow, and
  many-body localization},\ }\href
  {https://doi.org/10.1103/PhysRevB.106.214202} {\bibfield  {journal} {\bibinfo
   {journal} {Phys. Rev. B}\ }\textbf {\bibinfo {volume} {106}},\ \bibinfo
  {pages} {214202} (\bibinfo {year} {2022})}\BibitemShut {NoStop}%
\bibitem [{\citenamefont {Sierant}\ \emph
  {et~al.}(2023{\natexlab{a}})\citenamefont {Sierant}, \citenamefont
  {Lewenstein},\ and\ \citenamefont {Scardicchio}}]{sierant2023universality}%
  \BibitemOpen
  \bibfield  {author} {\bibinfo {author} {\bibfnamefont {P.}~\bibnamefont
  {Sierant}}, \bibinfo {author} {\bibfnamefont {M.}~\bibnamefont
  {Lewenstein}},\ and\ \bibinfo {author} {\bibfnamefont {A.}~\bibnamefont
  {Scardicchio}},\ }\bibfield  {title} {\bibinfo {title} {{Universality in
  Anderson localization on random graphs with varying connectivity}},\ }\href
  {https://doi.org/10.21468/SciPostPhys.15.2.045} {\bibfield  {journal}
  {\bibinfo  {journal} {SciPost Phys.}\ }\textbf {\bibinfo {volume} {15}},\
  \bibinfo {pages} {045} (\bibinfo {year} {2023}{\natexlab{a}})}\BibitemShut
  {NoStop}%
\bibitem [{\citenamefont {Valba}\ and\ \citenamefont
  {Gorsky}(2022)}]{valba2022mobility}%
  \BibitemOpen
  \bibfield  {author} {\bibinfo {author} {\bibfnamefont {O.}~\bibnamefont
  {Valba}}\ and\ \bibinfo {author} {\bibfnamefont {A.}~\bibnamefont {Gorsky}},\
  }\bibfield  {title} {\bibinfo {title} {Mobility edge in the anderson model on
  partially disordered random regular graphs},\ }\href@noop {} {\bibfield
  {journal} {\bibinfo  {journal} {JETP Letters}\ }\textbf {\bibinfo {volume}
  {116}},\ \bibinfo {pages} {398} (\bibinfo {year} {2022})}\BibitemShut
  {NoStop}%
\bibitem [{\citenamefont {Mirlin}\ and\ \citenamefont
  {Fyodorov}(1991{\natexlab{a}})}]{mirlin1991universality}%
  \BibitemOpen
  \bibfield  {author} {\bibinfo {author} {\bibfnamefont {A.}~\bibnamefont
  {Mirlin}}\ and\ \bibinfo {author} {\bibfnamefont {Y.~V.}\ \bibnamefont
  {Fyodorov}},\ }\bibfield  {title} {\bibinfo {title} {Universality of level
  correlation function of sparse random matrices},\ }\href@noop {} {\bibfield
  {journal} {\bibinfo  {journal} {Journal of Physics A: Mathematical and
  General}\ }\textbf {\bibinfo {volume} {24}},\ \bibinfo {pages} {2273}
  (\bibinfo {year} {1991}{\natexlab{a}})}\BibitemShut {NoStop}%
\bibitem [{\citenamefont {Fyodorov}\ and\ \citenamefont
  {Mirlin}(1991)}]{fyodorov1991localization}%
  \BibitemOpen
  \bibfield  {author} {\bibinfo {author} {\bibfnamefont {Y.~V.}\ \bibnamefont
  {Fyodorov}}\ and\ \bibinfo {author} {\bibfnamefont {A.~D.}\ \bibnamefont
  {Mirlin}},\ }\bibfield  {title} {\bibinfo {title} {Localization in ensemble
  of sparse random matrices},\ }\href@noop {} {\bibfield  {journal} {\bibinfo
  {journal} {Physical Review Letters}\ }\textbf {\bibinfo {volume} {67}},\
  \bibinfo {pages} {2049} (\bibinfo {year} {1991})}\BibitemShut {NoStop}%
\bibitem [{\citenamefont {Ghosh}\ \emph {et~al.}(2019)\citenamefont {Ghosh},
  \citenamefont {Acharya}, \citenamefont {Sahu},\ and\ \citenamefont
  {Mukerjee}}]{ghosh2019many-body}%
  \BibitemOpen
  \bibfield  {author} {\bibinfo {author} {\bibfnamefont {S.}~\bibnamefont
  {Ghosh}}, \bibinfo {author} {\bibfnamefont {A.}~\bibnamefont {Acharya}},
  \bibinfo {author} {\bibfnamefont {S.}~\bibnamefont {Sahu}},\ and\ \bibinfo
  {author} {\bibfnamefont {S.}~\bibnamefont {Mukerjee}},\ }\bibfield  {title}
  {\bibinfo {title} {Many-body localization due to correlated disorder in
  {Fock} space},\ }\href {https://doi.org/10.1103/PhysRevB.99.165131}
  {\bibfield  {journal} {\bibinfo  {journal} {Phys. Rev. B}\ }\textbf {\bibinfo
  {volume} {99}},\ \bibinfo {pages} {165131} (\bibinfo {year}
  {2019})}\BibitemShut {NoStop}%
\bibitem [{\citenamefont {Roy}\ and\ \citenamefont
  {Logan}(2020{\natexlab{b}})}]{Roy2020}%
  \BibitemOpen
  \bibfield  {author} {\bibinfo {author} {\bibfnamefont {S.}~\bibnamefont
  {Roy}}\ and\ \bibinfo {author} {\bibfnamefont {D.~E.}\ \bibnamefont
  {Logan}},\ }\bibfield  {title} {\bibinfo {title} {Fock-space correlations and
  the origins of many-body localization},\ }\href
  {https://doi.org/10.1103/PhysRevB.101.134202} {\bibfield  {journal} {\bibinfo
   {journal} {Phys. Rev. B}\ }\textbf {\bibinfo {volume} {101}},\ \bibinfo
  {pages} {134202} (\bibinfo {year} {2020}{\natexlab{b}})}\BibitemShut
  {NoStop}%
\bibitem [{\citenamefont {Gopalakrishnan}\ and\ \citenamefont
  {Huse}(2019{\natexlab{a}})}]{Gopalakrishnan2019a}%
  \BibitemOpen
  \bibfield  {author} {\bibinfo {author} {\bibfnamefont {S.}~\bibnamefont
  {Gopalakrishnan}}\ and\ \bibinfo {author} {\bibfnamefont {D.~A.}\
  \bibnamefont {Huse}},\ }\bibfield  {title} {\bibinfo {title} {Instability of
  many-body localized systems as a phase transition in a nonstandard
  thermodynamic limit},\ }\href {https://doi.org/10.1103/PhysRevB.99.134305}
  {\bibfield  {journal} {\bibinfo  {journal} {Phys. Rev. B}\ }\textbf {\bibinfo
  {volume} {99}},\ \bibinfo {pages} {134305} (\bibinfo {year}
  {2019}{\natexlab{a}})}\BibitemShut {NoStop}%
\bibitem [{\citenamefont {Goldschmidt}(1990)}]{Goldschmidt1990}%
  \BibitemOpen
  \bibfield  {author} {\bibinfo {author} {\bibfnamefont {Y.~Y.}\ \bibnamefont
  {Goldschmidt}},\ }\bibfield  {title} {\bibinfo {title} {Solvable model of the
  quantum spin glass in a transverse field},\ }\href
  {https://doi.org/10.1103/PhysRevB.41.4858} {\bibfield  {journal} {\bibinfo
  {journal} {Phys. Rev. B}\ }\textbf {\bibinfo {volume} {41}},\ \bibinfo
  {pages} {4858} (\bibinfo {year} {1990})}\BibitemShut {NoStop}%
\bibitem [{\citenamefont {Laumann}\ \emph {et~al.}(2014)\citenamefont
  {Laumann}, \citenamefont {Pal},\ and\ \citenamefont
  {Scardicchio}}]{laumann2014many-body}%
  \BibitemOpen
  \bibfield  {author} {\bibinfo {author} {\bibfnamefont {C.~R.}\ \bibnamefont
  {Laumann}}, \bibinfo {author} {\bibfnamefont {A.}~\bibnamefont {Pal}},\ and\
  \bibinfo {author} {\bibfnamefont {A.}~\bibnamefont {Scardicchio}},\
  }\bibfield  {title} {\bibinfo {title} {Many-body mobility edge in a
  mean-field quantum spin glass},\ }\href
  {https://doi.org/10.1103/PhysRevLett.113.200405} {\bibfield  {journal}
  {\bibinfo  {journal} {Phys. Rev. Lett.}\ }\textbf {\bibinfo {volume} {113}},\
  \bibinfo {pages} {200405} (\bibinfo {year} {2014})}\BibitemShut {NoStop}%
\bibitem [{\citenamefont {Baldwin}\ \emph {et~al.}(2016)\citenamefont
  {Baldwin}, \citenamefont {Laumann}, \citenamefont {Pal},\ and\ \citenamefont
  {Scardicchio}}]{baldwin2016the_many-body}%
  \BibitemOpen
  \bibfield  {author} {\bibinfo {author} {\bibfnamefont {C.~L.}\ \bibnamefont
  {Baldwin}}, \bibinfo {author} {\bibfnamefont {C.~R.}\ \bibnamefont
  {Laumann}}, \bibinfo {author} {\bibfnamefont {A.}~\bibnamefont {Pal}},\ and\
  \bibinfo {author} {\bibfnamefont {A.}~\bibnamefont {Scardicchio}},\
  }\bibfield  {title} {\bibinfo {title} {The many-body localized phase of the
  quantum random energy model},\ }\href
  {https://doi.org/10.1103/PhysRevB.93.024202} {\bibfield  {journal} {\bibinfo
  {journal} {Phys. Rev. B}\ }\textbf {\bibinfo {volume} {93}},\ \bibinfo
  {pages} {024202} (\bibinfo {year} {2016})}\BibitemShut {NoStop}%
\bibitem [{\citenamefont {Cugliandolo}\ and\ \citenamefont
  {M{\"u}ller}(2023)}]{Cugliandolo}%
  \BibitemOpen
  \bibfield  {author} {\bibinfo {author} {\bibfnamefont {L.~F.}\ \bibnamefont
  {Cugliandolo}}\ and\ \bibinfo {author} {\bibfnamefont {M.}~\bibnamefont
  {M{\"u}ller}},\ }\bibfield  {title} {\bibinfo {title} {Quantum glasses},\
  }in\ \href {https://doi.org/10.1142/9789811273926_0018} {\emph {\bibinfo
  {booktitle} {Spin Glass Theory and Far Beyond}}},\ \bibinfo {editor} {edited
  by\ \bibinfo {editor} {\bibfnamefont {P.}~\bibnamefont {Charbonneau}},
  \bibinfo {editor} {\bibfnamefont {E.}~\bibnamefont {Marinari}}, \bibinfo
  {editor} {\bibfnamefont {M.}~\bibnamefont {M{\'e}zard}}, \bibinfo {editor}
  {\bibfnamefont {G.}~\bibnamefont {Parisi}}, \bibinfo {editor} {\bibfnamefont
  {F.}~\bibnamefont {Ricci-Tersenghi}}, \bibinfo {editor} {\bibfnamefont
  {G.}~\bibnamefont {Sicuro}},\ and\ \bibinfo {editor} {\bibfnamefont
  {F.}~\bibnamefont {Zamponi}}}\ (\bibinfo  {publisher} {{World Scientific}},\
  \bibinfo {year} {2023})\ Chap.~\bibinfo {chapter} {18}, pp.\ \bibinfo {pages}
  {353--374}\BibitemShut {NoStop}%
\bibitem [{\citenamefont {Abou-Chacra}\ \emph {et~al.}(1973)\citenamefont
  {Abou-Chacra}, \citenamefont {Thouless},\ and\ \citenamefont
  {Anderson}}]{abou1973selfconsistent}%
  \BibitemOpen
  \bibfield  {author} {\bibinfo {author} {\bibfnamefont {R.}~\bibnamefont
  {Abou-Chacra}}, \bibinfo {author} {\bibfnamefont {D.}~\bibnamefont
  {Thouless}},\ and\ \bibinfo {author} {\bibfnamefont {P.}~\bibnamefont
  {Anderson}},\ }\bibfield  {title} {\bibinfo {title} {A selfconsistent theory
  of localization},\ }\href@noop {} {\bibfield  {journal} {\bibinfo  {journal}
  {Journal of Physics C: Solid State Physics}\ }\textbf {\bibinfo {volume}
  {6}},\ \bibinfo {pages} {1734} (\bibinfo {year} {1973})}\BibitemShut
  {NoStop}%
\bibitem [{\citenamefont {Mirlin}\ and\ \citenamefont
  {Fyodorov}(1991{\natexlab{b}})}]{mirlin1991localization}%
  \BibitemOpen
  \bibfield  {author} {\bibinfo {author} {\bibfnamefont {A.~D.}\ \bibnamefont
  {Mirlin}}\ and\ \bibinfo {author} {\bibfnamefont {Y.~V.}\ \bibnamefont
  {Fyodorov}},\ }\bibfield  {title} {\bibinfo {title} {Localization transition
  in the {A}nderson model on the {B}ethe lattice: spontaneous symmetry breaking
  and correlation functions},\ }\href@noop {} {\bibfield  {journal} {\bibinfo
  {journal} {Nuclear Physics B}\ }\textbf {\bibinfo {volume} {366}},\ \bibinfo
  {pages} {507} (\bibinfo {year} {1991}{\natexlab{b}})}\BibitemShut {NoStop}%
\bibitem [{Note2()}]{Note2}%
  \BibitemOpen
  \bibinfo {note} {{\protect \color {black} Specifically, this is a crossover
  between the random-matrix-theory regime (where eigenstates spread over all
  Fock-space sites) and the self-consistent golden rule regime (also ergodic
  but with eigenstates spreading over energy windows parametrically smaller
  than the total disorder-induced bandwidth), see Ref.~\cite {Herre2023}. These
  correspond to regimes I and II of Ref.~\cite {Monteiro2021}, whereas the MBL
  transition is located on the border of regimes III and IV in terminology of
  that paper.}}\BibitemShut {Stop}%
\bibitem [{\citenamefont {Ros}\ \emph {et~al.}(2015)\citenamefont {Ros},
  \citenamefont {M{\"u}ller},\ and\ \citenamefont {Scardicchio}}]{Ros2015a}%
  \BibitemOpen
  \bibfield  {author} {\bibinfo {author} {\bibfnamefont {V.}~\bibnamefont
  {Ros}}, \bibinfo {author} {\bibfnamefont {M.}~\bibnamefont {M{\"u}ller}},\
  and\ \bibinfo {author} {\bibfnamefont {A.}~\bibnamefont {Scardicchio}},\
  }\bibfield  {title} {\bibinfo {title} {Integrals of motion in the many-body
  localized phase},\ }\href {https://doi.org/10.1016/j.nuclphysb.2014.12.014}
  {\bibfield  {journal} {\bibinfo  {journal} {Nucl. Phys. B}\ }\textbf
  {\bibinfo {volume} {891}},\ \bibinfo {pages} {420 } (\bibinfo {year}
  {2015})}\BibitemShut {NoStop}%
\bibitem [{\citenamefont {Imbrie}(2016{\natexlab{a}})}]{imbrie16a}%
  \BibitemOpen
  \bibfield  {author} {\bibinfo {author} {\bibfnamefont {J.~Z.}\ \bibnamefont
  {Imbrie}},\ }\bibfield  {title} {\bibinfo {title} {Diagonalization and
  many-body localization for a disordered quantum spin chain},\ }\href@noop {}
  {\bibfield  {journal} {\bibinfo  {journal} {Physical Review Letters}\
  }\textbf {\bibinfo {volume} {117}},\ \bibinfo {pages} {027201} (\bibinfo
  {year} {2016}{\natexlab{a}})}\BibitemShut {NoStop}%
\bibitem [{\citenamefont {Imbrie}(2016{\natexlab{b}})}]{Imbrie2016JSP}%
  \BibitemOpen
  \bibfield  {author} {\bibinfo {author} {\bibfnamefont {J.~Z.}\ \bibnamefont
  {Imbrie}},\ }\bibfield  {title} {\bibinfo {title} {On many-body localization
  for quantum spin chains},\ }\href {https://doi.org/10.1007/s10955-016-1508-x}
  {\bibfield  {journal} {\bibinfo  {journal} {Journal of Statistical Physics}\
  }\textbf {\bibinfo {volume} {163}},\ \bibinfo {pages} {998} (\bibinfo {year}
  {2016}{\natexlab{b}})}\BibitemShut {NoStop}%
\bibitem [{\citenamefont {De~Roeck}\ and\ \citenamefont
  {Huveneers}(2017)}]{roeck17}%
  \BibitemOpen
  \bibfield  {author} {\bibinfo {author} {\bibfnamefont {W.}~\bibnamefont
  {De~Roeck}}\ and\ \bibinfo {author} {\bibfnamefont {F.}~\bibnamefont
  {Huveneers}},\ }\bibfield  {title} {\bibinfo {title} {Stability and
  instability towards delocalization in many-body localization systems},\
  }\href@noop {} {\bibfield  {journal} {\bibinfo  {journal} {Phys. Rev. B}\
  }\textbf {\bibinfo {volume} {95}},\ \bibinfo {pages} {155129} (\bibinfo
  {year} {2017})}\BibitemShut {NoStop}%
\bibitem [{\citenamefont {Thiery}\ \emph {et~al.}(2018)\citenamefont {Thiery},
  \citenamefont {Huveneers}, \citenamefont {M\"uller},\ and\ \citenamefont
  {De~Roeck}}]{Thiery2017a}%
  \BibitemOpen
  \bibfield  {author} {\bibinfo {author} {\bibfnamefont {T.}~\bibnamefont
  {Thiery}}, \bibinfo {author} {\bibfnamefont {F.}~\bibnamefont {Huveneers}},
  \bibinfo {author} {\bibfnamefont {M.}~\bibnamefont {M\"uller}},\ and\
  \bibinfo {author} {\bibfnamefont {W.}~\bibnamefont {De~Roeck}},\ }\bibfield
  {title} {\bibinfo {title} {Many-body delocalization as a quantum avalanche},\
  }\href@noop {} {\bibfield  {journal} {\bibinfo  {journal} {Phys. Rev. Lett.}\
  }\textbf {\bibinfo {volume} {121}},\ \bibinfo {pages} {140601} (\bibinfo
  {year} {2018})}\BibitemShut {NoStop}%
\bibitem [{\citenamefont {Gopalakrishnan}\ and\ \citenamefont
  {Huse}(2019{\natexlab{b}})}]{gopalakrishnan2019instability}%
  \BibitemOpen
  \bibfield  {author} {\bibinfo {author} {\bibfnamefont {S.}~\bibnamefont
  {Gopalakrishnan}}\ and\ \bibinfo {author} {\bibfnamefont {D.~A.}\
  \bibnamefont {Huse}},\ }\bibfield  {title} {\bibinfo {title} {Instability of
  many-body localized systems as a phase transition in a nonstandard
  thermodynamic limit},\ }\href@noop {} {\bibfield  {journal} {\bibinfo
  {journal} {Physical Review B}\ }\textbf {\bibinfo {volume} {99}},\ \bibinfo
  {pages} {134305} (\bibinfo {year} {2019}{\natexlab{b}})}\BibitemShut
  {NoStop}%
\bibitem [{\citenamefont {Doggen}\ \emph {et~al.}(2020)\citenamefont {Doggen},
  \citenamefont {Gornyi}, \citenamefont {Mirlin},\ and\ \citenamefont
  {Polyakov}}]{doggen2020slow}%
  \BibitemOpen
  \bibfield  {author} {\bibinfo {author} {\bibfnamefont {E.~V.}\ \bibnamefont
  {Doggen}}, \bibinfo {author} {\bibfnamefont {I.~V.}\ \bibnamefont {Gornyi}},
  \bibinfo {author} {\bibfnamefont {A.~D.}\ \bibnamefont {Mirlin}},\ and\
  \bibinfo {author} {\bibfnamefont {D.~G.}\ \bibnamefont {Polyakov}},\
  }\bibfield  {title} {\bibinfo {title} {Slow many-body delocalization beyond
  one dimension},\ }\href@noop {} {\bibfield  {journal} {\bibinfo  {journal}
  {Physical Review Letters}\ }\textbf {\bibinfo {volume} {125}},\ \bibinfo
  {pages} {155701} (\bibinfo {year} {2020})}\BibitemShut {NoStop}%
\bibitem [{\citenamefont {Luitz}\ \emph {et~al.}(2015)\citenamefont {Luitz},
  \citenamefont {Laflorencie},\ and\ \citenamefont {Alet}}]{luitz2015many}%
  \BibitemOpen
  \bibfield  {author} {\bibinfo {author} {\bibfnamefont {D.~J.}\ \bibnamefont
  {Luitz}}, \bibinfo {author} {\bibfnamefont {N.}~\bibnamefont {Laflorencie}},\
  and\ \bibinfo {author} {\bibfnamefont {F.}~\bibnamefont {Alet}},\ }\bibfield
  {title} {\bibinfo {title} {Many-body localization edge in the random-field
  {H}eisenberg chain},\ }\href@noop {} {\bibfield  {journal} {\bibinfo
  {journal} {Phys. Rev. B}\ }\textbf {\bibinfo {volume} {91}},\ \bibinfo
  {pages} {081103} (\bibinfo {year} {2015})}\BibitemShut {NoStop}%
\bibitem [{\citenamefont {Laflorencie}\ \emph {et~al.}(2020)\citenamefont
  {Laflorencie}, \citenamefont {Lemari\'e},\ and\ \citenamefont
  {Mac\'e}}]{Laflorencie2020chain}%
  \BibitemOpen
  \bibfield  {author} {\bibinfo {author} {\bibfnamefont {N.}~\bibnamefont
  {Laflorencie}}, \bibinfo {author} {\bibfnamefont {G.}~\bibnamefont
  {Lemari\'e}},\ and\ \bibinfo {author} {\bibfnamefont {N.}~\bibnamefont
  {Mac\'e}},\ }\bibfield  {title} {\bibinfo {title} {Chain breaking and
  {Kosterlitz-Thouless} scaling at the many-body localization transition in the
  random-field heisenberg spin chain},\ }\href
  {https://doi.org/10.1103/PhysRevResearch.2.042033} {\bibfield  {journal}
  {\bibinfo  {journal} {Phys. Rev. Res.}\ }\textbf {\bibinfo {volume} {2}},\
  \bibinfo {pages} {042033} (\bibinfo {year} {2020})}\BibitemShut {NoStop}%
\bibitem [{\citenamefont {Doggen}\ \emph {et~al.}(2018)\citenamefont {Doggen},
  \citenamefont {Schindler}, \citenamefont {Tikhonov}, \citenamefont {Mirlin},
  \citenamefont {Neupert}, \citenamefont {Polyakov},\ and\ \citenamefont
  {Gornyi}}]{Doggen2018a}%
  \BibitemOpen
  \bibfield  {author} {\bibinfo {author} {\bibfnamefont {E.~V.~H.}\
  \bibnamefont {Doggen}}, \bibinfo {author} {\bibfnamefont {F.}~\bibnamefont
  {Schindler}}, \bibinfo {author} {\bibfnamefont {K.~S.}\ \bibnamefont
  {Tikhonov}}, \bibinfo {author} {\bibfnamefont {A.~D.}\ \bibnamefont
  {Mirlin}}, \bibinfo {author} {\bibfnamefont {T.}~\bibnamefont {Neupert}},
  \bibinfo {author} {\bibfnamefont {D.~G.}\ \bibnamefont {Polyakov}},\ and\
  \bibinfo {author} {\bibfnamefont {I.~V.}\ \bibnamefont {Gornyi}},\ }\bibfield
   {title} {\bibinfo {title} {Many-body localization and delocalization in
  large quantum chains},\ }\href@noop {} {\bibfield  {journal} {\bibinfo
  {journal} {Phys. Rev. B}\ }\textbf {\bibinfo {volume} {98}},\ \bibinfo
  {pages} {174202} (\bibinfo {year} {2018})}\BibitemShut {NoStop}%
\bibitem [{\citenamefont {Sierant}\ \emph
  {et~al.}(2020{\natexlab{a}})\citenamefont {Sierant}, \citenamefont
  {Lewenstein},\ and\ \citenamefont {Zakrzewski}}]{Sierant2020b}%
  \BibitemOpen
  \bibfield  {author} {\bibinfo {author} {\bibfnamefont {P.}~\bibnamefont
  {Sierant}}, \bibinfo {author} {\bibfnamefont {M.}~\bibnamefont
  {Lewenstein}},\ and\ \bibinfo {author} {\bibfnamefont {J.}~\bibnamefont
  {Zakrzewski}},\ }\bibfield  {title} {\bibinfo {title} {Polynomially filtered
  exact diagonalization approach to many-body localization},\ }\href
  {https://doi.org/10.1103/PhysRevLett.125.156601} {\bibfield  {journal}
  {\bibinfo  {journal} {Phys. Rev. Lett.}\ }\textbf {\bibinfo {volume} {125}},\
  \bibinfo {pages} {156601} (\bibinfo {year} {2020}{\natexlab{a}})}\BibitemShut
  {NoStop}%
\bibitem [{\citenamefont {Weiner}\ \emph {et~al.}(2019)\citenamefont {Weiner},
  \citenamefont {Evers},\ and\ \citenamefont {Bera}}]{Weiner2019slow}%
  \BibitemOpen
  \bibfield  {author} {\bibinfo {author} {\bibfnamefont {F.}~\bibnamefont
  {Weiner}}, \bibinfo {author} {\bibfnamefont {F.}~\bibnamefont {Evers}},\ and\
  \bibinfo {author} {\bibfnamefont {S.}~\bibnamefont {Bera}},\ }\bibfield
  {title} {\bibinfo {title} {Slow dynamics and strong finite-size effects in
  many-body localization with random and quasiperiodic potentials},\ }\href
  {https://doi.org/10.1103/PhysRevB.100.104204} {\bibfield  {journal} {\bibinfo
   {journal} {Phys. Rev. B}\ }\textbf {\bibinfo {volume} {100}},\ \bibinfo
  {pages} {104204} (\bibinfo {year} {2019})}\BibitemShut {NoStop}%
\bibitem [{\citenamefont {Kiefer-Emmanouilidis}\ \emph
  {et~al.}(2020)\citenamefont {Kiefer-Emmanouilidis}, \citenamefont {Unanyan},
  \citenamefont {Fleischhauer},\ and\ \citenamefont {Sirker}}]{Kiefer2020a}%
  \BibitemOpen
  \bibfield  {author} {\bibinfo {author} {\bibfnamefont {M.}~\bibnamefont
  {Kiefer-Emmanouilidis}}, \bibinfo {author} {\bibfnamefont {R.}~\bibnamefont
  {Unanyan}}, \bibinfo {author} {\bibfnamefont {M.}~\bibnamefont
  {Fleischhauer}},\ and\ \bibinfo {author} {\bibfnamefont {J.}~\bibnamefont
  {Sirker}},\ }\bibfield  {title} {\bibinfo {title} {Evidence for unbounded
  growth of the number entropy in many-body localized phases},\ }\href
  {https://doi.org/10.1103/PhysRevLett.124.243601} {\bibfield  {journal}
  {\bibinfo  {journal} {Phys. Rev. Lett.}\ }\textbf {\bibinfo {volume} {124}},\
  \bibinfo {pages} {243601} (\bibinfo {year} {2020})}\BibitemShut {NoStop}%
\bibitem [{\citenamefont {Kiefer-Emmanouilidis}\ \emph
  {et~al.}(2021{\natexlab{a}})\citenamefont {Kiefer-Emmanouilidis},
  \citenamefont {Unanyan}, \citenamefont {Fleischhauer},\ and\ \citenamefont
  {Sirker}}]{Kiefer2020b}%
  \BibitemOpen
  \bibfield  {author} {\bibinfo {author} {\bibfnamefont {M.}~\bibnamefont
  {Kiefer-Emmanouilidis}}, \bibinfo {author} {\bibfnamefont {R.}~\bibnamefont
  {Unanyan}}, \bibinfo {author} {\bibfnamefont {M.}~\bibnamefont
  {Fleischhauer}},\ and\ \bibinfo {author} {\bibfnamefont {J.}~\bibnamefont
  {Sirker}},\ }\bibfield  {title} {\bibinfo {title} {Unlimited growth of
  particle fluctuations in many-body localized phases},\ }\href
  {https://doi.org/https://doi.org/10.1016/j.aop.2021.168481} {\bibfield
  {journal} {\bibinfo  {journal} {Annals of Physics}\ }\textbf {\bibinfo
  {volume} {435}},\ \bibinfo {pages} {168481} (\bibinfo {year}
  {2021}{\natexlab{a}})}\BibitemShut {NoStop}%
\bibitem [{\citenamefont {Kiefer-Emmanouilidis}\ \emph
  {et~al.}(2021{\natexlab{b}})\citenamefont {Kiefer-Emmanouilidis},
  \citenamefont {Unanyan}, \citenamefont {Fleischhauer},\ and\ \citenamefont
  {Sirker}}]{Kiefer2021a}%
  \BibitemOpen
  \bibfield  {author} {\bibinfo {author} {\bibfnamefont {M.}~\bibnamefont
  {Kiefer-Emmanouilidis}}, \bibinfo {author} {\bibfnamefont {R.}~\bibnamefont
  {Unanyan}}, \bibinfo {author} {\bibfnamefont {M.}~\bibnamefont
  {Fleischhauer}},\ and\ \bibinfo {author} {\bibfnamefont {J.}~\bibnamefont
  {Sirker}},\ }\bibfield  {title} {\bibinfo {title} {Slow delocalization of
  particles in many-body localized phases},\ }\href
  {https://doi.org/10.1103/PhysRevB.103.024203} {\bibfield  {journal} {\bibinfo
   {journal} {Phys. Rev. B}\ }\textbf {\bibinfo {volume} {103}},\ \bibinfo
  {pages} {024203} (\bibinfo {year} {2021}{\natexlab{b}})}\BibitemShut
  {NoStop}%
\bibitem [{\citenamefont {Sierant}\ and\ \citenamefont
  {Zakrzewski}(2022)}]{Sierant2022challenges}%
  \BibitemOpen
  \bibfield  {author} {\bibinfo {author} {\bibfnamefont {P.}~\bibnamefont
  {Sierant}}\ and\ \bibinfo {author} {\bibfnamefont {J.}~\bibnamefont
  {Zakrzewski}},\ }\bibfield  {title} {\bibinfo {title} {Challenges to
  observation of many-body localization},\ }\href
  {https://doi.org/10.1103/PhysRevB.105.224203} {\bibfield  {journal} {\bibinfo
   {journal} {Phys. Rev. B}\ }\textbf {\bibinfo {volume} {105}},\ \bibinfo
  {pages} {224203} (\bibinfo {year} {2022})}\BibitemShut {NoStop}%
\bibitem [{\citenamefont {Evers}\ \emph {et~al.}(2023)\citenamefont {Evers},
  \citenamefont {Modak},\ and\ \citenamefont {Bera}}]{Evers2023internal}%
  \BibitemOpen
  \bibfield  {author} {\bibinfo {author} {\bibfnamefont {F.}~\bibnamefont
  {Evers}}, \bibinfo {author} {\bibfnamefont {I.}~\bibnamefont {Modak}},\ and\
  \bibinfo {author} {\bibfnamefont {S.}~\bibnamefont {Bera}},\ }\bibfield
  {title} {\bibinfo {title} {Internal clock of many-body delocalization},\
  }\href {https://doi.org/10.1103/PhysRevB.108.134204} {\bibfield  {journal}
  {\bibinfo  {journal} {Phys. Rev. B}\ }\textbf {\bibinfo {volume} {108}},\
  \bibinfo {pages} {134204} (\bibinfo {year} {2023})}\BibitemShut {NoStop}%
\bibitem [{\citenamefont {Luitz}\ and\ \citenamefont
  {Lev}(2020)}]{Luitz2020absence}%
  \BibitemOpen
  \bibfield  {author} {\bibinfo {author} {\bibfnamefont {D.~J.}\ \bibnamefont
  {Luitz}}\ and\ \bibinfo {author} {\bibfnamefont {Y.~B.}\ \bibnamefont
  {Lev}},\ }\bibfield  {title} {\bibinfo {title} {Absence of slow particle
  transport in the many-body localized phase},\ }\href
  {https://doi.org/10.1103/PhysRevB.102.100202} {\bibfield  {journal} {\bibinfo
   {journal} {Phys. Rev. B}\ }\textbf {\bibinfo {volume} {102}},\ \bibinfo
  {pages} {100202} (\bibinfo {year} {2020})}\BibitemShut {NoStop}%
\bibitem [{\citenamefont {Ghosh}\ and\ \citenamefont {\ifmmode \check{Z}\else
  \v{Z}\fi{}nidari\ifmmode~\check{c}\else
  \v{c}\fi{}}(2022)}]{Ghosh2022resonance}%
  \BibitemOpen
  \bibfield  {author} {\bibinfo {author} {\bibfnamefont {R.}~\bibnamefont
  {Ghosh}}\ and\ \bibinfo {author} {\bibfnamefont {M.}~\bibnamefont {\ifmmode
  \check{Z}\else \v{Z}\fi{}nidari\ifmmode~\check{c}\else \v{c}\fi{}}},\
  }\bibfield  {title} {\bibinfo {title} {Resonance-induced growth of number
  entropy in strongly disordered systems},\ }\href
  {https://doi.org/10.1103/PhysRevB.105.144203} {\bibfield  {journal} {\bibinfo
   {journal} {Phys. Rev. B}\ }\textbf {\bibinfo {volume} {105}},\ \bibinfo
  {pages} {144203} (\bibinfo {year} {2022})}\BibitemShut {NoStop}%
\bibitem [{\citenamefont {Panda}\ \emph {et~al.}(2020)\citenamefont {Panda},
  \citenamefont {Scardicchio}, \citenamefont {Schulz}, \citenamefont {Taylor},\
  and\ \citenamefont {{\v{Z}}nidari{\v{c}}}}]{Panda2020a}%
  \BibitemOpen
  \bibfield  {author} {\bibinfo {author} {\bibfnamefont {R.~K.}\ \bibnamefont
  {Panda}}, \bibinfo {author} {\bibfnamefont {A.}~\bibnamefont {Scardicchio}},
  \bibinfo {author} {\bibfnamefont {M.}~\bibnamefont {Schulz}}, \bibinfo
  {author} {\bibfnamefont {S.~R.}\ \bibnamefont {Taylor}},\ and\ \bibinfo
  {author} {\bibfnamefont {M.}~\bibnamefont {{\v{Z}}nidari{\v{c}}}},\
  }\bibfield  {title} {\bibinfo {title} {Can we study the many-body
  localisation transition?},\ }\href@noop {} {\bibfield  {journal} {\bibinfo
  {journal} {Europhys. Lett.}\ }\textbf {\bibinfo {volume} {128}},\ \bibinfo
  {pages} {67003} (\bibinfo {year} {2020})}\BibitemShut {NoStop}%
\bibitem [{\citenamefont {\ifmmode~\check{S}\else \v{S}\fi{}untajs}\ \emph
  {et~al.}(2020)\citenamefont {\ifmmode~\check{S}\else \v{S}\fi{}untajs},
  \citenamefont {Bon\ifmmode~\check{c}\else \v{c}\fi{}a}, \citenamefont
  {Prosen},\ and\ \citenamefont {Vidmar}}]{Suntajs2020a}%
  \BibitemOpen
  \bibfield  {author} {\bibinfo {author} {\bibfnamefont {J.}~\bibnamefont
  {\ifmmode~\check{S}\else \v{S}\fi{}untajs}}, \bibinfo {author} {\bibfnamefont
  {J.}~\bibnamefont {Bon\ifmmode~\check{c}\else \v{c}\fi{}a}}, \bibinfo
  {author} {\bibfnamefont {T.}~\bibnamefont {Prosen}},\ and\ \bibinfo {author}
  {\bibfnamefont {L.}~\bibnamefont {Vidmar}},\ }\bibfield  {title} {\bibinfo
  {title} {Quantum chaos challenges many-body localization},\ }\href
  {https://doi.org/10.1103/PhysRevE.102.062144} {\bibfield  {journal} {\bibinfo
   {journal} {Phys. Rev. E}\ }\textbf {\bibinfo {volume} {102}},\ \bibinfo
  {pages} {062144} (\bibinfo {year} {2020})}\BibitemShut {NoStop}%
\bibitem [{\citenamefont {Sels}\ and\ \citenamefont
  {Polkovnikov}(2021)}]{Sels2021dynamical}%
  \BibitemOpen
  \bibfield  {author} {\bibinfo {author} {\bibfnamefont {D.}~\bibnamefont
  {Sels}}\ and\ \bibinfo {author} {\bibfnamefont {A.}~\bibnamefont
  {Polkovnikov}},\ }\bibfield  {title} {\bibinfo {title} {Dynamical obstruction
  to localization in a disordered spin chain},\ }\href
  {https://doi.org/10.1103/PhysRevE.104.054105} {\bibfield  {journal} {\bibinfo
   {journal} {Phys. Rev. E}\ }\textbf {\bibinfo {volume} {104}},\ \bibinfo
  {pages} {054105} (\bibinfo {year} {2021})}\BibitemShut {NoStop}%
\bibitem [{\citenamefont {Abanin}\ \emph {et~al.}(2021)\citenamefont {Abanin},
  \citenamefont {Bardarson}, \citenamefont {De~Tomasi}, \citenamefont
  {Gopalakrishnan}, \citenamefont {Khemani}, \citenamefont {Parameswaran},
  \citenamefont {Pollmann}, \citenamefont {Potter}, \citenamefont {Serbyn},\
  and\ \citenamefont {Vasseur}}]{abanin2019distinguishing}%
  \BibitemOpen
  \bibfield  {author} {\bibinfo {author} {\bibfnamefont {D.}~\bibnamefont
  {Abanin}}, \bibinfo {author} {\bibfnamefont {J.}~\bibnamefont {Bardarson}},
  \bibinfo {author} {\bibfnamefont {G.}~\bibnamefont {De~Tomasi}}, \bibinfo
  {author} {\bibfnamefont {S.}~\bibnamefont {Gopalakrishnan}}, \bibinfo
  {author} {\bibfnamefont {V.}~\bibnamefont {Khemani}}, \bibinfo {author}
  {\bibfnamefont {S.}~\bibnamefont {Parameswaran}}, \bibinfo {author}
  {\bibfnamefont {F.}~\bibnamefont {Pollmann}}, \bibinfo {author}
  {\bibfnamefont {A.}~\bibnamefont {Potter}}, \bibinfo {author} {\bibfnamefont
  {M.}~\bibnamefont {Serbyn}},\ and\ \bibinfo {author} {\bibfnamefont
  {R.}~\bibnamefont {Vasseur}},\ }\bibfield  {title} {\bibinfo {title}
  {Distinguishing localization from chaos: challenges in finite-size systems},\
  }\href@noop {} {\bibfield  {journal} {\bibinfo  {journal} {Annals of
  Physics}\ }\textbf {\bibinfo {volume} {427}},\ \bibinfo {pages} {168415}
  (\bibinfo {year} {2021})}\BibitemShut {NoStop}%
\bibitem [{\citenamefont {Sierant}\ \emph
  {et~al.}(2020{\natexlab{b}})\citenamefont {Sierant}, \citenamefont
  {Delande},\ and\ \citenamefont {Zakrzewski}}]{Sierant2020a}%
  \BibitemOpen
  \bibfield  {author} {\bibinfo {author} {\bibfnamefont {P.}~\bibnamefont
  {Sierant}}, \bibinfo {author} {\bibfnamefont {D.}~\bibnamefont {Delande}},\
  and\ \bibinfo {author} {\bibfnamefont {J.}~\bibnamefont {Zakrzewski}},\
  }\bibfield  {title} {\bibinfo {title} {Thouless time analysis of {A}nderson
  and many-body localization transitions},\ }\href
  {https://doi.org/10.1103/PhysRevLett.124.186601} {\bibfield  {journal}
  {\bibinfo  {journal} {Phys. Rev. Lett.}\ }\textbf {\bibinfo {volume} {124}},\
  \bibinfo {pages} {186601} (\bibinfo {year} {2020}{\natexlab{b}})}\BibitemShut
  {NoStop}%
\bibitem [{\citenamefont {Morningstar}\ \emph {et~al.}(2022)\citenamefont
  {Morningstar}, \citenamefont {Colmenarez}, \citenamefont {Khemani},
  \citenamefont {Luitz},\ and\ \citenamefont
  {Huse}}]{Morningstar2022avalanches}%
  \BibitemOpen
  \bibfield  {author} {\bibinfo {author} {\bibfnamefont {A.}~\bibnamefont
  {Morningstar}}, \bibinfo {author} {\bibfnamefont {L.}~\bibnamefont
  {Colmenarez}}, \bibinfo {author} {\bibfnamefont {V.}~\bibnamefont {Khemani}},
  \bibinfo {author} {\bibfnamefont {D.~J.}\ \bibnamefont {Luitz}},\ and\
  \bibinfo {author} {\bibfnamefont {D.~A.}\ \bibnamefont {Huse}},\ }\bibfield
  {title} {\bibinfo {title} {Avalanches and many-body resonances in many-body
  localized systems},\ }\href {https://doi.org/10.1103/PhysRevB.105.174205}
  {\bibfield  {journal} {\bibinfo  {journal} {Phys. Rev. B}\ }\textbf {\bibinfo
  {volume} {105}},\ \bibinfo {pages} {174205} (\bibinfo {year}
  {2022})}\BibitemShut {NoStop}%
\bibitem [{\citenamefont {Long}\ \emph {et~al.}(2023)\citenamefont {Long},
  \citenamefont {Crowley}, \citenamefont {Khemani},\ and\ \citenamefont
  {Chandran}}]{long2023phenomenology}%
  \BibitemOpen
  \bibfield  {author} {\bibinfo {author} {\bibfnamefont {D.~M.}\ \bibnamefont
  {Long}}, \bibinfo {author} {\bibfnamefont {P.~J.~D.}\ \bibnamefont
  {Crowley}}, \bibinfo {author} {\bibfnamefont {V.}~\bibnamefont {Khemani}},\
  and\ \bibinfo {author} {\bibfnamefont {A.}~\bibnamefont {Chandran}},\
  }\bibfield  {title} {\bibinfo {title} {Phenomenology of the prethermal
  many-body localized regime},\ }\href
  {https://doi.org/10.1103/PhysRevLett.131.106301} {\bibfield  {journal}
  {\bibinfo  {journal} {Phys. Rev. Lett.}\ }\textbf {\bibinfo {volume} {131}},\
  \bibinfo {pages} {106301} (\bibinfo {year} {2023})}\BibitemShut {NoStop}%
\bibitem [{\citenamefont {Chávez}\ \emph {et~al.}(2023)\citenamefont
  {Chávez}, \citenamefont {Artiaco}, \citenamefont {Kvorning}, \citenamefont
  {Herviou},\ and\ \citenamefont {Bardarson}}]{chavez2023ultraslow}%
  \BibitemOpen
  \bibfield  {author} {\bibinfo {author} {\bibfnamefont {D.~A.}\ \bibnamefont
  {Chávez}}, \bibinfo {author} {\bibfnamefont {C.}~\bibnamefont {Artiaco}},
  \bibinfo {author} {\bibfnamefont {T.~K.}\ \bibnamefont {Kvorning}}, \bibinfo
  {author} {\bibfnamefont {L.}~\bibnamefont {Herviou}},\ and\ \bibinfo {author}
  {\bibfnamefont {J.~H.}\ \bibnamefont {Bardarson}},\ }\href@noop {} {\bibinfo
  {title} {Ultraslow growth of number entropy in an l-bit model of many-body
  localization}} (\bibinfo {year} {2023}),\ \Eprint
  {https://arxiv.org/abs/2312.13420} {arXiv:2312.13420 [cond-mat.dis-nn]}
  \BibitemShut {NoStop}%
\bibitem [{\citenamefont {Biroli}\ \emph {et~al.}(2023)\citenamefont {Biroli},
  \citenamefont {Hartmann},\ and\ \citenamefont
  {Tarzia}}]{biroli2023largedeviation}%
  \BibitemOpen
  \bibfield  {author} {\bibinfo {author} {\bibfnamefont {G.}~\bibnamefont
  {Biroli}}, \bibinfo {author} {\bibfnamefont {A.~K.}\ \bibnamefont
  {Hartmann}},\ and\ \bibinfo {author} {\bibfnamefont {M.}~\bibnamefont
  {Tarzia}},\ }\href@noop {} {\bibinfo {title} {Large-deviation analysis of
  rare resonances for the many-body localization transition}} (\bibinfo {year}
  {2023}),\ \Eprint {https://arxiv.org/abs/2312.14873} {arXiv:2312.14873
  [cond-mat.dis-nn]} \BibitemShut {NoStop}%
\bibitem [{\citenamefont {Sierant}\ \emph
  {et~al.}(2023{\natexlab{b}})\citenamefont {Sierant}, \citenamefont
  {Lewenstein}, \citenamefont {Scardicchio},\ and\ \citenamefont
  {Zakrzewski}}]{Sierant2023stability}%
  \BibitemOpen
  \bibfield  {author} {\bibinfo {author} {\bibfnamefont {P.}~\bibnamefont
  {Sierant}}, \bibinfo {author} {\bibfnamefont {M.}~\bibnamefont {Lewenstein}},
  \bibinfo {author} {\bibfnamefont {A.}~\bibnamefont {Scardicchio}},\ and\
  \bibinfo {author} {\bibfnamefont {J.}~\bibnamefont {Zakrzewski}},\ }\bibfield
   {title} {\bibinfo {title} {Stability of many-body localization in {Floquet}
  systems},\ }\href {https://doi.org/10.1103/PhysRevB.107.115132} {\bibfield
  {journal} {\bibinfo  {journal} {Phys. Rev. B}\ }\textbf {\bibinfo {volume}
  {107}},\ \bibinfo {pages} {115132} (\bibinfo {year}
  {2023}{\natexlab{b}})}\BibitemShut {NoStop}%
\bibitem [{\citenamefont {Oganesyan}\ and\ \citenamefont
  {Huse}(2007)}]{Oganesyan2007}%
  \BibitemOpen
  \bibfield  {author} {\bibinfo {author} {\bibfnamefont {V.}~\bibnamefont
  {Oganesyan}}\ and\ \bibinfo {author} {\bibfnamefont {D.~A.}\ \bibnamefont
  {Huse}},\ }\bibfield  {title} {\bibinfo {title} {Localization of interacting
  fermions at high temperature},\ }\href
  {https://doi.org/10.1103/PhysRevB.75.155111} {\bibfield  {journal} {\bibinfo
  {journal} {Phys. Rev. B}\ }\textbf {\bibinfo {volume} {75}},\ \bibinfo
  {pages} {155111} (\bibinfo {year} {2007})}\BibitemShut {NoStop}%
\bibitem [{\citenamefont {Pal}\ and\ \citenamefont {Huse}(2010)}]{Pal2010a}%
  \BibitemOpen
  \bibfield  {author} {\bibinfo {author} {\bibfnamefont {A.}~\bibnamefont
  {Pal}}\ and\ \bibinfo {author} {\bibfnamefont {D.~A.}\ \bibnamefont {Huse}},\
  }\bibfield  {title} {\bibinfo {title} {Many-body localization phase
  transition},\ }\href {https://doi.org/10.1103/PhysRevB.82.174411} {\bibfield
  {journal} {\bibinfo  {journal} {Phys. Rev. B}\ }\textbf {\bibinfo {volume}
  {82}},\ \bibinfo {pages} {174411} (\bibinfo {year} {2010})}\BibitemShut
  {NoStop}%
\bibitem [{\citenamefont {Giraud}\ \emph {et~al.}(2022)\citenamefont {Giraud},
  \citenamefont {Mac\'e}, \citenamefont {Vernier},\ and\ \citenamefont
  {Alet}}]{giraud2022probing}%
  \BibitemOpen
  \bibfield  {author} {\bibinfo {author} {\bibfnamefont {O.}~\bibnamefont
  {Giraud}}, \bibinfo {author} {\bibfnamefont {N.}~\bibnamefont {Mac\'e}},
  \bibinfo {author} {\bibfnamefont {E.}~\bibnamefont {Vernier}},\ and\ \bibinfo
  {author} {\bibfnamefont {F.}~\bibnamefont {Alet}},\ }\bibfield  {title}
  {\bibinfo {title} {Probing symmetries of quantum many-body systems through
  gap ratio statistics},\ }\href {https://doi.org/10.1103/PhysRevX.12.011006}
  {\bibfield  {journal} {\bibinfo  {journal} {Phys. Rev. X}\ }\textbf {\bibinfo
  {volume} {12}},\ \bibinfo {pages} {011006} (\bibinfo {year}
  {2022})}\BibitemShut {NoStop}%
\bibitem [{\citenamefont {Atas}\ \emph {et~al.}(2013)\citenamefont {Atas},
  \citenamefont {Bogomolny}, \citenamefont {Giraud},\ and\ \citenamefont
  {Roux}}]{Atas2013}%
  \BibitemOpen
  \bibfield  {author} {\bibinfo {author} {\bibfnamefont {Y.~Y.}\ \bibnamefont
  {Atas}}, \bibinfo {author} {\bibfnamefont {E.}~\bibnamefont {Bogomolny}},
  \bibinfo {author} {\bibfnamefont {O.}~\bibnamefont {Giraud}},\ and\ \bibinfo
  {author} {\bibfnamefont {G.}~\bibnamefont {Roux}},\ }\bibfield  {title}
  {\bibinfo {title} {Distribution of the ratio of consecutive level spacings in
  random matrix ensembles},\ }\href
  {https://doi.org/10.1103/PhysRevLett.110.084101} {\bibfield  {journal}
  {\bibinfo  {journal} {Phys. Rev. Lett.}\ }\textbf {\bibinfo {volume} {110}},\
  \bibinfo {pages} {084101} (\bibinfo {year} {2013})}\BibitemShut {NoStop}%
\bibitem [{Note3()}]{Note3}%
  \BibitemOpen
  \bibinfo {note} {{\protect \color {black} This definition of the finite-size
  critical disorder allows us to determine $W_c(n)$ numerically in an unbiased
  way, i.e., without using any information on the scaling of $W_c$ with $n$.
  This is essential in models in which $W_c$ exhibits a power-law growth with
  $n$, as in most of the models considered in this paper.}}\BibitemShut {Stop}%
\bibitem [{\citenamefont {Berke}\ \emph {et~al.}(2022)\citenamefont {Berke},
  \citenamefont {Varvelis}, \citenamefont {Trebst}, \citenamefont {Altland},\
  and\ \citenamefont {DiVincenzo}}]{Berke2022transmon}%
  \BibitemOpen
  \bibfield  {author} {\bibinfo {author} {\bibfnamefont {C.}~\bibnamefont
  {Berke}}, \bibinfo {author} {\bibfnamefont {E.}~\bibnamefont {Varvelis}},
  \bibinfo {author} {\bibfnamefont {S.}~\bibnamefont {Trebst}}, \bibinfo
  {author} {\bibfnamefont {A.}~\bibnamefont {Altland}},\ and\ \bibinfo {author}
  {\bibfnamefont {D.~P.}\ \bibnamefont {DiVincenzo}},\ }\bibfield  {title}
  {\bibinfo {title} {Transmon platform for quantum computing challenged by
  chaotic fluctuations},\ }\href@noop {} {\bibfield  {journal} {\bibinfo
  {journal} {Nature Communications}\ }\textbf {\bibinfo {volume} {13}},\
  \bibinfo {pages} {2495} (\bibinfo {year} {2022})}\BibitemShut {NoStop}%
\bibitem [{\citenamefont {Qian}\ \emph {et~al.}(2023)\citenamefont {Qian},
  \citenamefont {Xu}, \citenamefont {Zhao}, \citenamefont {Li},\ and\
  \citenamefont {Liu}}]{qian2023mitigating}%
  \BibitemOpen
  \bibfield  {author} {\bibinfo {author} {\bibfnamefont {P.}~\bibnamefont
  {Qian}}, \bibinfo {author} {\bibfnamefont {H.-Z.}\ \bibnamefont {Xu}},
  \bibinfo {author} {\bibfnamefont {P.}~\bibnamefont {Zhao}}, \bibinfo {author}
  {\bibfnamefont {X.}~\bibnamefont {Li}},\ and\ \bibinfo {author}
  {\bibfnamefont {D.~E.}\ \bibnamefont {Liu}},\ }\href@noop {} {\bibinfo
  {title} {Mitigating crosstalk and residual coupling errors in superconducting
  quantum processors using many-body localization}} (\bibinfo {year} {2023}),\
  \Eprint {https://arxiv.org/abs/2310.06618} {arXiv:2310.06618 [quant-ph]}
  \BibitemShut {NoStop}%
\end{thebibliography}%

\end{document}